\documentclass[twocolumn,tighten]{aastex631}

\usepackage{physics,bm}	
\usepackage{chemformula} 
\usepackage[mathscr]{euscript}

\defcitealias{Anders_2022}{Anders22}
\defcitealias{Gaia_Collaboration_2021b}{\textit{Gaia}-EDR3}
\defcitealias{Green_2019}{G19}
\defcitealias{Marshall_2006}{M06}
\defcitealias{Planck-Collaboration_2016}{PC16}
\defcitealias{Lallement_2019}{\textsc{Stilism}}

\received{March 17, 2023}
\accepted{June 5, 2023}

\shorttitle{Interstellar Polarization Survey III}
\shortauthors{Angarita et al.}

\begin{document}

\title{Interstellar Polarization Survey III: \\ Relation Between Optical Polarization and Reddening in the General Interstellar Medium}

\author[0000-0001-5016-5645]{Y. Angarita}
\affiliation{Department of Astrophysics/IMAPP, Radboud University, PO Box 9010, 6500 GL Nijmegen, The Netherlands}

\author[0000-0003-0400-8846]{M.J.F. Versteeg}
\affiliation{Department of Astrophysics/IMAPP, Radboud University, PO Box 9010, 6500 GL Nijmegen, The Netherlands}

\author[0000-0002-5288-312X]{M. Haverkorn}
\affiliation{Department of Astrophysics/IMAPP, Radboud University, PO Box 9010, 6500 GL Nijmegen, The Netherlands}

\author[0000-0002-9459-043X]{C.V. Rodrigues}
\affiliation{Divis\~ao de Astrof\'isica, Instituto Nacional de Pesquisas Espaciais (INPE/MCTI), Av. dos Astronautas, 1758, S\~ao Jos\'e dos Campos, SP, Brazil}

\author[0000-0002-1580-0583]{A.M. Magalh\~aes}
\affiliation{Instituto de Astronomia, Geof\'isica e Ci\^encias Atmosf\'ericas, Universidade de S\~ao Paulo, R. do Mat\~ao, 1226, S\~ao Paulo, SP 05508-090, Brazil}

\author[0000-0001-6880-4468]{R. Santos-Lima }
\affiliation{Instituto de Astronomia, Geof\'isica e Ci\^encias Atmosf\'ericas, Universidade de S\~ao Paulo, R. do Mat\~ao, 1226, S\~ao Paulo, SP 05508-090, Brazil}

\author[0000-0001-6099-9539]{Koji S. Kawabata}
\affiliation{Hiroshima Astrophysical Science Center, Hiroshima University, Kagamiyama, Higashi-Hiroshima, Hiroshima, 739-8526, Japan}

\begin{abstract}
    Optical starlight can be partially polarized while propagating through the dusty, magnetized interstellar medium. The polarization efficiency describes the polarization intensity fraction per reddening unit, \mbox{P$_V$/E(\bv)}, related to the interstellar dust grains and magnetic field properties. The maximum value observed, \mbox{[P$_V$/E(\bv)]$_{\mathrm{max}}$}, is thus achieved under optimal polarizing conditions of the interstellar medium. Therefore, the analysis of polarization efficiency observations across the Galaxy contributes to the study of magnetic field topology, small-scale magnetic fluctuations, grain-alignment efficiency, and composition. Infrared observations from \textit{Planck} satellite have set \mbox{[P$_V$/E(\bv)]$_{\mathrm{max}}$} to \mbox{13$\%$~mag$^{-1}$}. However, recent optical polarization observations in \textit{Planck}'s highly polarized regions showed polarization efficiency values between \mbox{13.6$\%$~mag$^{-1}$} and \mbox{18.2$\%$~mag$^{-1}$} (depending on the extinction map used), indicating that \mbox{[P$_V$/E(\bv)]$_{\mathrm{max}}$} is not well constrained yet. We used \textit{V}-band polarimetry of the Interstellar Polarization Survey (consisting of $\sim$10500 high-quality observations distributed in 34 fields of \mbox{$0.3\degr\times0.3\degr$}) to accurately estimate the polarization efficiency in the interstellar medium. We estimated the upper limit of \mbox{P$_V$/E(\bv)} with the weighted $99th$ percentile of the field. In five regions, the polarization efficiency upper limit is above \mbox{13$\%$~mag$^{-1}$}. Furthermore, we found \mbox{[P$_V$/E(\bv)]$_{\mathrm{max}} = 15.8^{+1.3}_{-0.9}$\%~mag$^{-1}$} using diffuse intermediate latitude ($|b|>7.5\degr$) regions with apparently strong regular Galactic magnetic field in the plane-of-sky. We studied the variations of \mbox{P$_V$/E(\bv)} across the sky and tested toy models of polarization efficiency with Galactic longitude that showed some correspondence with a uniform spiral magnetic field. 
\end{abstract}


\section{Introduction} \label{sec:intro}

    The dichroic extinction of starlight from elongated dust grains aligned with the Galactic magnetic field (GMF) causes interstellar optical linear polarization parallel to the rotation axis of the grains. The optical polarization is, therefore, tied not only to the properties of the GMF but also to the properties of the dust grains in the interstellar medium (ISM). Together, optical polarization and dust extinction are essential for studying the plane-of-sky component of the GMF at small and large scales, in addition to other properties of the ISM, such as grain alignment and composition \citep[see e.g.~reviews of][]{Beck_2013,Andersson_2015}. 

    The ratio between \mbox{\textit{V}-band} polarization fraction and the color excess, \mbox{P$_V$/E(\bv)}, is known as polarization efficiency. The distribution of \mbox{P$_V$/E(\bv)} may contain relevant information about mechanisms that reduce the polarization efficiency, such as magnetohydrodynamic (MHD) turbulence, which is known to permeate the different phases of the ISM \citep{Beresnyak_2019}. The maximum polarization efficiency, in turn, represents the highest degree of polarization possible for a given color excess, which might be achieved under optimal polarizing conditions of the ISM, i.e.~elongated grains, full grain alignment, and uniform magnetic field with lines oriented parallel to the plane-of-sky.
    
    \cite{Serkowski_1975} proposed one of the first empirical estimates of the maximum polarization efficiency as \mbox{P$_V$/E(\bv) $\leq$ 9\%~mag$^{-1}$}, using the ratio of total to selective extinction \mbox{R$_{\lambda_{max}}$ = 3.0} and the maximum polarization fraction at $\lambda_{max}$, the wavelength of maximum polarization. Many years later, \cite{Fosalba_2002} found a power-law, \mbox{P$_V$ = 3.5 E(\bv)$^{0.8}$}, that describes the median trend of polarization with reddening for the whole sky using the \citet{Heiles_2000} optical polarimetry database. Recently, \cite{Planck-Collaboration_2018_20} released thermal dust polarization maps and established a new indirect measurement of polarization efficiency using highly polarized regions in diffuse ISM at high latitudes. They found \mbox{P$_V$/E(\bv)=13\%~mag$^{-1}$}, where the polarization fraction in absorption, P$_V$, is obtained from the correlation with the submillimeter polarization in emission and the so-called  \textit{polarization ratios} \citep{Planck-Collaboration_2015}. They assumed a ratio of total to selective extinction of \mbox{R$_V$ = 3.1} \citep{Fitzpatrick_2004} and used the optical depth and extinction relation \mbox{$\tau_V =$ A$_V$/1.086}. Furthermore, \cite{Panopoulou_2019} did an optical polarimetric analysis (in the \textit{R}-band) in some highly polarized regions of \textit{Planck} maps. They found \mbox{P$_V$/E(\bv)} values between \mbox{13.6\%~mag$^{-1}$} and \mbox{18.2\%~mag$^{-1}$} depending on the extinction map used and proposed the former as the lower limit of the maximum polarization efficiency. The determination of the true maximum value is, therefore, highly dependent on the systematic uncertainties of the dust map chosen. There is then a large uncertainty in the maximum polarization efficiency value for the general ISM.

    Optical starlight polarization allows studying the ISM properties at a higher spatial resolution (pc and sub-pc scales) compared to \textit{Planck}'s submillimeter observations (with $\sim5\arcmin$ resolution at intermediate latitude, \citealt{Planck-Collaboration_2015}). The Interstellar Polarization Survey (IPS), (\citealt{Magalhaes_2005}; 2023, in preparation), is a pilot survey, already completed, that illustrates the potential of future high-resolution optical polarimetry surveys such as \mbox{SOUTHPOL} \citep{Magalhaes_2012} and \mbox{PASIPHAE} \citep{Tassis_2018}. In combination with accurate distances to the stars from \citet[hereafter \citetalias{Gaia_Collaboration_2021b}]{Gaia_Collaboration_2021b}, the IPS catalog in the general ISM, IPS-GI \citep{Versteeg_2023}, will let us reconstruct the information from the magnetic field and absorption structures along the line of sight. 
    
    We use high-quality polarimetry and photometry measurements of $\sim10500$ stars of the IPS-GI catalog \citep{Versteeg_2023}, distributed in 34 high spatial density regions, to study the GMF and dust grain properties through the \mbox{P$_V$/E(\bv)} relation under the same ISM conditions in each field observed. Contrary, \cite{Fosalba_2002} and \cite{Planck-Collaboration_2018_20} calculated the maximum polarization efficiency using all sorts of lines-of-sight (LOS) across the entire sky. The observations and data are described in detail in Section~\ref{sec:Observ_data}. In Section~\ref{sec:Ebv}, we describe the \textit{V}-band extinction of \citet[hereafter \citetalias{Anders_2022}]{Anders_2022}, used to obtain the reddening and the polarization efficiency. We also compare \citetalias{Anders_2022} reddening with other measurements available. In Section~\ref{sec:P_vs_E} we show the relation between the degree of polarization and the reddening under different ISM conditions, as well as the method to calculate its upper envelope. Our results on the \mbox{P$_V$/E(\bv)} upper limit calculation are presented in Section~\ref{sec:Results}. Furthermore, we discuss the variations of the polarization efficiency observed across the Southern sky, the possible explanation with toy models, the dependency on the dust map used, and the correlation with the dispersion of polarization angle in Section~\ref{sec:Discu}. Finally, we summarize our findings in Section~\ref{sec:Conclu}.

\section{Polarimetric Observations and Data} 
    \label{sec:Observ_data}

    In this section, we describe the data acquisition and reduction, the cross-match with auxiliary databases, and the selection of high-quality objects to study polarization efficiency.

    \subsection{Interstellar Polarization Survey}
    \label{subsec:IPs_Survey}
    
        One of the scientific aims of the Interstellar Polarization Survey (IPS) is to improve our knowledge of the magnetic field structure in the general ISM in relation to its dust properties \citep{Magalhaes_2005}. For that purpose, different types of sources such as open clusters, nearby dark clouds, and the general ISM were observed between 2000 and 2003 with the IAGPOL polarimeter developed at the \textit{Instituto de Astronomia, Geof\'isica e Ci\^encias Atmosf\'ericas} of the \textit{Universidade de S\~ao Paulo}  (IAG-USP), Brazil \citep{Magalhaes_2005}. This  optical/NIR imaging polarimeter \citep[see][]{Magalhaes_1996} is equipped with a Savart prism and a rotating half-wave plate and is mounted ahead of a CCD detector. The assembly was installed onto the Cassegrain focus of the IAG Boller \& Chivens 61 cm telescope at the Observat\'orio do Pico dos Dias (OPD) in Brazil. The Savart prism forms the extraordinary and ordinary beams, i.e.~two perpendicularly polarized images of every source in the field-of-view. Such a dual-beam polarimeter 
        allows to cancel out the sky polarization through the superposition of the extraordinary and ordinary images. Additionally, the differential photometry technique used allows observations even in non-photometric conditions. Further information about the IPS project, the polarimetry, and observations can be found in (\citealt{Magalhaes_2005}; 2023, in preparation; \citealt{Versteeg_2023}).

    \subsection{General ISM data}
    \label{subsec:IPS-GI}
     
        Our research focuses on the study of the general ISM fields observed in the IPS project (IPS-GI). A new photometric and polarimetric catalog with precise measurements of the degree of polarization (P), polarization angle ($\theta$), and magnitudes in the \textit{V}-band is presented in detail by \cite{Versteeg_2023}. The new stellar catalog contains data from 38 fields of approximately $0.3\degr \times 0.3\degr$ in size, carefully chosen in different locations near and within the Galactic disk in the Southern sky.

    \subsection{Data Reduction}
    \label{subsec:Data_reduction}
        
        Polarimetric and photometric information was obtained with the reduction pipeline SOLVEPOL developed by \cite{Ramirez_2017}. The algorithm, written in Interactive Data Language (IDL), calculates the Stokes parameters Q and U from the modulation of the intensity in terms of the ordinary and extraordinary fluxes. In this process, Q and U are normalized by the intensity, i.e.~the I Stokes parameter. The degree of linear polarization and polarization angle are calculated as:
        \begin{equation}
            P = \sqrt{Q^2 + U^2} ,
            \label{eq:P}
        \end{equation} 
        \begin{equation}
            \theta = \frac{1}{2}\tan^{-1}\frac{U}{Q} ,
            \label{eq:theta}
        \end{equation}
        Following \cite{Magalhaes_1984} and \cite{Naghizadeh_Khouei_1993}, \mbox{$\sigma_U\approx\sigma_Q\approx\sigma_P$} is assumed. The uncertainties are calculated as in \citet{Ramirez_2017}:
        %
        \begin{equation}
            \sigma_P = \frac{1}{\sqrt{\mu - 2}}\sqrt{\frac{2}{\mu} \sum_{i}^{\mu}{z_i^2 - Q^2 - U^2}} ,
            \label{eq:dp}
        \end{equation}
        \begin{equation}
            \sigma_{\theta} = 28.65\degr\frac{\sigma_P}{P} ,
            \label{eq:dtheta}
        \end{equation}
        where $\mu$ is the number of positions of the half-wave plate and $z_i$ is the ratio between the difference and the sum of the ordinary and extraordinary beams of the $i$-th orientation of the half-wave plate \citep[see][]{Ramirez_2017}.
        
        The degree of polarization measured is affected by Ricean bias, especially at low signal-to-noise ratios ($P/\sigma_P\leq3$, according to \citealt{Clarke_Stewart_1986}). The SOLVEPOL pipeline itself purposely does not correct for the bias. However, following \citet{Ramirez_2017} recommendations, we used only high signal-to-noise measurements with $P/\sigma_P>5$ which are not significantly affected by the bias (\citealt{Simmons&Stewart_1985}, also see the quality filters in Section~\ref{subsec:Qflags}). Furthermore, since the average instrumental polarization is $0.07\%$ \citep[see][]{Versteeg_2023}, far below the median polarization error, there is no need to correct instrumental errors. 

    \subsection{Auxiliary data}
    \label{subsec:ancillary_Data}
        
        We used the IPS-GI catalog \citep{Versteeg_2023} containing $G$, $G_{BP}$, $G_{RP}$, and \textit{V} bands  parameters from the \citetalias{Gaia_Collaboration_2021b} and \citetalias{Anders_2022} catalogs. The latter used \citetalias{Gaia_Collaboration_2021b} parallaxes and photometry 
        to estimate accurate distances and \textit{V}-band extinction of millions of stars. Firstly, around 44000 IPS-GI stars with photometric and polarimetric measurements in the \textit{V}-band were cross-matched with the \citetalias{Gaia_Collaboration_2021b} database using a cone search with 3$''$ and 2~mag margins in \textsc{Topcat} \citep{Taylor_2005}. Subsequently, we used \citetalias{Gaia_Collaboration_2021b}'s \textit{source ID} parameter of the successful matches to cross-match with \citetalias{Anders_2022} database. Finally, we kept approximately $36000$ stars from the original IPS-GI catalog that have measurements of distance and \textit{V}-band extinction.
       
    \subsection{Quality filters}
    \label{subsec:Qflags}
    
        In order to use the IPS-GI data together with known parameters from \citetalias{Anders_2022} and \citetalias{Gaia_Collaboration_2021b} databases, certain filters must be defined to ensure a high quality of the final data set and the reliability of its parameters. Below we will explain each of the quality filters in order of importance and the number of sources removed or remaining after the flagging.
        
        \subsubsection{Signal-to-noise ratio and polarization uncertainty}
        \label{subsec:Qflags-dp}
        
            The bias of the degree of polarization mainly depends on the signal-to-noise ratios \citep{Clarke_Stewart_1986}. The lower the signal-to-noise, the larger the polarization bias. A signal-to-noise cut of P$_V$/$\sigma_{P_V} >$ 5 was defined through the SOLVEPOL pipeline to eliminate low signal-to-noise measurements. This filter discards $\sim$46\% of the initial sample, ending up with $\sim$16800 sources for which the degree of polarization is still biased. However, \cite{Versteeg_2023} demonstrated that the difference between the biased and debiased polarization fraction after filtering is of the order of 10$^{-3}\%$. Consequently, we see no need to debias the degree of polarization of the filtered sample.
            
            We observed an increase in the degree of polarization and its uncertainty with higher V magnitudes in the IPS-GI data. Although this is expected, an anomalous increasing noise with magnitude may originate in the data reduction process. \cite{Ramirez_2017} reported that the procedure followed by IDL within the SOLVEPOL pipeline to calculate the sky flux can yield a different estimation of the total flux in comparison with other methods (e.g.~IRAF; \citealt{Tody_IRAF1_1986,Tody_IRAF2_1993}). The difference in flux propagates through the pipeline affecting the final polarization and its uncertainty, especially for low-count objects, which are often the faintest stars. Therefore, we only included stars with $\sigma_{P_V} \leq$ 0.8$\%$ to avoid untrustworthy and noisy high polarization values, reducing the sample size to $\sim$15700 stars.
            
        \subsubsection{Fidelity}
        \label{subsec:Qflags-fidelity}
        
            The \textit{fidelity} is a metric defined by \cite{Rybizki_2022} using neural-networks and all astrometric columns available from \citetalias{Gaia_Collaboration_2021b}. It summarizes the goodness of the astrometric solutions. Following \cite{Rybizki_2022} and \citetalias{Anders_2022} recommendations, we use stars with \textit{fidelity} $> 0.5$. This leaves us with $\sim$14700 stars.
            
        \subsubsection{Color excess factor}
        \label{subsec:Qflags-color_excess}
        
            The corrected $G_{BP}-G_{RP}$ flux excess factor, $C^*$, is a correction applied to the ratio between the total $G_{BP}-G_{RP}$ flux and the $G$-band flux. It identifies inconsistencies between $G$, $G_{BP}$, and $G_{RP}$ photometry due to background flux from different sources \citep{Riello_2021}. The \textit{color excess factor} is defined as $|C^*|/\sigma_{C^*}$, where $\sigma_{C^*}$ is a function of $G$ magnitude described in Equation 18 from \cite{Riello_2021}. It might reveal some objects whose photometry measurements are complicated, including variable stars, extended sources, and multiple stars. Still, the \textit{color excess factor} is not intended to identify all of them properly as it cannot establish the difference between data affected by processing problems and variable sources.
            \citetalias{Anders_2022} suggest to use $|C^*|/\sigma_{C^*} < 5$; with this limit, we end up with $\sim$14300 sources.
            
        \subsubsection{Output flag}
        \label{subsec:Qflags-SH_Qflag}
        
            \citetalias{Anders_2022} database includes the \textsc{SH\_OUTFLAG}, an output parameter with 4 digits that denotes the fidelity of the algorithm in calculating the output parameters. The first digit represents the \textit{low number of consistent models}, a digit different from ``0'' means that the number of stellar models is too low (less than 30) to be trusted. The second digit is a flag for significantly negative extinctions in which the accepted values are marked also with a ``0''. The third and fourth digits are for very large and very small uncertainties respectively in all parameters, a ``0'' digit in both cases means acceptable parameters. We require a data set with the best quality possible by choosing stars with \textsc{SH\_OUTFLAG}$ = 0000$, remaining with $\sim$11500 stars after this step.
            
        \subsubsection{Negative extinctions}
        \label{subsec:Qflags-Av}
        
            The Bayesian nature of the \citetalias{Anders_2022} extinction, A$_{V}$, causes some sources to end up with negative values. As mentioned above, \citetalias{Anders_2022} included in their \textsc{SH\_OUTFLAG} parameter a flag for unreliable extinctions, even so, their definition rejects only negative values over $2 \sigma$, i.e.~A$_{V95} < 0$. Therefore, we chose to use only positive median values of the extinction, hereafter A$_{V50}$, with A$_{V50} < 0$, rejecting the only remaining star with negative extinction. 
            
        \subsubsection{Poor IPS-GI fields}
        \label{subsec:Qflags-Fields}
        
            Fields that end up with very few stars ($<$ 15) after applying all previous filters -- i.e.~fields \textit{C34}, \textit{C38}, \textit{C44} and \textit{C53} (see Table~\ref{tab:PE_uplim}) as defined in \cite{Versteeg_2023} -- are not considered since any statistic analysis carried out on these fields would be poor and unreliable. We reject here 32 stars in total. 
            
        \subsubsection{Catalogs match duplicates}
        \label{subsec:Qflags-duplicates}
        
            The number of detections and precision of \citetalias{Gaia_Collaboration_2021b} has increased in the new release. This means that a cross-match with other catalogs becomes more difficult and less reliable in crowded regions. The cross-match process of the IPS-GI catalog with \citetalias{Anders_2022} and \citetalias{Gaia_Collaboration_2021b} databases returned multiple matches for some stars. Despite having used a second matching criterion, the \textit{V}-band magnitude \citep[see][for more details]{Versteeg_2023}, it is difficult to obtain a unique reliable match for the duplicates. An identifier was given in the column \mbox{\textit{GroupSize}} to those groups of stars with multiple matches. We decided not to consider stars with more than one match. This leaves us with $\sim$10500 high-quality stars, spread in 34 IPS-GI fields, that make up our final data set.

    \subsection{Variability of the IPS-GI sources} \label{subsec:Outliers}
               
        Intrinsic polarization originating in the environment around variable sources, e.g.~due to the scattering of radiation from circumstellar dust, may contaminate our sample with information that does not represent the general ISM. Knowing the stellar classification of the IPS-GI stars would help to assess their variability and whether they could have intrinsic polarization. Unfortunately, we do not know the exact stellar type or variability status of all IPS-GI sources; even if we naively look at \textit{Gaia}'s H-R diagram, it would only give us a rough stellar classification. 
        
        One solution to identify variable stars within the IPS-GI sample is to cross-match our catalog with variable stars databases. We considered ASAS-SN \citep{Jayasinghe_2018, Jayasinghe_2019a, Jayasinghe_2019b, Jayasinghe_2020}, ATLAS-VAR \citep{Heinze_2018}, and the \textit{Gaia}-DR3 variable stars catalog \citep{GaiaDR3_Collaboration_2022}. We identified 923 variable stars inside our data set, of which 294 passed all quality filters (Section~\ref{subsec:Qflags}). The variable stars in IPS-GI did not show unusual values in their polarimetry properties (see e.g.~Figure~\ref{fig:dev_P_theta_all_IPS}, left). They are therefore kept in our analysis while acknowledging their existence. The details about the IPS-GI variable stars are in Appendix~\ref{appex:known_varia}.

\section{Reddening} \label{sec:Ebv}
   
    To calculate the polarization to reddening ratio, we need to estimate the reddening, E(\bv), of the IPS-GI sources and its uncertainty. We describe and compare the different dust maps considered in the following sections. We also justify our choice of \citetalias{Anders_2022}'s all-sky database, which includes multi-wavelength photometry, distance, and optical extinction, among other parameters, for millions of stars.

    \subsection{Dust maps considered} \label{subsec:dustmaps}
    
        The most widely used interstellar dust tracer is the extinction. However, few databases have large samples of stars with reliable measurements of \textit{V}-band extinction (or reddening) and distances. Moreover, there are even fewer with information in the Southern sky -- or at least in the regions observed by IPS -- making it hard to find a suitable dust extinction database covering our entire data set. We considered the following dust maps queried through the Python library \textsc{Dustmaps}\footnote{\url{https://dustmaps.readthedocs.io/en/latest/\#}} \citep{Green_dustmaps_2018} using IPS-GI Equatorial coordinates and  \citetalias{Anders_2022}'s distances (for comparison reasons):
        \begin{itemize}
            \item The dust extinction map from \citet[hereafter \citetalias{Marshall_2006}]{Marshall_2006} used the \textit{Besançon} model of the Galaxy \citep{Robin_2003} and the comparison of 2MASS colors \citep{Skrutskie_2MASS_2006} to estimate the distances and the \textit{K}-band extinction along binned beams of $15\arcmin\times15\arcmin$. The uncertainty of the estimates is the mean absolute deviation (MAD) from the median extinction in the bin. We converted the \textit{K}-band extinction and its error to the \textit{V}-band with the relative extinction value A$_K$/A$_V$ = 0.078 \citep[see Table 3 from][]{Wang_2019}.
            \item The Generalized Needlet Internal Linear Combination (GNILC) optical depth map from \citet[hereafter \citetalias{Planck-Collaboration_2016}]{Planck-Collaboration_2016} has a resolution of $\sim5\arcmin$ in the IPS-GI regions. We use the \mbox{E(\bv) = (1.49$\times10^4$~mag)$\tau_{353}$} convention to convert the optical depth ($\tau_{353}$, measured at 353 GHz) to reddening with the respective observed error. The \citetalias{Planck-Collaboration_2016} reddening is then the integrated value throughout the entire Galaxy.
            \item The tridimensional dust map of the local interstellar matter from \cite{Capitanio_2017} and \citet[hereafter \citetalias{Lallement_2019}]{Lallement_2019}\footnote{\url{https://stilism.obspm.fr/}} gives the reddening calculated by Bayesian inversion of individual reddening estimates towards 71000 stars (see \citealt{Lallement_2018} and \citealt{Lallement_2019} for more details). The resolution near the Sun is 25 pc, and the uncertainty is the standard deviation among the nearest extinction density values within the box. 
            \item The Bayesian dust reddening map from \citet[hereafter \citetalias{Green_2019}]{Green_2019} has a resolution on the order of $7\arcmin$. The reddening is reported in arbitrary units. So, we follow the authors' recommendation and use Equations 29 and 30 from \citetalias{Green_2019} to convert the reddening to the \textit{V}-band with their closest definition to \citetalias{Anders_2022}-like data (for comparison reasons). We also used the suggested filter for \textit{reliable distance} and calculate the uncertainties from the posteriors of the reddening samples, i.e.~the \textit{84th} and \textit{16th} percentiles.
            \item The all-sky stellar catalog from \citetalias{Anders_2022} was created with the Bayesian algorithm \textsc{StarHorse} \citep{Queiroz_2018}. The algorithm uses \citetalias{Gaia_Collaboration_2021b} parallaxes and multi-band photometry (cross-matched with different surveys) to estimate the posteriors of the \textit{V}-band extinctions and distances, among other parameters, of millions of stars with high accuracy and precision. Similarly to \citetalias{Green_2019}, the errors of the estimates are calculated from the posteriors. The resolution of \citetalias{Anders_2022}'s optical extinction is based on starlight measurements of point-like sources, i.e.~in the order of $\sim1 \arcsec$.
        \end{itemize}
        
        To convert the \textit{V}-band extinction, A$_V$, to reddening, E(\bv), of any dust map or catalog, we used the total to selective extinction ratio of \mbox{R$_V$ = A$_V$/E(\bv) = 3.1} \citep{Savage_1979,Fitzpatrick_2004}.
        %
        \begin{figure*}[t!]
        \gridline{
                  \fig{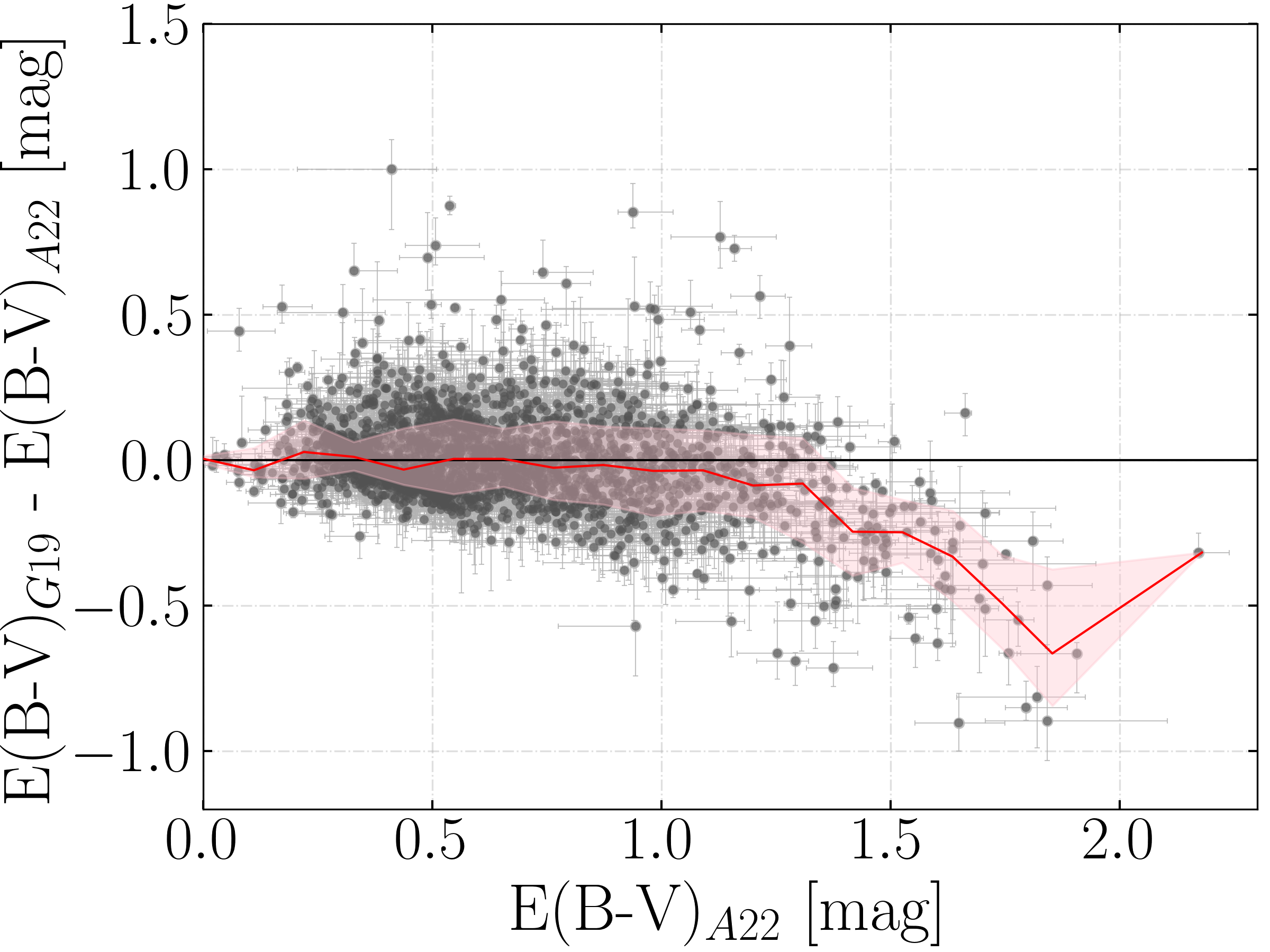}{0.33\linewidth}{}
                  \fig{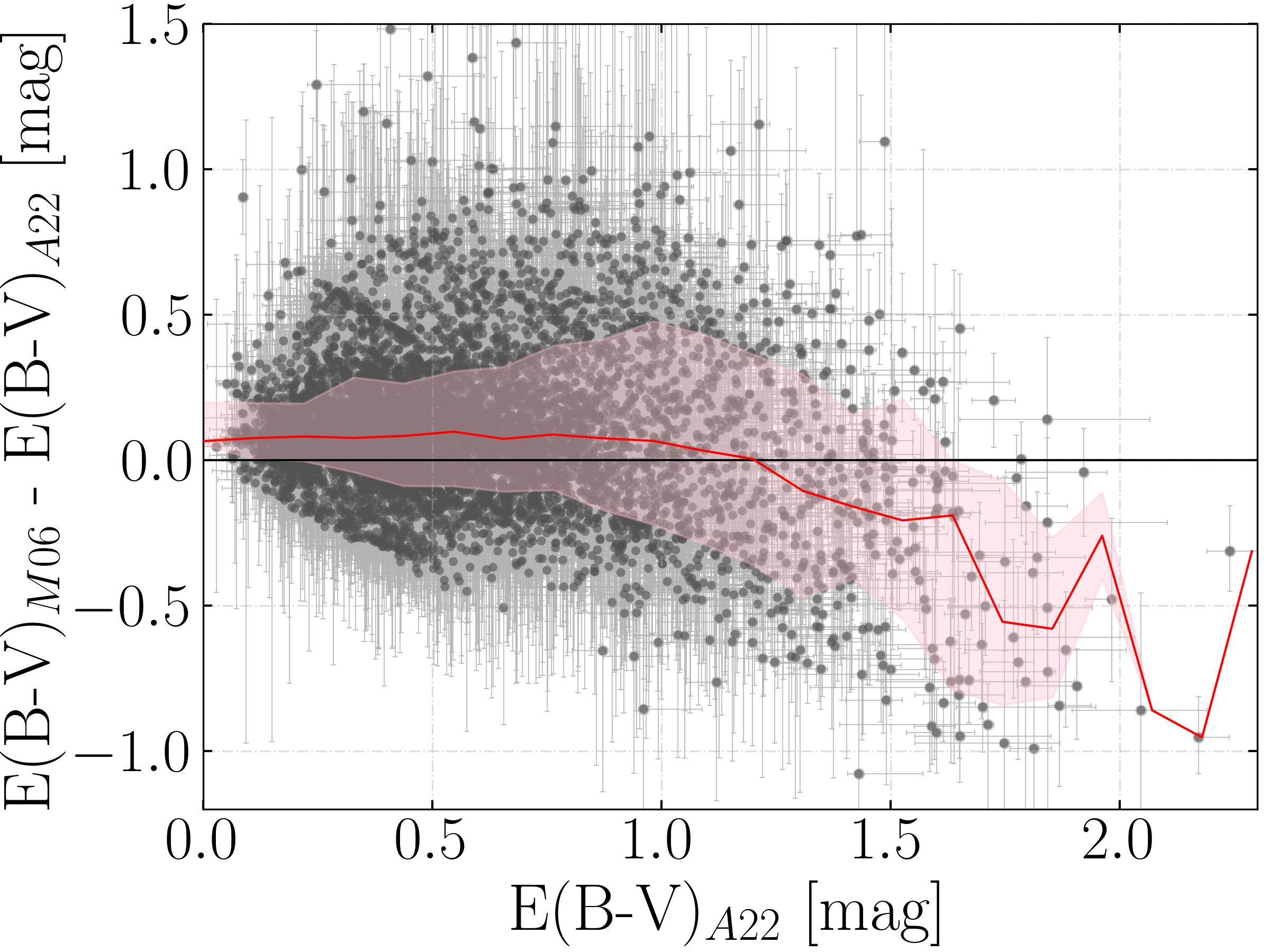}{0.33\linewidth}{}
                  \fig{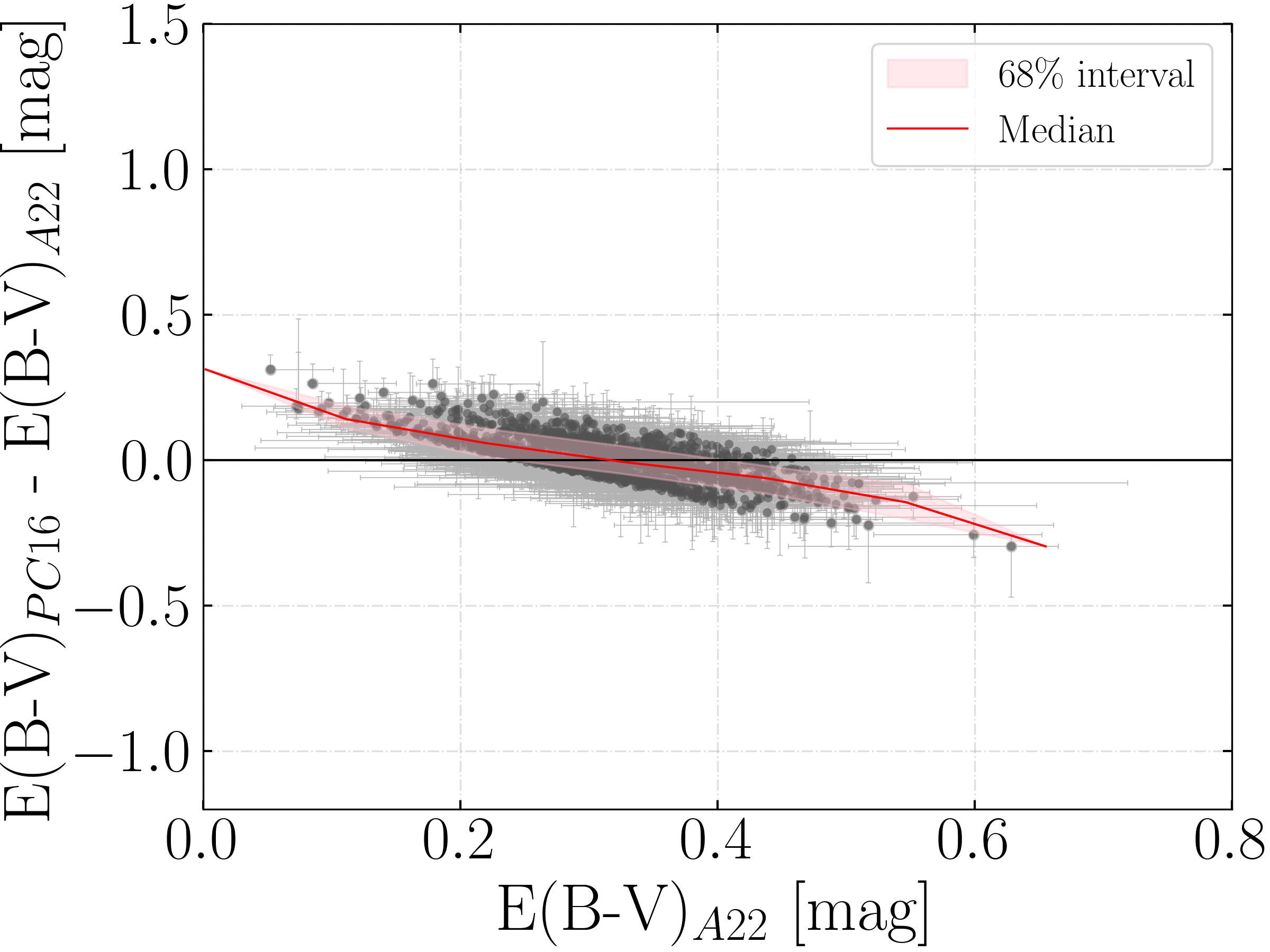}{0.33\linewidth}{}
                  }
        \gridline{
                  \fig{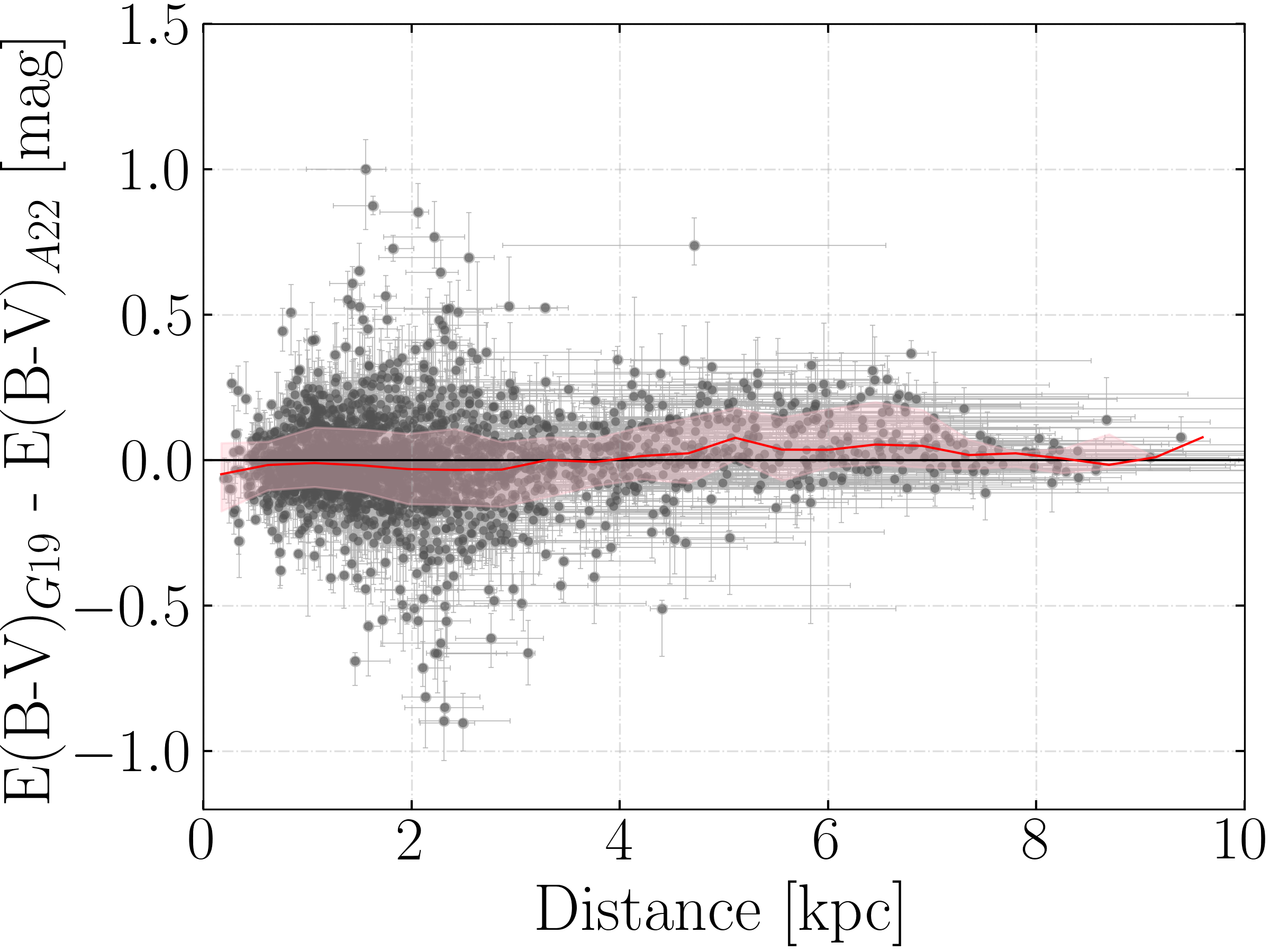}{0.33\linewidth}{}
                  \fig{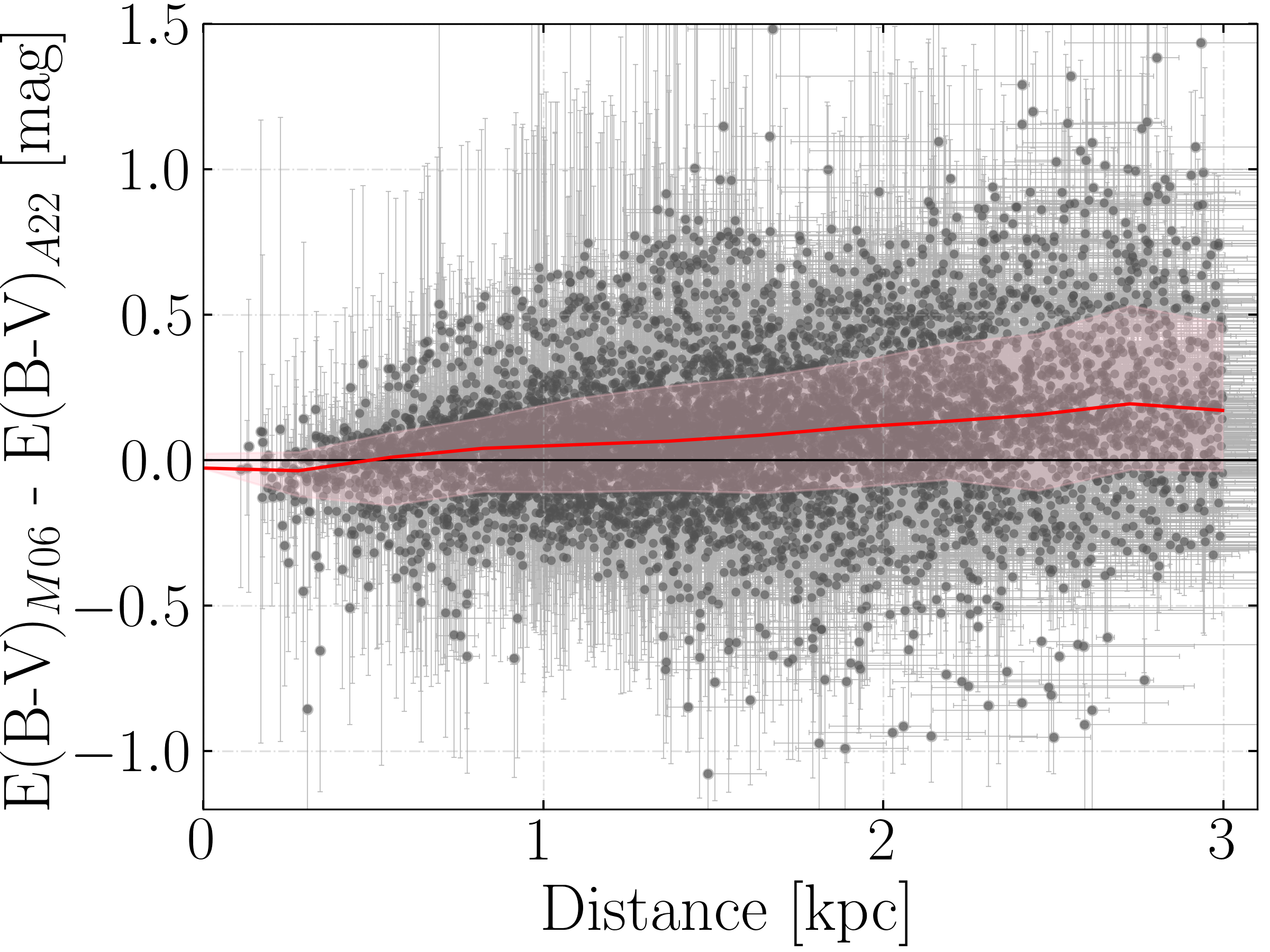}{0.33\linewidth}{}
                  \fig{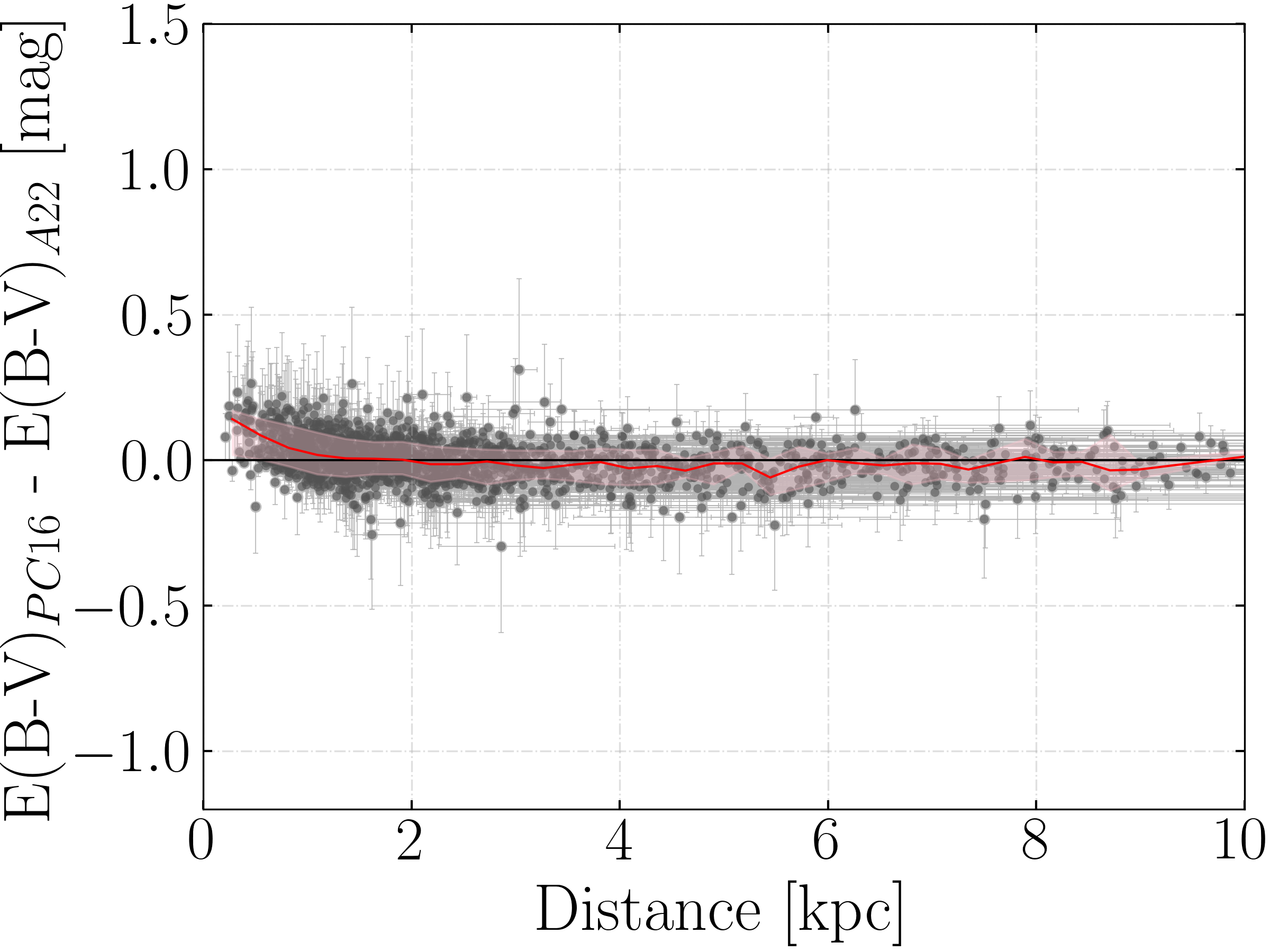}{0.33\linewidth}{}
                  }
        \caption{Systematic reddening difference of \citetalias{Green_2019} (left), \citetalias{Marshall_2006} (middle), and \citetalias{Planck-Collaboration_2016} (right) with \citetalias{Anders_2022} shown as a function of \citetalias{Anders_2022}'s reddening (top row) and distance (bottom row). Only the diffuse fields, i.e.~\textit{C2}, \textit{C16}, \textit{C36}, and \textit{C37}, are used in the \citetalias{Planck-Collaboration_2016} comparison (right) (see Table~\ref{tab:PE_uplim} and the explanation of Section \ref{subsec:Ebv_comp}). Red solid lines represent the median value of the difference, and the pink shaded area is the $\sim1\sigma$ confidence interval delimited by the $16th$ and $84th$ percentiles.
        \label{fig:Green-Stil_vs_SH}}
        \end{figure*}

    \subsection{Dust maps comparison} \label{subsec:Ebv_comp}
        
        The \citetalias{Anders_2022} stellar database has multiple advantages over others due to the characteristics of our data set and the precision of their estimates. For instance, \citetalias{Green_2019} and \citetalias{Lallement_2019} used parallaxes from \citet[][\textit{Gaia}-DR2]{Gaia_Collaboration_2018}, while \citetalias{Anders_2022} used more precise measurements from \citetalias{Gaia_Collaboration_2021b}, making their new results more accurate. \citetalias{Green_2019} lacks information in a large portion of the Southern sky, including many of the IPS-GI regions. However, \citetalias{Green_2019} is one of the dust maps used by \citetalias{Anders_2022} to calculate their extinction prior. Therefore, they should be very consistent. \citetalias{Lallement_2019} and \citetalias{Marshall_2006} have reliable measurements only up to \mbox{$\sim$2-3~kpc} due to the resolution and characteristics of their methods. \citetalias{Marshall_2006} only covers the sky at \mbox{$-10\degr \geq b \geq 10\degr$}. Hence, intermediate Galactic latitude IPS-GI fields ($|b| > 10\degr$) are missing in \citetalias{Marshall_2006}. In addition, the semi-empirical nature of \citetalias{Marshall_2006}'s method makes the results largely dependent on the model parameters. \citetalias{Anders_2022}, on the other hand, covers the entire sky and has accurate distances up to above 3~kpc. Finally, \citetalias{Planck-Collaboration_2016}'s submillimeter observations probe the thermal dust emission of the pathlength throughout the Galaxy, not only up to the star distance as \citetalias{Anders_2022}. Moreover, the small angular size of the IPS-GI fields challenges the \citetalias{Planck-Collaboration_2016} map resolution, i.e.~many sources are assigned equal E(\bv) measurement from a single pixel losing information from small and dense dust structures. Hence, for completeness, we only compare and analyze \citetalias{Planck-Collaboration_2016} reddening in intermediate latitude fields (\textit{C2},  \textit{C16}, \textit{C36}, and \textit{C37}, see Table~\ref{tab:PE_uplim}), where the LOS for IPS-GI and \citetalias{Planck-Collaboration_2016} are assumed to be comparable.

        We computed the systematic difference between the reddening of \citetalias{Green_2019}, \citetalias{Marshall_2006}, and \citetalias{Planck-Collaboration_2016} with the median value of \citetalias{Anders_2022} reddening, E(\bv)$_{A22}$ (see Figure~\ref{fig:Green-Stil_vs_SH}). The error bars in Figure~\ref{fig:Green-Stil_vs_SH} result from the propagation of errors from the measurements. We decided not to consider the \citetalias{Lallement_2019} map for further analysis since the difference between \citetalias{Lallement_2019} and \citetalias{Anders_2022} reddening -- only considering stars below 2.5~kpc -- has systematic errors that increase with E(\bv)$_{A22}$ up to a maximum of $\sim$1.6~mag for highly extinct LOS. Furthermore, \citetalias{Marshall_2006} reddening estimates are only reliable up to a limited distance as the column density increases and sources in the \textit{K}-band no longer have counterparts in the \textit{J}-band. The model presented by \citetalias{Marshall_2006} then has limitations at large distances where the calculations depend on the parameters and assumptions of the authors. We, therefore, decide to take the safe path and use \citetalias{Marshall_2006} reddening estimates only up to a distance of 3~kpc. 
        
        We found that \citetalias{Green_2019} is consistent with \citetalias{Anders_2022} within the 68$\%$ confidence interval (approximately $1 \sigma$ margin error of 0.12~mag), up to E(\bv)$_{A22} \sim$ 1.3~mag (Figure~\ref{fig:Green-Stil_vs_SH}, top-left). The systematic difference scatters mostly below 0.3~mag for the entire IPS-GI sample, with very few stars scaling to a maximum of \mbox{$\sim$1~mag}. Nevertheless, the systematic difference depends on the individual fields, e.g.~we discuss in Section \ref{subsec:Discu_PE_SH22_G19_M06} the systematic reddening error of field \textit{C2} observed between 0.14~mag and 0.16~mag depending on the dust map. 
        
        \citetalias{Marshall_2006}, on the other hand, is consistent with \citetalias{Anders_2022} within $1 \sigma$ margin error of $\sim$0.2~mag up to \mbox{E(\bv)$_{A22}\sim$1.6~mag} (Figure~\ref{fig:Green-Stil_vs_SH}, top-middle). Moreover, \citetalias{Marshall_2006} reddening is often higher at \mbox{E(\bv)$_{A22} \lesssim$ 1.2~mag}, the scatter scales up to 1~mag, and the uncertainties are extremely large. All this results from the limitations of the method explained above.
        
        The difference with \citetalias{Planck-Collaboration_2016} (top-right panel of Figure~\ref{fig:Green-Stil_vs_SH}) shows a systematic error mostly below 0.2~mag throughout the entire range of reddening values. At E(\bv)$_{A22}<$ 0.3~mag, \citetalias{Planck-Collaboration_2016} reddening is higher because it is integrated along longer pathlengths. At \mbox{E(\bv)$_{A22}>$ 0.3~mag}, \citetalias{Planck-Collaboration_2016} reddening is often lower, which in addition to the increasing systematic error found above \mbox{E(\bv)$_{A22} \sim$ 1.3~mag} in the \citetalias{Green_2019} and \citetalias{Marshall_2006} comparison, demonstrates that \citetalias{Anders_2022} may be overestimating the reddening in highly extinct LOS. 
        %
        \begin{figure*}[t!]
            \epsscale{0.9}
            \plottwo{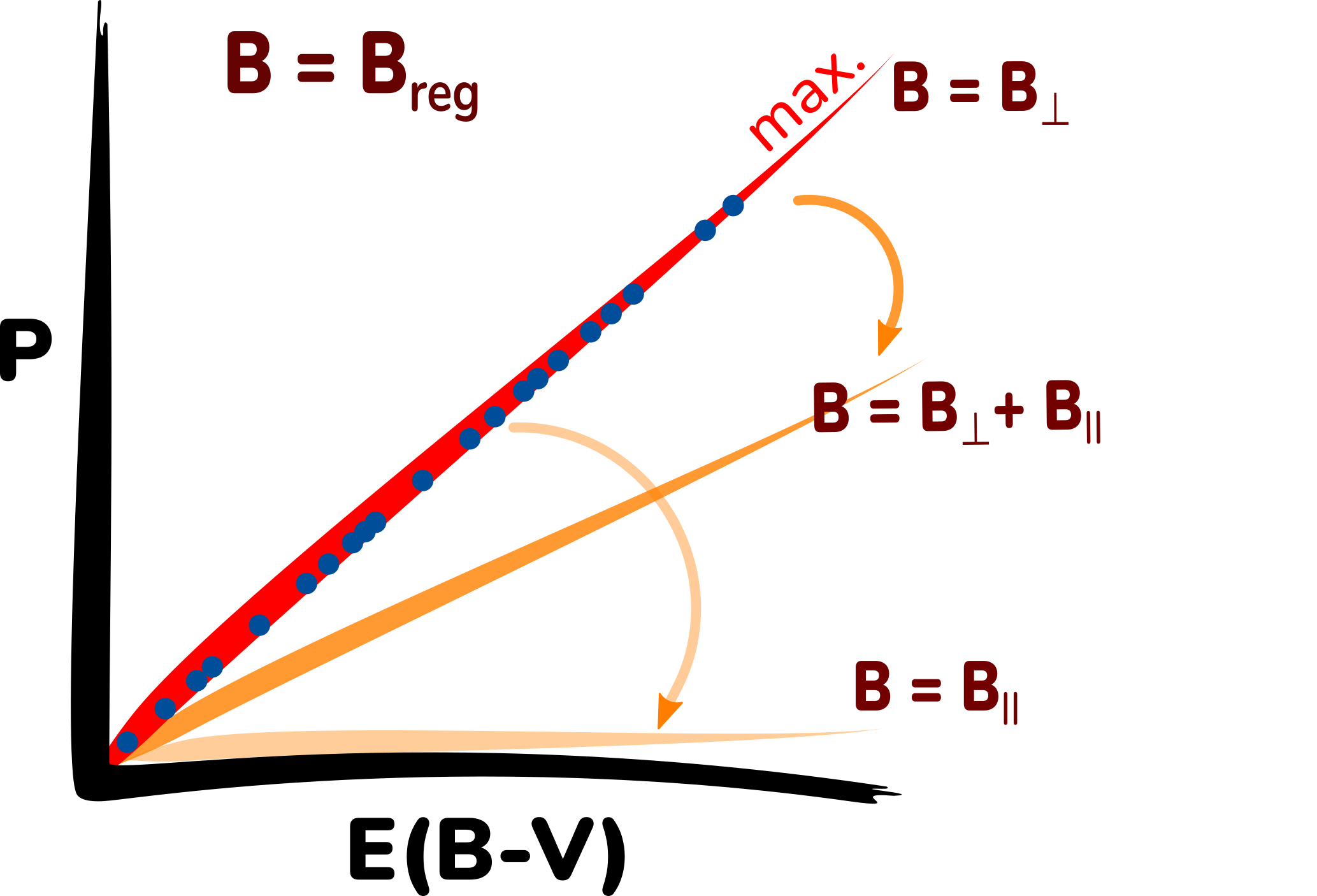}{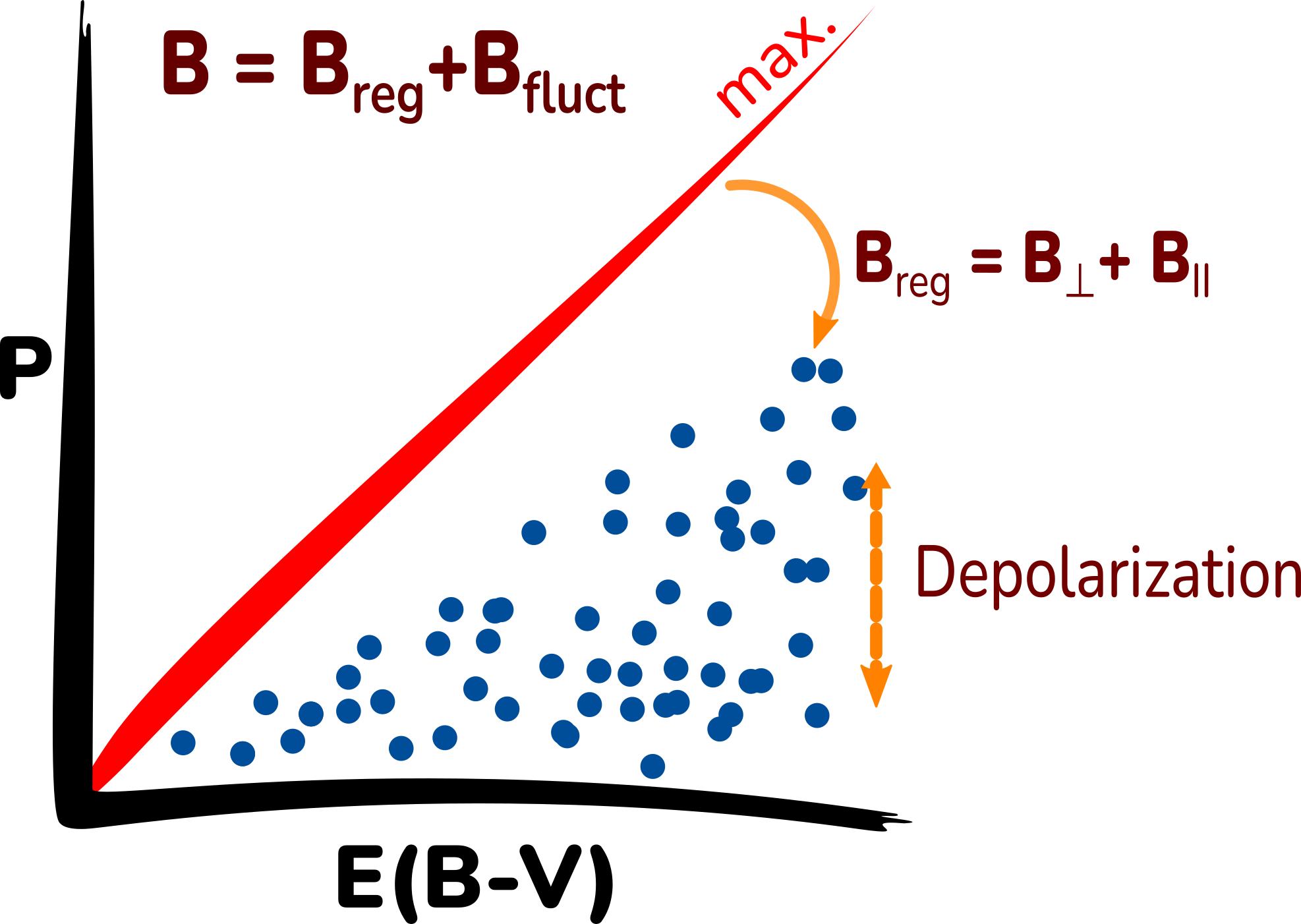}
            \caption{Polarization as a function of the reddening under different conditions of the ISM. \textit{Left:} Uniform magnetic field and constant/uniform dust properties. \textit{Right:} Magnetic field with uniform and fluctuating components, dust properties vary.
            \label{fig:PvsE_cartoon}}
        \end{figure*}

        The reddening difference as a function of distance in Figure~\ref{fig:Green-Stil_vs_SH} (bottom row) shows a median difference of \citetalias{Green_2019} (left), \citetalias{Marshall_2006} (middle), and \citetalias{Planck-Collaboration_2016} (right), with  \citetalias{Anders_2022} close to zero. The 1$\sigma$ confidence interval of the difference is below $\sim0.1$~mag, $\sim0.2$~mag, and $\sim0.06$~mag on average, respectively. \citetalias{Marshall_2006} reddening is systematically higher for longer LOS and has significant errors. \citetalias{Planck-Collaboration_2016} reddening is up to $\sim0.16$~mag higher at most diffuse LOS ($d<1$~kpc), which may be due to the highest column density proven throughout the Galaxy. The scatter of \citetalias{Green_2019}'s difference around 2~kpc may be due to the systemic errors of the individual fields.
        
        In summary, despite the different methods and bandwidth used, the limitations on distance, and sky coverage, the differences of \citetalias{Green_2019} and \citetalias{Planck-Collaboration_2016} with \citetalias{Anders_2022} using all fields together are small (\mbox{E(\bv)$_{A22}<$ 0.3~mag}). Although the difference scatters up to 0.2-0.3~mag ($3\sigma$ of the reddening difference) using the entire data set, it is important to mention that the systematic error observed in diffuse regions (e.g.~\textit{C2}, \textit{C16}, and \textit{C37}) is below $0.14-0.2$~mag depending on the field and the dust map (see Section \ref{subsec:Discu_PE_SH22_G19_M06} for more details). Consequently, we must account for uncertainties of $<$ 0.2~mag out to E(\bv)$_{A22} \sim 1.3$~mag depending on the field, and possibly a systematic error of maximum 1~mag only at high reddening, i.e.~E(\bv)$_{A22} \gtrsim$ 1.3~mag. Finally, from here on, we will refer to the \citetalias{Anders_2022} reddening simply as E(\bv).

    \subsection{Variations of the extinction curve} \label{subsec:Ebv_var_curve}
        
        The extinction curve varies across the sky and along each sight-line \citep{Serkowski_1975}, as the total to selective extinction ratio map, R$_V$, presented by \cite{Schlafly_2016}, shows. Unfortunately, the R$_V$ map, built with APOGEE spectroscopic data \citep{Majewski_2017}, only covers the Northern sky. A reasonable assumption, however, is to take a constant value R$_V$ = 3.1 \citep{Savage_1979,Fitzpatrick_2004}, as we did to find the reddening using \textit{V}-band extinction. 
        
        Nonetheless, going a step further, we can use \cite{Schlafly_2016} map to have a rough idea of how different values of R$_V$ affect the polarization efficiency using the trends observed in the Northern sky. By doing so, we found that in the Northernmost fields, one can use \mbox{$3.0 <$ R$_V < 3.3$}, whereas, in fields located at \mbox{$360\degr< l <210\degr$}, the R$_V$ values may be higher, \mbox{$3.3 <$ R$_V < 3.7$}. Assuming an extreme R$_V$ value for each case, we end up with higher polarization efficiencies in the South (i.e.~an increase of $\sim$20\% of the value calculated with  R$_V$ = 3.1) and lower values in the North (i.e.~a decrease of $\sim$5\% of the value calculated with  R$_V$ = 3.1). 

\section{The relation between reddening and polarization} \label{sec:P_vs_E}

    \subsection{Basic Scenarios in the General ISM} \label{subsec:P_vs_E_scenarios}
    
        Graphs of the linear polarization as a function of the reddening can be used to quantify the maximum starlight polarizing efficiency in the dusty ISM. In particular, the reddening increases monotonically with the interstellar dust column density, but this is not necessarily true for polarization. For instance, spherical and carbonaceous dust grains produce no polarization at all \citep{Draine&Fraisse_2009}, nor does dust in a magnetic field oriented along the line of sight. Moreover, small-scale and meso-scale structures, like turbulence or discrete objects, might change the magnetic field properties along the sight-line. Consequently, the polarization produced by far interstellar dust clouds might be canceled out by foreground structures with different dust and magnetic field properties (i.e.~depolarization) because the starlight polarization observed is a vector quantity averaged along the sight-line. Therefore, the polarization efficiency depends on both the interstellar dust grains and the GMF properties.
        
        Let us first consider the general diffuse ISM as an ideal scenario with uniform distribution and properties of elongated polarizing dust grains and a completely regular magnetic field, $\bm{B} = \bm{B}_{reg}$, that does not change along the sight-line. Thus, one would expect a continuous linear increase in reddening and polarization with distance. Consequently, the relation between the degree of polarization and the reddening becomes a straight line, as in Figure~\ref{fig:PvsE_cartoon} (red line in the left panel), with a steep slope for a GMF perpendicular to the sight-line, $\bm{B}_{reg} = \bm{B}_{\bot}$, that decreases until the GMF become parallel to the sight-line, as $\bm{B}_{reg} = \bm{B}_{\|}$.

        As mentioned above, dust properties could vary, and the magnetic field can have different strengths, orientations, \mbox{$\bm{B}_{reg} = \bm{B}_{\|} + \bm{B}_{\bot}$}, and random fluctuations, \mbox{$\bm{B} = \bm{B}_{reg} + \bm{B}_{fluct}$} (i.e.~components with coherent scales smaller than the regular magnetic field scales), in reality. These variations can happen along the sight-line and between LOS themselves. All of this adds dispersion to the \mbox{P$_V$/E(\bv)} relation. In other words, the variations in dust and magnetic field properties can cause smaller values of polarization relative to the optimum situation, or no polarization at all \citep[see e.g.~models from][]{Jones_1989,Jones_1992}. The polarization efficiency straight line case (\mbox{$\bm{B} = \bm{B}_{reg} = \bm{B}_{\bot}$}) then becomes an upper limit or envelope, [P$_V$/E(\bv)]$_{max}$, that covers the measurements, as in Figure~\ref{fig:PvsE_cartoon}, right. 
        
        We describe two general scenarios in Figure~\ref{fig:PvsE_cartoon} but more complex ones may exist. For instance: 
        \begin{itemize}
            \item P and E(\bv) may be constant along the sight-line, which happens when we run out of dust grains at some distance. In this case, the polarimetry, dust grains, and GMF properties belong to some nearby polarizing dust structure that makes up all our observed polarization and reddening.
            \item Only P is constant in some regions along the sight-line, whereas the reddening increases. So either there are no interstellar polarizing dust grains in some sections of the sight-line, there is a significant parallel component of the GMF to the sight-line, \mbox{$\bm{B}_{reg}$ = $\bm{B}_{\|}$}, or there is a substantial fluctuating component of the large-scale GMF. All of these conditions can occur combined as well. 
        \end{itemize}
        
        In essence, the degree of polarization as a function of the reddening in dense and turbulent ISM regions, e.g.~molecular clouds, is different from the general diffuse ISM. Along highly extincted LOS, the dust and GMF properties vary considerably \citep[see e.g.~][]{Andersson_2015}, and polarization efficiency can constrain dust grain alignment efficiency with the local magnetic field \citep[see e.g.][]{Whittet_2008,jones_whittet_2015,Draine_2021}. In contrast, the high polarization efficiency observed in different wavelengths in diffuse ISM regions \citep{Pereyra_Magalhaes_2007, Andersson_Potter_2007, Skalidis_2018, Planck-Collaboration_2018_20, Panopoulou_2019}, and complex dust population models \citep{Kim_Martin_1995} are all best explained with high alignment efficiency. In this case, if we can assume uniform/constant elongated dust grain properties and complete grain alignment in the general diffuse ISM, most variations in the polarization efficiency may be due to changes in the magnetic field intensity and geometry. For instance, a low polarization efficiency can be caused by GMF lines approximately parallel to the sight-line, depolarization due to a fluctuating component of the GMF, or a combination of both. Meanwhile, high polarization efficiency would mean the presence of a significant regular GMF and a favorable alignment of the GMF with the plane-of-sky.

    \subsection{Calculating the upper envelope of the polarization efficiency} \label{subsec:P_vs_E_Cal_PE_uplim}
        
        The upper limit of the polarization efficiency is calculated as the weighted $99th$ percentile, or the weighted $0.99$ quantile (Section \ref{subsec:Wquantile}), of the \mbox{P$_V$/E(\bv)} distribution (see e.g.~Figure~\ref{fig:PE_hist}) using only high-quality data (Section~\ref{subsec:Qflags}). This limit covers $99\%$ of the weighted measurements, as shown for example in Figure~\ref{fig:PE_hist}, which is approximately a $3\sigma$ limit. The non-weighted $99th$ percentile (the orange dotted line in the same figure) is often higher and very sensitive to outliers. A weighted calculation, on the other hand, can account for uncertainties in the measurements and performs better with outliers (see e.g.~robust estimation, chapter 15, \citealt{Press_FortranStats_1992}). However, Bayesian methods often present their results with asymmetric errors. These errors arise from a non-linear dependency of the results in a nuisance parameter \citep{Barlow_2003}, in our case, E(\bv). Hence, any statistical calculation becomes non-trivial, and very few formulations exist for asymmetric errors \citep{Barlow_2003, Erdim_SOAD_2019}. Furthermore, there is no straightforward method to estimate weighted statistics, such as quantiles or linear fits. Thus, more complicated techniques such as bootstrapping or Monte Carlo simulations are needed.

        \subsubsection{Error propagation} \label{subsec:err_propag}
        
            In the propagation of asymmetric errors, the typical solution is to add in quadrature the two different errors separately, but there is no statistical justification for this. \cite{Barlow_2003} presented a method to add asymmetric uncertainties and accurately calculate some (few) statistics on the linear combinations of quantities. Still, the procedure may be not the same when the operation is different from an addition, e.g.~a ratio. Unfortunately, \cite{Barlow_2003} does not have a solution for our specific problem. Lacking a more precise method, we calculate our errors through standard error propagation as
            \begin{equation}
            \label{eq:dPE}
                \sigma^{\pm}_{x/y} = \frac{x}{y}\sqrt{\left(\frac{\sigma^{\pm}_x}{x}\right)^2 +\left(\frac{\sigma^{\pm}_y}{y}\right)^2} .
            \end{equation}
        
        \subsubsection{Quantile estimator} \label{subsec:quantile}
        
            The quantiles divide the probability distribution function (PDF) of a particular range of measurements into $p$ intervals with equal probability. The quantile function, also known as the percentile function or the inverse cumulative distribution function (CDF), allows finding the data value where the CDF crosses the $k$-th quantile (with $k = 1,2,...,p-1$), i.e.~the real value which probability is the same as the $k$-th quantile. Most statistical software packages\footnote{For instance R, NumPy, Sci-Py, among others.} use the method 7 (or Type 7) defined by \cite{Hyndman_1996} as the quantile estimator. This quantile estimator can be expressed as \citep[see e.g.~][]{Hyndman_1996, Akinshin_2022, Akinshin_2023}:
            \begin{equation} 
            \label{eq:T7_q_estim}
                \begin{split}
                Q(x,q_k) 
                &= \sum_{i=1}^{n}{W_{n,i}\cdot x_i} ,  \\
                &= x_{(\lfloor h \rfloor)} + (h - \lfloor h \rfloor)(x_{(\lfloor h \rfloor + 1)} - x_{(\lfloor h \rfloor)}) ,
                \end{split}
            \end{equation}
            where: $x_{i}$ is the $i$-th measurement or order statistic\footnote{The order statistics are, in simple words, the sorted measurements.}, with $i = 1, 2, ..., n$ for a total number of measurements $n$; $q_k$ is the $k$-th quantile with values on [0,1]; $W_{n,i}$ is the area of the $k$-th interval of the PDF; and $\lfloor h \rfloor$ denotes the floor of the parameter $h$, which is the real value index to which the $q_k$ quantile probability corresponds,
            \begin{equation}
            \label{eq:h}
                h = q_k(n-1) + 1~.
            \end{equation}
            
            One can define a CDF function ($F_7$) such that,
            \begin{equation}
            \label{eq:T7_cdf_func}
                F_7(u) = 
                    \begin{cases} 
                        0 ,    & \text{for } ~~~~~~~~~~~~~~~~~ u < (h-1)/n, \\
                        nu-h+1 , & \text{for } (h-1)/n \leq u \leq h/n,  \\
                        1  ,   & \text{for } ~~~~~~~~ h/n  < u, ~~~~
                    \end{cases}
            \end{equation}
            with the corresponding PDF $f_7(u) = F'_7(u)$, as the derivative of the CDF. Then,
            \begin{equation}
            \label{eq:W_T7}
                W_{n,i} = F_7(r_i) - F_7(l_i)\ .
            \end{equation}
            In the non-weighted quantile estimation, we choose the edges of the $k$-th interval to be
            \begin{equation}
            \label{eq:fragm_edges}
            l_i = \frac{i-1}{n} \ \ \ \ \mathrm{and}\ \ \ \
            r_i = \frac{i}{n} ,
            \end{equation}
            so all fragments have equal width $1/n$. Now, we define the $m$-th moment as
            \begin{equation}
            \label{eq:wT7_moment}
                Q_m(x,q_k) = \sum_{i=1}^{n}{W_{n,i}\cdot x^m_i} ,
            \end{equation}
            where $m = 1, 2, ...$ denotes the moments, and the first moment is the quantile \citep[see][for more details]{Akinshin_2023}. The uncertainty of the quantile can be obtained from the first and second moments. So, the standard error of the calculation is \footnote{Here we used the property of the variance of a variable $X$: $ \operatorname{Var}(X) = \operatorname{E}\left[X^2 \right] - \operatorname{E}[X]^2$}
            \begin{equation}
            \label{eq:wT7_std_err}
                \sigma_{Q(x,q_k)} = \sqrt{Q_2(x,q_k) - Q_1^2(x,q_k)}~.
            \end{equation}
            
        \subsubsection{Weighted quantile estimator} \label{subsec:Wquantile}

            The weighted T7 quantile can be considered the generalization of the non-weighted estimator (Equations~\ref{eq:W_T7},~\ref{eq:fragm_edges}, and~\ref{eq:wT7_moment}), which would be the particular case when all the weights are equal to 1 \citep{Akinshin_2023}. Firstly, the total sum of the weights defined as $S_{n}(w) = \sum_{i=1}^{n}{w_i} > 0$, for $i = 1, 2, ..., n$, is always positive and different from zero (assuming $w_i\geq0$). Secondly, the partial sum of weights $S_i(w)$ is defined as the cumulative sum of the order statistics until the $i$-th measurement. Hence, the interval of width proportional to the weights, i.e.~equal to $w_i/S_n(w)$, should be between
            \begin{equation}
            \label{eq:w_fragm_edges}
            l_i^* = \frac{S_{i-1}(w)}{S_n(w)} , ~~~
            r_i^* = \frac{S_{i}(w)}{S_n(w)} ,
            \end{equation}
            where $S_0(w)=0$ is assumed. When all weights are equal (i.e.~$w=1$), then, $l_i=l_i^*$ and $r_i=r_i^*$. Furthermore, we consider $n^*$ as the weighted or effective sample size \citep{Wiegand_1968}:
            \begin{equation}
            \label{eq:w_sample_size}
            n^* = \frac{(\sum_{i=1}^{n}{w_i})^2}{\sum_{i=1}^{n}{w_i^2}} ,
            \end{equation}
            which redefines the $h^*$ parameter from Equation~\ref{eq:h} and the CDF $F_7^*(u)$ in Equation~\ref{eq:T7_cdf_func}. Then, we evaluate the weighted interval edges in the latter equation to calculate the area of the weighted interval, $W_{n,i}^*$, as in Equation~\ref{eq:W_T7}. Finally, we obtain the weighted quantile estimation $Q^*(x,q_k)$ with Equation~\ref{eq:wT7_moment} (m=1). The standard error of the weighted quantile can be computed with the weighted first and second moments as in Equation~\ref{eq:wT7_std_err} \citep[see][for more details]{Akinshin_2023}. 
                
        \subsubsection{Weighted quantile with asymmetric errors} \label{subsec:wQ_asym_err}
    
            The weighted quantile estimation presented in Section~\ref{subsec:Wquantile} assumes symmetric uncertainties, but this is not our case. \mbox{P$_V$/E(\bv)} has asymmetric errors as in Equation~\ref{eq:dPE}, therefore, we used the bootstrap technique to calculate the weighted $0.99$ quantile, or $99th$ percentile.
            %
            \begin{figure*}[t!]
            \epsscale{1.15}
            \plottwo{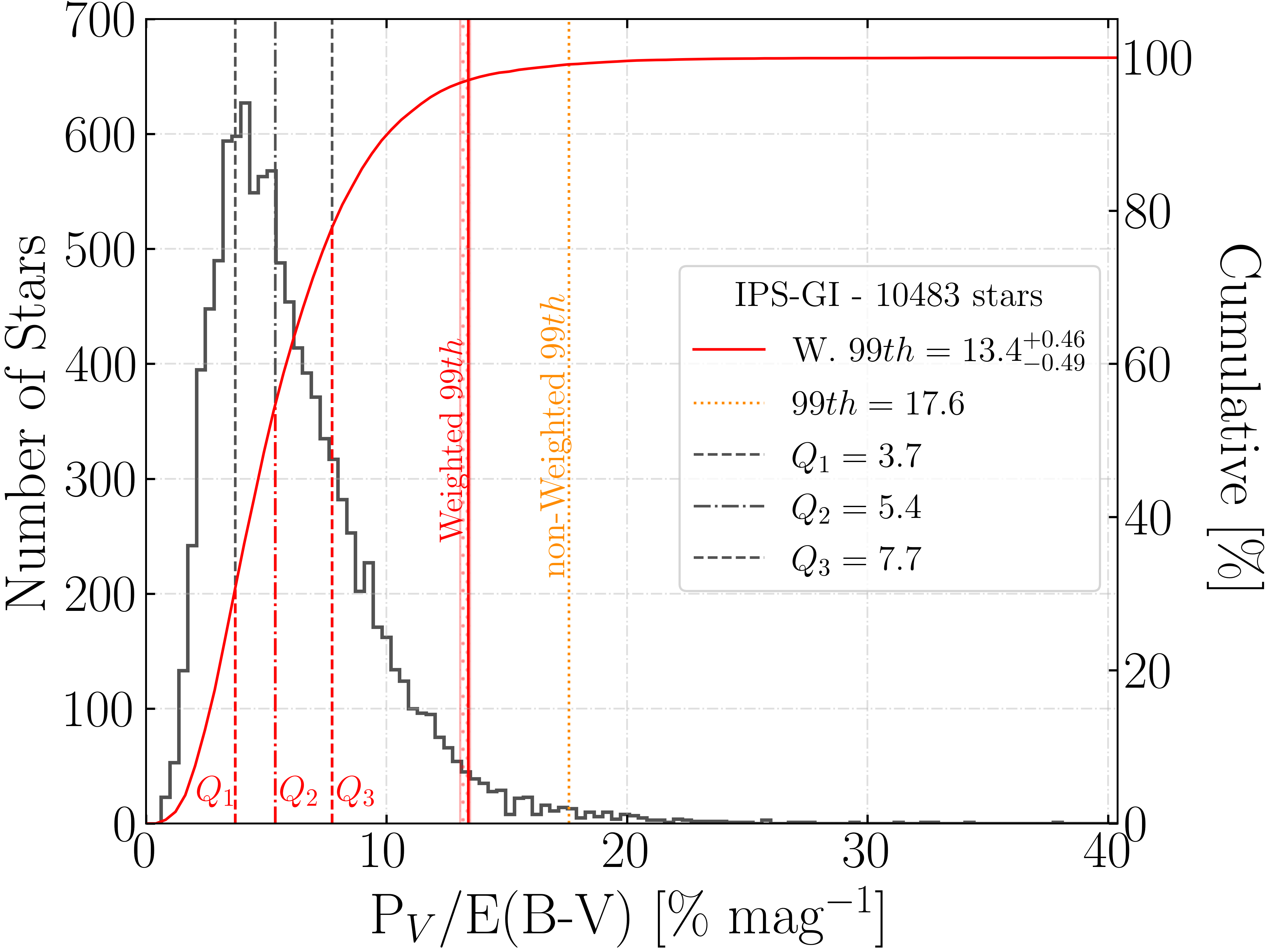}{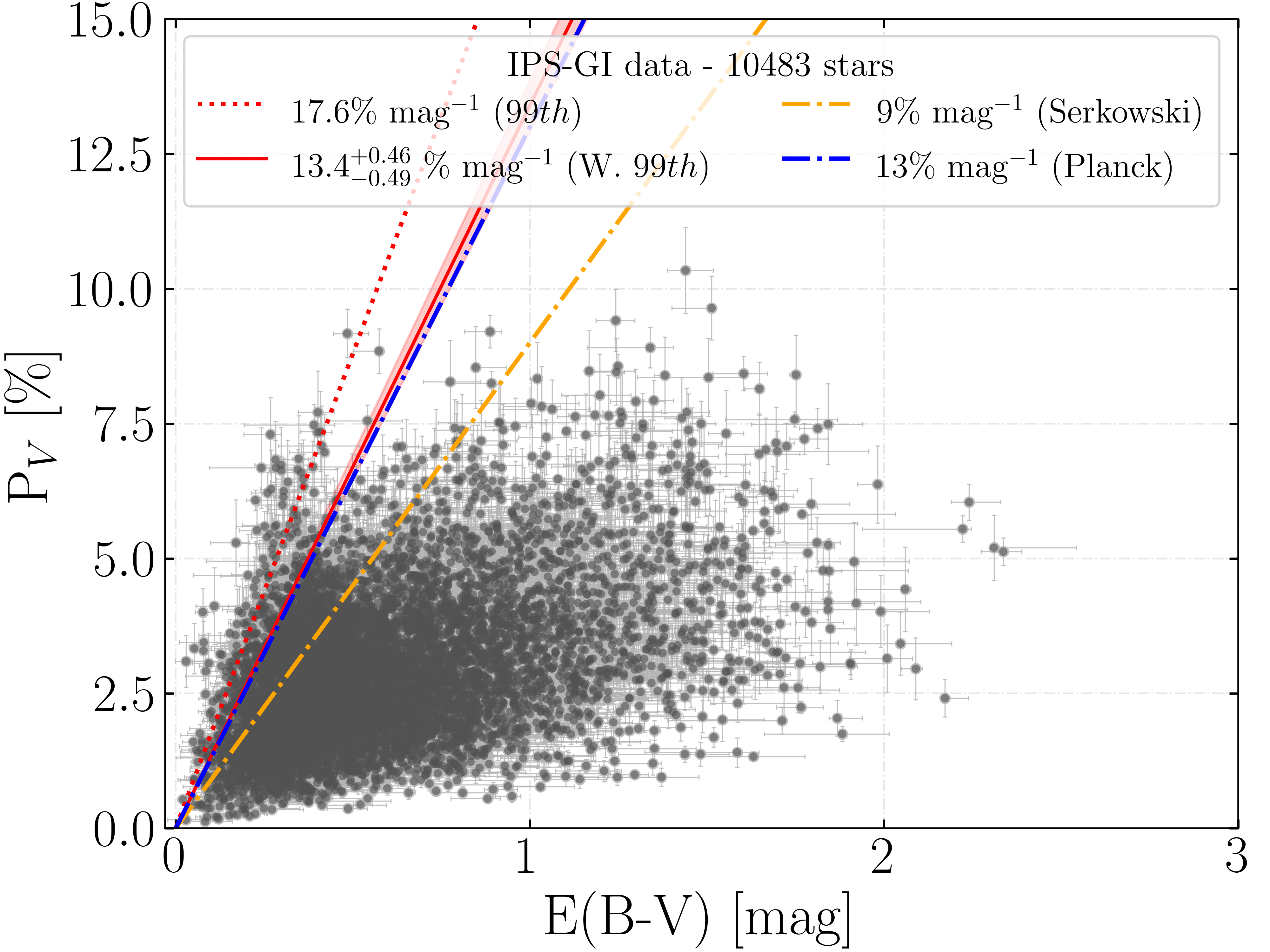}
            \caption{\textit{Left}: Polarization efficiency histogram of high-quality IPS-GI data. The cumulative percentage function is shown by the red curve. Dashed lines are the 0.25 ($Q_1$) and 0.75 ($Q_3$) quantiles, while the dot-dashed line shows the 0.5 ($Q_2$) quantile or median value. \textit{Right}: Polarization degree as a function of the reddening. Upper limits of \textit{Serkowski} and \textit{Planck} are represented by the dot-dashed lines in orange and blue respectively. The red dotted line is the non-weighted $99th$ percentile and the red solid line, with the underlying shaded region, is the weighted calculation with the $68\%$ confidence interval.
            \label{fig:PE_hist}}
            \end{figure*}
            
            The bootstrap method uses the original sample to re-calculate the weighted statistic in several iterations (e.g.~we use $N=10000$) with different weights each time. It should not be confused with re-sampling as we are not using a sub-sample of the original data. We are rather re-calculating our statistic $N$ times with a measurement error re-sampled from a two-sided exponential probability density function (PDF) defined for each data point:
            \begin{equation}
            \label{eq:two_par_exp}
                \frac{1} {2\lambda_{1,2}} exp\biggl\{\frac{-\mid x - \mu \mid}{\lambda_{1,2}}\biggr\} ~,
            \end{equation}
            with $\mu=0$ and $\lambda_{1,2} = \sigma_{\pm}/\sqrt{2}$, where $\sigma_{\pm}$ are the asymmetric uncertainties and the discontinuity at zero is avoided when re-sampling. We used the inverse of the measured errors, $1/\sigma$ \citep{Press_FortranStats_1992}, as weights for the quantile estimation (Section~\ref{subsec:Wquantile}), where $\sigma$ is the re-sampled error. In our specific case, a two-sided exponential PDF and the respective inverse of the error weight proved to be more robust against outliers than an asymmetric Gaussian PDF and the inverse of the squared error weight, i.e.~$w=1/\sigma^2$ (see e.g.~the comparison between Gaussian and exponential probability errors in section 15.7, chapter 15, of \citealt{Press_FortranStats_1992}). The latter approach turned out to be a heavy weight on the highest signal-to-noise measurements that yielded underestimated upper envelopes.
            
            Finally, the outcome of the iterations is a distribution from which we can obtain some metrics, such as the median value and its $68\%$ confidence interval. These metrics are computed for the entire IPS-GI sample or any sub-sample (i.e.~individual fields) to determine our results on the polarization efficiency upper limit. The confidence interval is useful to identify potential outliers in polarization efficiency, i.e.~high \mbox{P$_V$/E(\bv)} values that can affect our statistics and increase the uncertainty of the calculations, especially in small distributions. The variability of potential outliers is assessed in Section~\ref{subsec:Outliers}, and Appendixes~\ref{appex:known_varia} and~\ref{appex:varia_proxy}.

\section{Results} \label{sec:Results}
     %
    \begin{figure}[t!]
        \epsscale{1.19}
        \plotone{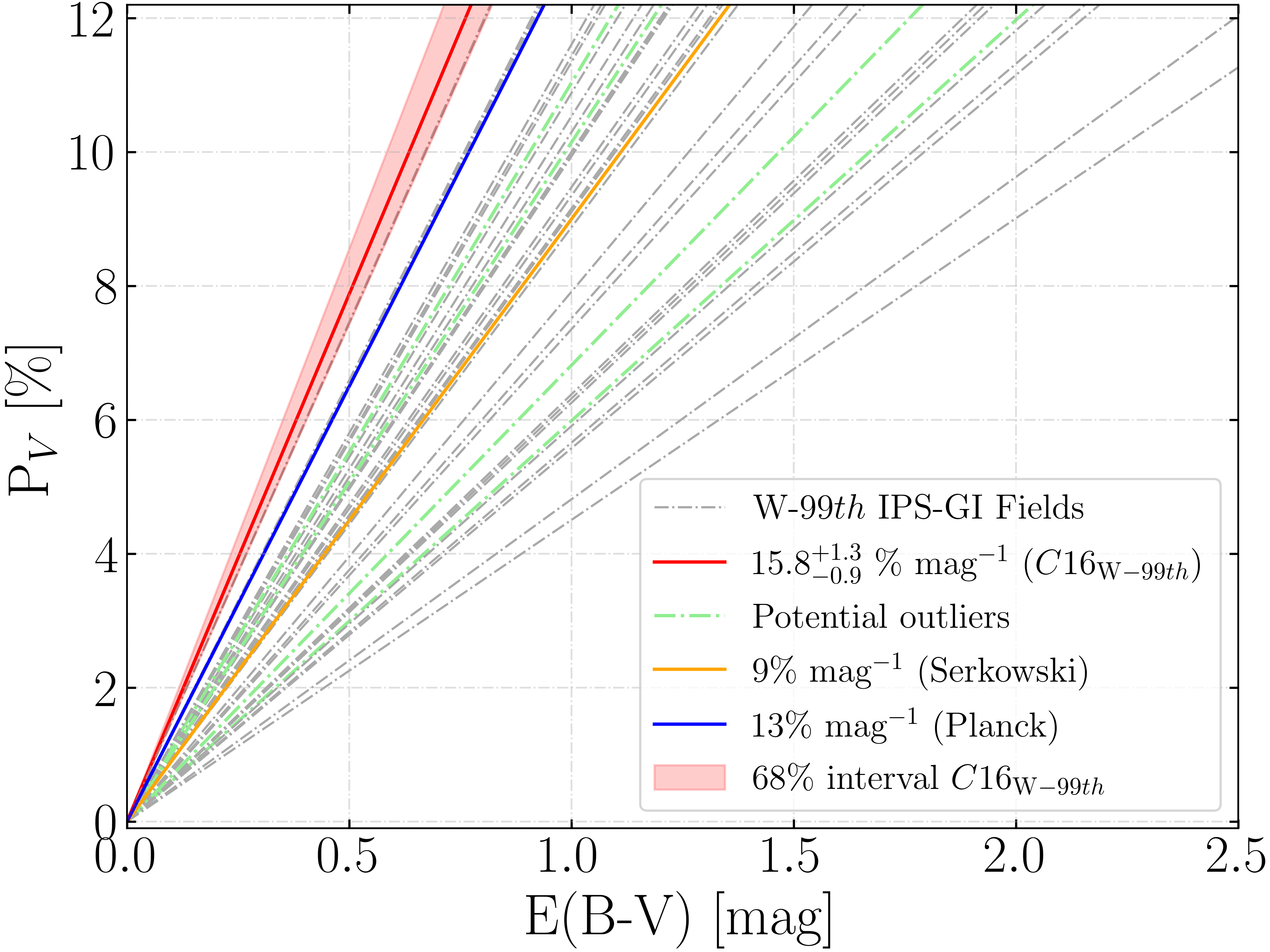}
        \caption{Weighted \mbox{P$_V$/E(\bv)} upper limit per IPS-GI field (gray dot-dashed lines). The red solid line and pink area are the weighted polarization efficiency upper limit and confidence interval in field \textit{C16}. The light-green dot-dashed lines are slopes affected by potential outliers in polarization efficiency (see the figure set available online for more details). The blue and  orange solid lines are, respectively, \textit{Planck}'s and \textit{Serkowski}'s limits. 
        \label{fig:PE_slopes}}
    \end{figure}
    We present our results on the polarization efficiency for the entire IPS-GI sample and individual fields using \citetalias{Anders_2022} reddening. Additionally, we show results on the relation between the polarization efficiency and Galactic coordinates.

    %
    \begin{deluxetable*}{lcccccccccccccc}
        \tablecaption{Polarization efficiency upper limit statistics per IPS-GI field. \label{tab:PE_uplim}}
        \tabletypesize{\scriptsize}
        \tablehead{
        \colhead{Field} & \colhead{No.} & \colhead{$\langle l \rangle$} & \colhead{$\langle b \rangle$} & \colhead{99th} & \colhead{Weighted} & \colhead{84th} & \colhead{16th} & \colhead{G19} & \colhead{G19} & \colhead{G19} & \colhead{M06} & \colhead{M06 Std$_{err}$} & \colhead{PC16} & \colhead{PC16 Std$_{err}$}
        \\
        \colhead{ } & \colhead{Stars} & \colhead{} & \colhead{} & \colhead{} & \colhead{99th} & \colhead{} & \colhead{} & \colhead{W-99th}& \colhead{84th} & \colhead{16th}& \colhead{W-99th} & \colhead{W-99th}& \colhead{W-99th} & \colhead{W-99th}
        }
        \colnumbers
        \startdata
        \textit{C0} & 397 & 327.57 & -0.83 & 16.1 & 10.9 & 13.1 & 10.0 & ... & ... & ... & 10.16 & 0.21 & ... & ... \\
        \textit{C1} & 545 & 330.42 & 4.59 & 13.7 & 10.0 & 10.8 & 9.4 & ... & ... & ... & 8.17 & 0.04 & ... & ... \\
        \textit{C2} & 589 & 359.34 & 13.47 & 20.0 & 14.9 & 17.4 & 13.9 & 14.7 & 16.6 & 13.9 & ... & ... & 17.01 & 0.23 \\
        \textit{C3} & 449 & 339.49 & -0.42 & 16.4 & 9.1 & 10.6 & 8.2 & ... & ... & ... & 9.03 & 0.08 & ... & ... \\
        \textit{C4} & 507 & 18.63 & -4.46 & 9.8 & 8.9 & 9.6 & 8.4 & 9.3 & 10.3 & 9.0 & 10.34 & 0.29 & ... & ... \\
        \textit{C5} & 421 & 25.06 & -0.74 & 6.6 & 5.7 & 6.4 & 5.4 & 6.2 & 6.8 & 5.7 & 5.41 & 0.12 & ... & ... \\
        \textit{C6} & 210 & 298.61 & 0.64 & 15.3 & 10.3 & 12.5 & 9.2 & ... & ... & ... & 9.52 & 0.39 & ... & ... \\
        \textit{C7} & 340 & 41.68 & 3.39 & 7.9 & 6.4 & 7.7 & 5.8 & 6.8 & 8.1 & 6.1 & 5.45 & 0.05 & ... & ... \\
        \textit{C11} & 946 & 307.53 & 1.33 & 17.8 & 11.3 & 13.0 & 10.4 & ... & ... & ... & 6.90 & 0.22 & ... & ... \\
        \textit{C12} & 881 & 331.04 & -4.70 & 19.2 & 13.1 & 15.7 & 12.3 & ... & ... & ... & 10.54 & 0.19 & ... & ... \\
        \textit{C13} & 287 & 333.24 & 3.75 & 14.2 & 9.9 & 11.9 & 9.4 & ... & ... & ... & 10.49 & 0.22 & ... & ... \\
        \textit{C14} & 622 & 312.87 & 5.48 & 18.4 & 13.2 & 15.5 & 12.6 & ... & ... & ... & 27.61 & 0.90 & ... & ... \\
        \textit{C15} & 458 & 304.71 & -0.17 & 12.3 & 7.5 & 8.9 & 7.0 & ... & ... & ... & 4.64 & 0.07 & ... & ... \\
        \textit{C16} & 561 & 301.97 & -8.77 & 19.7 & 15.8 & 17.1 & 14.8 & ... & ... & ... & ... & ... & 12.73 & 0.12 \\
        \textit{C30} & 37 & 222.54 & -1.63 & 18.3 & 6.0 & 8.0 & 4.9 & 7.2 & 8.5 & 5.8 & ... & ... & ... & ... \\
        \textit{C35} & 144 & 271.45 & -1.07 & 14.2 & 10.5 & 11.7 & 9.7 & ... & ... & ... & 12.89 & 0.28 & ... & ... \\
        \textit{C36} & 191 & 343.29 & 11.97 & 19.1 & 10.0 & 11.9 & 8.8 & ... & ... & ... & ... & ... & 14.50 & 0.66 \\
        \textit{C37} & 539 & 331.16 & 7.62 & 20.2 & 14.9 & 17.2 & 13.8 & ... & ... & ... & ... & ... & 16.46 & 0.15 \\
        \textit{C39} & 309 & 303.07 & 1.63 & 17.7 & 9.5 & 11.8 & 8.7 & ... & ... & ... & 7.43 & 0.15 & ... & ... \\
        \textit{C40} & 264 & 302.16 & -1.21 & 11.3 & 7.9 & 9.1 & 7.2 & ... & ... & ... & 7.11 & 0.16 & ... & ... \\
        \textit{C41} & 116 & 257.34 & -0.48 & 10.2 & 7.4 & 8.6 & 6.6 & ... & ... & ... & ... & ... & ... & ... \\
        \textit{C42} & 87 & 245.48 & -0.11 & 43.5 & 11.0 & 14.9 & 8.5 & 13.3 & 18.8 & 10.8 & ... & ... & ... & ... \\
        \textit{C43} & 70 & 273.17 & -0.82 & 19.9 & 10.1 & 13.3 & 8.6 & ... & ... & ... & 14.35 & 1.97 & ... & ... \\
        \textit{C45} & 97 & 21.81 & 0.71 & 5.6 & 4.5 & 5.3 & 4.2 & 4.7 & 5.5 & 4.2 & 5.20 & 0.22 & ... & ... \\
        \textit{C46} & 36 & 44.21 & 2.66 & 6.1 & 4.8 & 5.3 & 4.3 & 4.8 & 5.3 & 4.5 & 5.61 & 0.54 & ... & ... \\
        \textit{C47} & 240 & 320.49 & -1.23 & 15.4 & 11.6 & 12.8 & 10.9 & ... & ... & ... & 9.53 & 0.07 & ... & ... \\
        \textit{C50} & 255 & 20.27 & 1.04 & 9.0 & 6.3 & 6.7 & 5.9 & 6.0 & 7.0 & 5.7 & 5.73 & 0.26 & ... & ... \\
        \textit{C52} & 177 & 318.76 & 2.78 & 15.1 & 9.3 & 11.6 & 7.7 & ... & ... & ... & 7.78 & 0.30 & ... & ... \\
        \textit{C54} & 137 & 305.17 & 1.31 & 13.1 & 9.2 & 10.6 & 8.4 & ... & ... & ... & 5.70 & 0.53 & ... & ... \\
        \textit{C55} & 278 & 15.15 & 1.68 & 10.2 & 6.3 & 7.8 & 5.7 & 6.2 & 7.5 & 5.6 & 8.06 & 0.34 & ... & ... \\
        \textit{C56} & 55 & 14.97 & -0.96 & 20.5 & 6.8 & 11.1 & 5.8 & 9.3 & 16.6 & 6.8 & 9.30 & 1.75 & ... & ... \\
        \textit{C57} & 343 & 40.59 & 4.14 & 6.7 & 5.9 & 6.4 & 5.7 & 5.8 & 6.9 & 5.3 & 6.08 & 0.18 & ... & ... \\
        \textit{C58} & 193 & 351.32 & 0.60 & 8.5 & 5.6 & 6.4 & 5.2 & ... & ... & ... & 6.59 & 0.29 & ... & ... \\
        \textit{C61} & 292 & 0.48 & 2.19 & 14.8 & 11.4 & 12.5 & 11.0 & 11.6 & 12.0 & 11.2 & 12.31 & 0.49 & ... & ...  \\
        \enddata
        \tablecomments{Columns: \textit{(1)} Field name; \textit{(2)} Total number of high-quality stars; \textit{(3 and 4)} Mean Galactic longitude and latitude of the IPS-GI field in degrees; \textit{(5)} non-weighted \mbox{P$_V$/E(\bv)} $99th$ percentile; \textit{(6, 7, and 8)} The $50th$, $84th$, and $16th$ percentiles of the weighted \mbox{[P$_V$/E(\bv)]$_{99th}$} distribution; \textit{(9, 10, and 11)} The $50th$, $84th$, and $16th$ percentiles of the weighted [P$_V$/E(\bv)]$_{\mathrm{G19-}99th}$ distribution using \citetalias{Green_2019} reddening; \textit{(12 and 13)} The median and standard error of the weighted [P$_V$/E(\bv)]$_{\mathrm{M06-}99th}$ distribution using \citetalias{Marshall_2006} reddening; \textit{(14 and 15)} The median and standard error of the weighted [P$_V$/E(\bv)]$_{\mathrm{PC16-}99th}$ distribution using \citetalias{Planck-Collaboration_2016} reddening. All \mbox{P$_V$/E(\bv)} values have units of \mbox{[\%~mag$^{-1}$]}.
        }
    \end{deluxetable*}
    %

    \subsection{P\texorpdfstring{$_V$}{V}/E(\texorpdfstring{\bv}{B-V}) upper envelope of all IPS-GI data} \label{subsec:Resul_PE_uplim_allIPS}
        
        Initially, the polarization efficiency upper envelope and its confidence interval were calculated including all IPS-GI objects having good quality data. Figure~\ref{fig:PE_hist} (right) shows the degree of polarization as a function of the reddening for all IPS-GI data. The upper envelope of the polarization efficiency, i.e.~the weighted $99th$ percentile, is \mbox{$13.4^{+0.46}_{-0.49}\%$~mag$^{-1}$}. This result is in good agreement with the maximum value from \cite{Planck-Collaboration_2018_20}, \mbox{$13\%$~mag$^{-1}$}, and far above the value from \cite{Serkowski_1975}, \mbox{$9\%$~mag$^{-1}$}, (both studies using all-sky samples). Nonetheless, the weighted $99th$ percentile found with the entire IPS-GI sample (Figure~\ref{fig:PE_hist}) should not be considered an average for the highest polarization efficiency since, towards each field (i.e.~each sight-line), the GMF and the interstellar dust have different properties. The upper limit of the polarization efficiency should, therefore, vary across the Galaxy. We must then study the IPS-GI fields individually as follows. 
    
    \subsection{P\texorpdfstring{$_V$}{V}/E(\texorpdfstring{\bv}{B-V}) upper envelope per IPS-GI field} \label{subsec:Resul_PE_up_fields}
        
\begin{figure*}[t!]
            \epsscale{1.19}
            \plotone{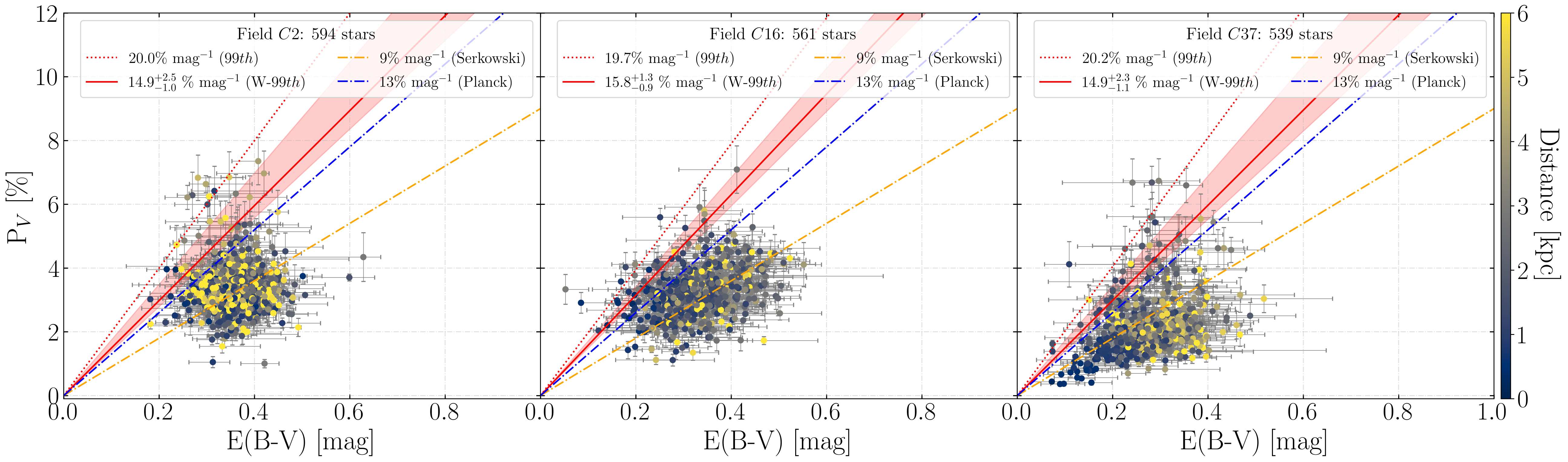}
            \caption{Degree of Polarization as a function of reddening in fields \textit{C2} (left), \textit{C16} (middle), and \textit{C37} (right). Stars are colored by distance with the color range truncated at 6~kpc for better visualization. The solid red lines and the pink areas are the weighted \mbox{P$_V$/E(\bv)} upper limits with their respective $68\%$ confidence interval. The dotted red lines are the non-weighted upper limits. The blue and orange dot-dashed lines show \textit{Planck}'s and \textit{Serkowski}'s upper limits respectively. The complete figure set (31 images) will be available on the journal website.
            \label{fig:Fields_PvsE}}
        \end{figure*}

        We present the weighted polarization efficiency upper limits calculated for each of the 34 IPS-GI fields in \mbox{Figure~\ref{fig:PE_slopes}}. The results are also shown in Table~\ref{tab:PE_uplim} and Figure~\ref{fig:Fields_PvsE} (the complete figure set is available online). A total of 20 IPS-GI fields have polarization efficiencies similar to, or above, the \mbox{$9\%$~mag$^{-1}$} limit. Five of them are higher than the \cite{Planck-Collaboration_2018_20} maximum upper limit of \mbox{$13\%$~mag$^{-1}$}: two slopes are within the red-shaded area, near the highest value, in Figure~\ref{fig:PE_slopes}, and other two are very close to the blue slope in the same figure, see Table~\ref{tab:PE_uplim} for more details. 
        %
        \figsetstart
        \figsetnum{5}
        \figsettitle{Polarization degree as a function of the reddening per IPS-GI field}

        \figsetend
        %
        \begin{figure*}[ht!]
            \gridline{
                      \fig{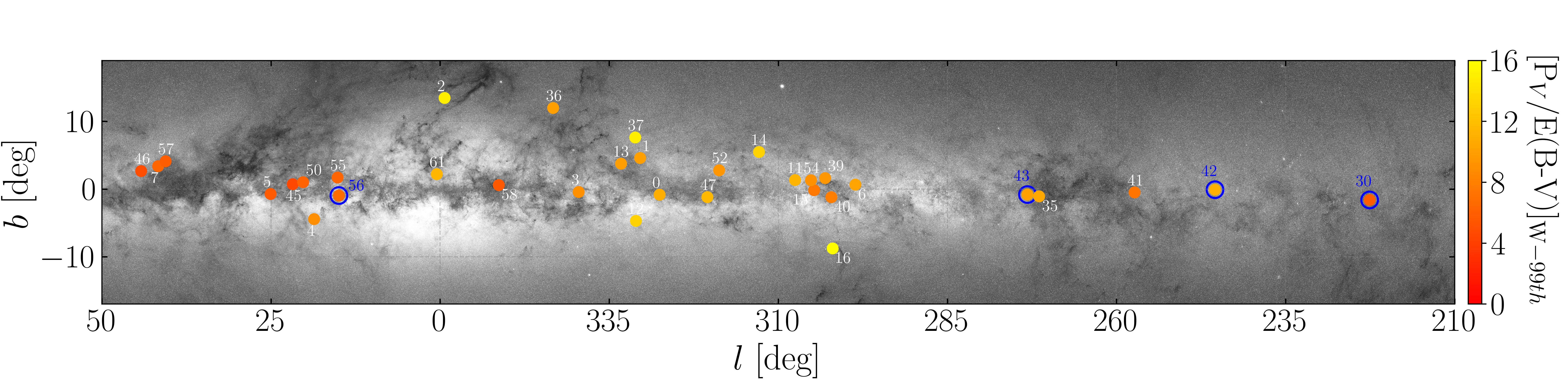}{\linewidth}{}
                      }
            \gridline{
                      \fig{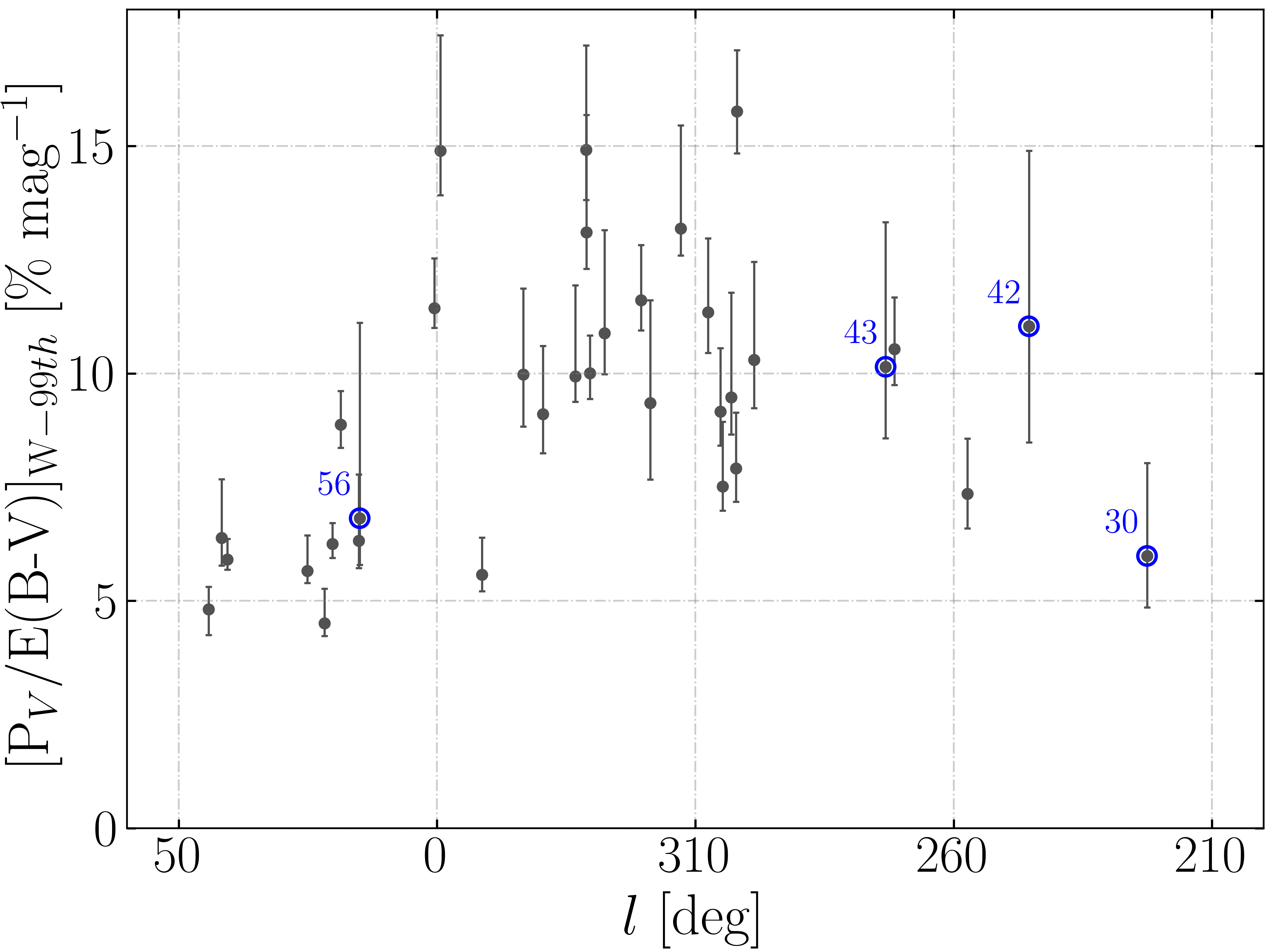}{0.42\linewidth}{}
                      \fig{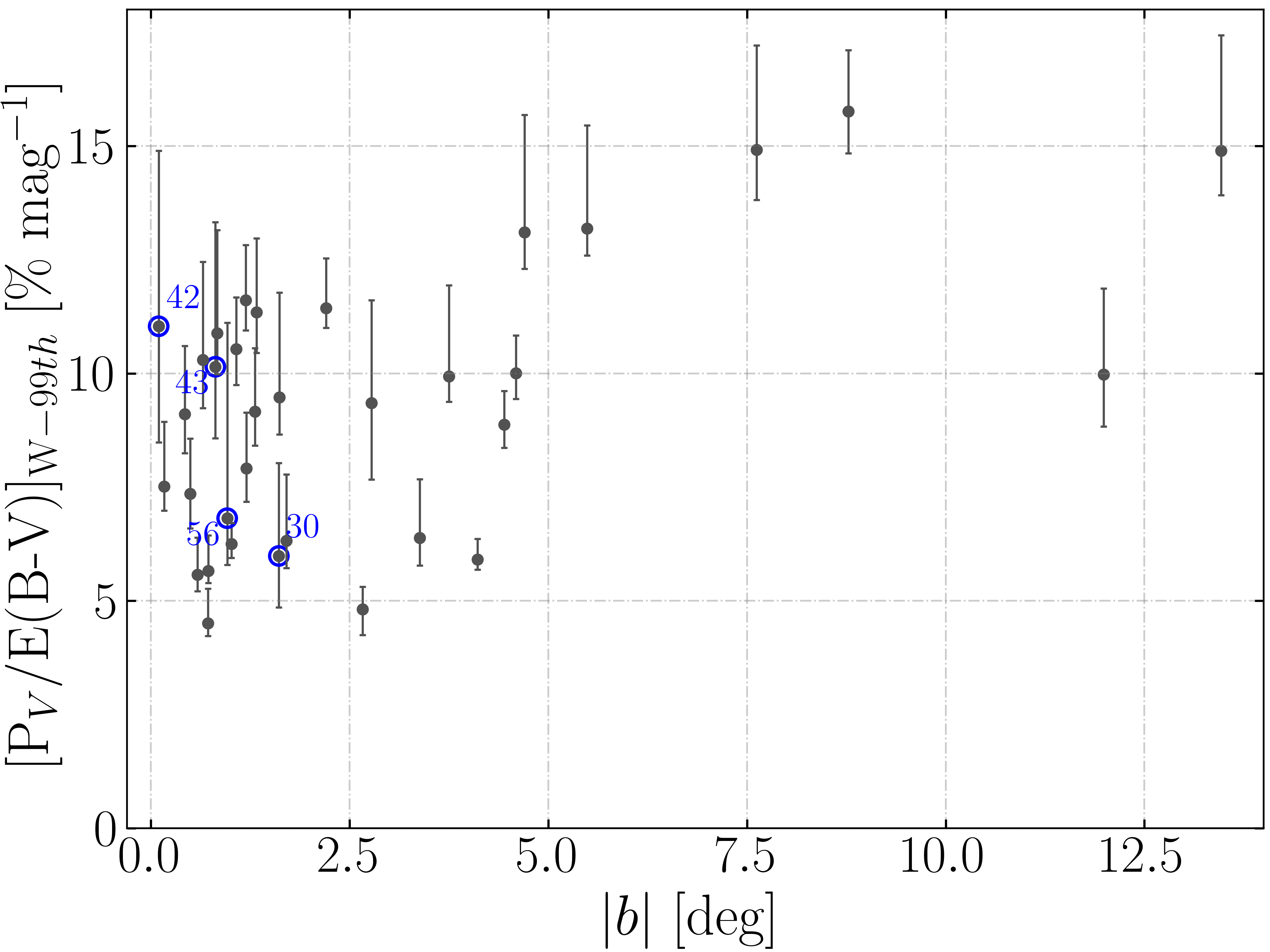}{0.42\linewidth}{}
                      }
            \caption{\textit{Top}: Sky plot showing the IPS-GI fields location colored by the weighted \mbox{P$_V$/E(\bv)} upper limits. The turquoise circles are unreliable upper limits due to potential outliers. The background image is the optical emission of the Galaxy adapted from ESA/Gaia/DPAC. \textit{Bottom}: Weighted \mbox{P$_V$/E(\bv)} upper limits with Galactic longitude (left), and absolute Galactic latitude (right). Error bars show the $68\%$ confidence interval of the weighted $99th$ percentile (Section~\ref{subsec:P_vs_E_Cal_PE_uplim}).
            \label{fig:PE_sky}}
        \end{figure*}
        
        We mitigated the impact of potential outliers in our results by using weighted quantile estimations (as described in Section~\ref{subsec:P_vs_E_Cal_PE_uplim}). In small samples, however, we still see some steep slopes with significant uncertainties (the light-green dot-dashed lines in \mbox{Figure~\ref{fig:PE_slopes}}, and \mbox{Table~\ref{tab:PE_uplim}}) that may be affected by poor statistics (e.g.~see the plots of \textit{C30}, \textit{C42}, \textit{C43}, and \textit{C56} in the figure set available online for more detail). In Section~\ref{subsec:Discu_PE_SH22_G19_M06} we discuss how the same fields present similar issues in the calculations of the upper limits with \citetalias{Green_2019} and \citetalias{Marshall_2006} reddening.
        
    \subsection{Maximum upper limit of P\texorpdfstring{$_V$}{V}/E(\texorpdfstring{\bv}{B-V})} \label{subsec:Resul_PEff_max}
    
        Figure~\ref{fig:Fields_PvsE} shows the degree of polarization as a function of the reddening of fields \textit{C2}, \textit{C16}, and \textit{C37}. These fields have the highest upper limits of polarization efficiency in our sample, around $\sim$15\%~mag$^{-1}$ (see Table~\ref{tab:PE_uplim} and Figure~\ref{fig:PE_slopes}). Their distributions cluster in a very narrow range of low reddening, between \mbox{0.1~mag $\lesssim$ E(\bv) $\lesssim$ 0.5~mag}. Meanwhile, the degree of polarization expands in a long range between $\sim$0.5\% and $\sim$7\%. The maximum weighted polarization efficiencies are estimated using more than 500 high-quality measurements in each of these fields (see Table~\ref{tab:PE_uplim}). This makes our results very robust, even against potential outliers in polarization efficiency (corresponding to $\sim$2\% of the stars on each field). Hence, we adopt the weighted \mbox{P$_V$/E(\bv)} upper limit observed in field \textit{C16}, \mbox{[P$_V$/E(\bv)]$_{C16} = 15.8^{+1.3}_{-0.9}\%$~mag$^{-1}$}, as the maximum polarization efficiency in our sample.  
  
    \subsection{P\texorpdfstring{$_V$}{V}/E(\texorpdfstring{\bv}{B-V}) upper envelope as a function of Galactic coordinates} \label{subsec:Resul_PE_Glat_Glog}
        
        Significant variations in polarization efficiency are observed with Galactic latitude and longitude (Figure~\ref{fig:PE_sky}). IPS-GI fields at intermediate latitudes ($|b|>7.5\degr$) and within Galactic longitudes $270\degr-360\degr$ have the highest polarization efficiency observed (see top panel of Figure~\ref{fig:PE_sky}). On the other hand, low polarization efficiencies are found towards Galactic longitudes $\sim 15\degr-45\degr$, and perhaps $l \sim 220\degr-275\degr$. Fields with a low number of stars and potential outliers in polarization efficiency (Section~\ref{subsec:P_vs_E_Cal_PE_uplim}), such as \textit{C30}, \textit{C42}, \textit{C43}, and \textit{C56}, likely have overestimated \mbox{P$_V$/E(\bv)} upper limits with significant uncertainties, as visible in the bottom row of Figure~\ref{fig:PE_sky} (see also Table~\ref{tab:PE_uplim}). So, their polarization efficiency may be significantly lower and consistent with nearby fields. \cite{Versteeg_2023} showed that in the corresponding Galactic longitudes, the degree of polarization of IPS-GI data is also high and low, respectively (see e.g.~their Figure~11). In fact, \citet{Fosalba_2002} observed similar trends in their average optical polarization with LOS across the entire sky.

        %
        \begin{figure}[ht!]
            \epsscale{1.15}
            \plotone{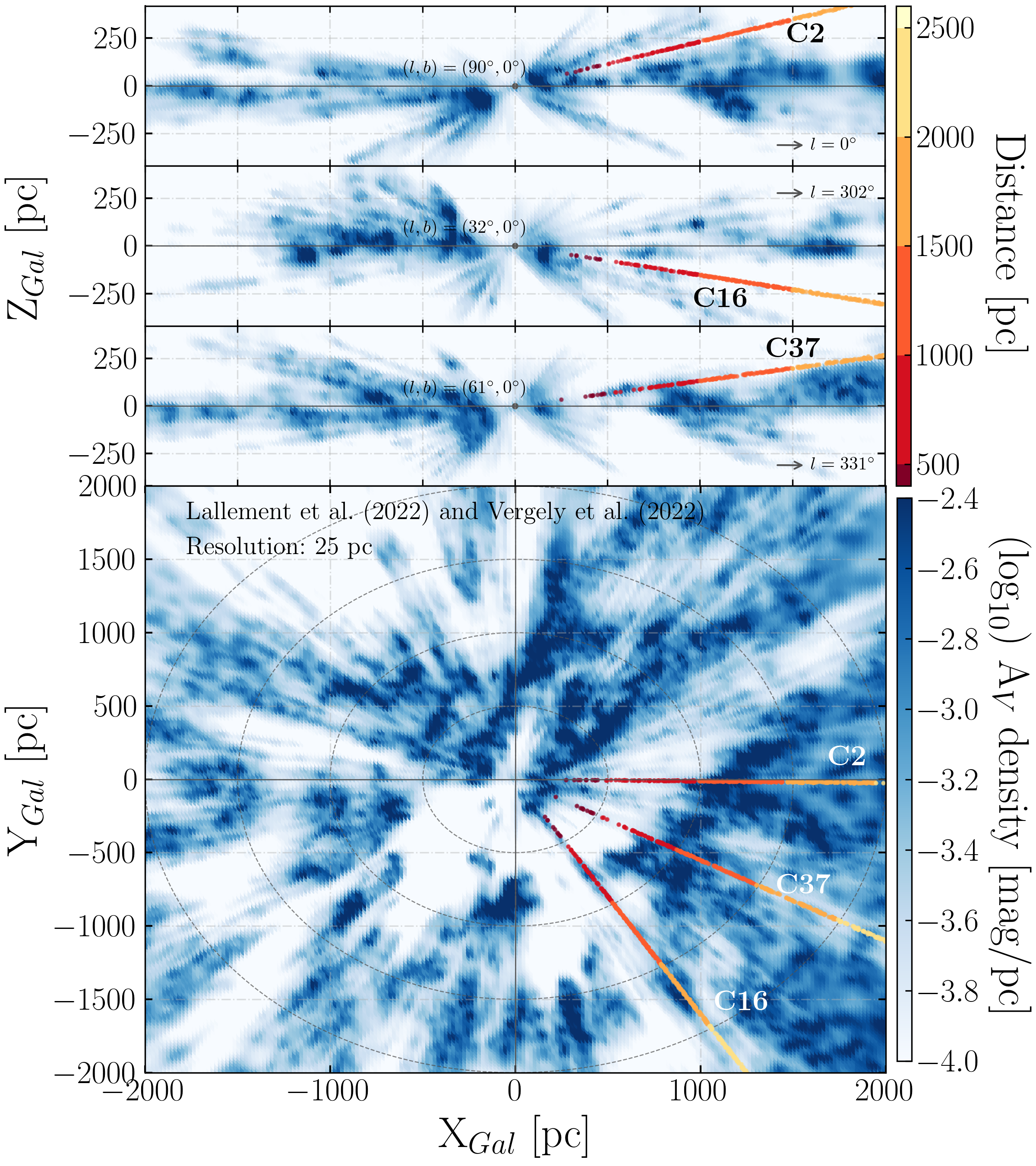}
            \caption{Intermediate Galactic latitude LOS (\textit{C2},  \textit{C16}, and  \textit{C37}) on top of \cite{Lallement_2022,Vergely_2022} dust map, at 25 pc resolution, of the nearby Galaxy ($d<2$~kpc). The blue color bar is the \textit{V}-band extinction density in log space. The red color bar is the distance of the stars in parsecs. \textit{Top}: Edge-on views of the Galaxy centered at different $(l,b)$ coordinates depending on the sight-line. \textit{Bottom}: Face-on view of the Galaxy centered in the Sun at $(0,0)$ pc. The Galactic center, $(l,b) = (0,0)$, is at $(8200,0)$ pc, towards the right, in all maps.  
            \label{fig:int_b_LOS}}
        \end{figure}
        %
        \begin{figure}[ht!]
            \epsscale{1.15}
            \plotone{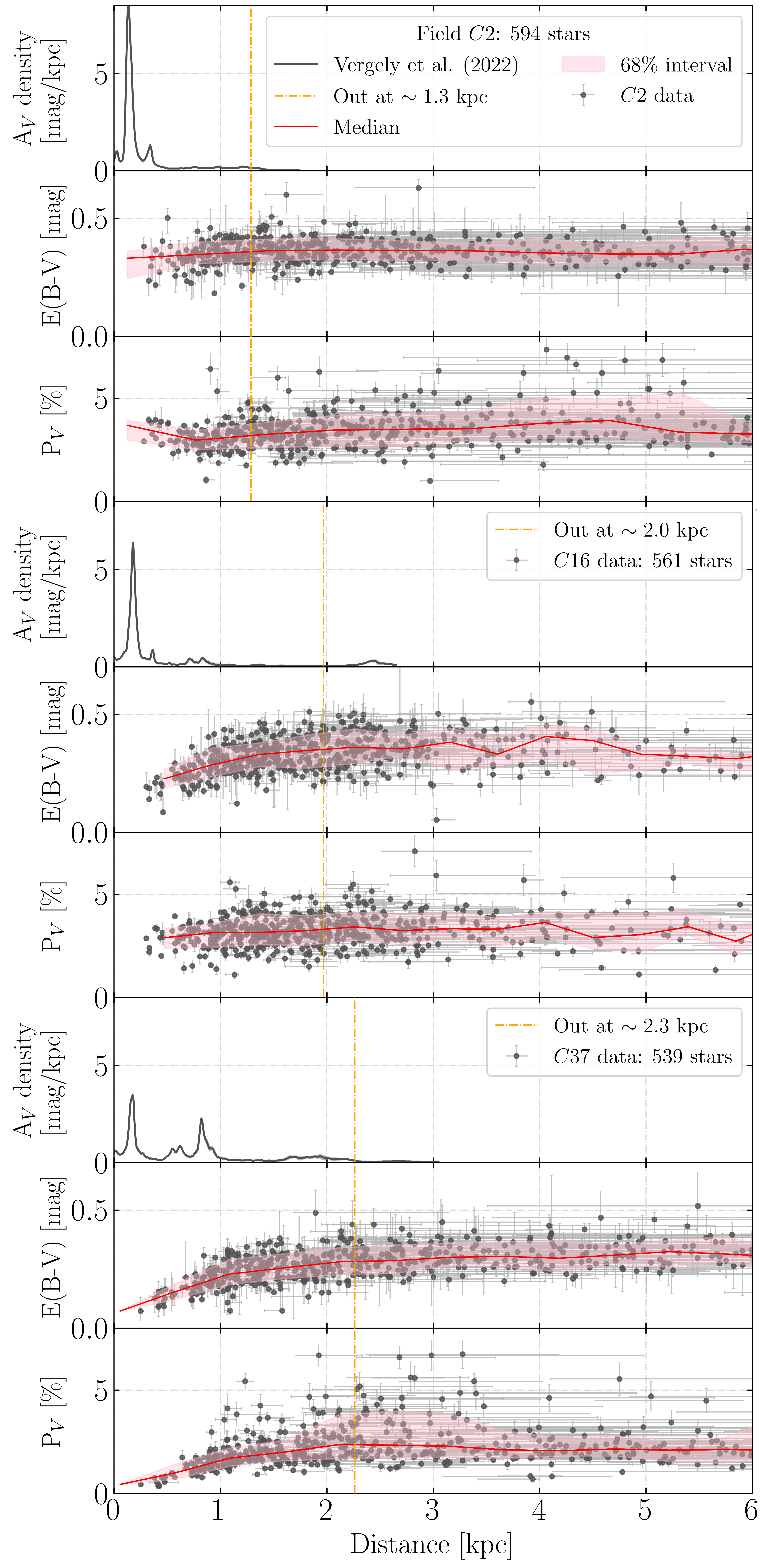}
            \caption{Differential optical extinction profile (25 pc resolution) from \citet{Vergely_2022} and \citet{Lallement_2022} (top row), reddening (middle row), and polarization intensity (bottom row), as a function of distance in fields \textit{C2} (left), \textit{C16} (middle), and \textit{C37} (right). The red solid line is the median value. The pink shaded area is the $16th-84th$ confidence interval. The vertical orange dot-dashed line is the approximated distance at which the sight-line leaves the Galactic thin disk, assuming a scale height of 300 pc.
            \label{fig:field_dist_plot}}
        \end{figure}
        %
        \begin{figure}[ht!]
            \epsscale{2.4}
            \plottwo{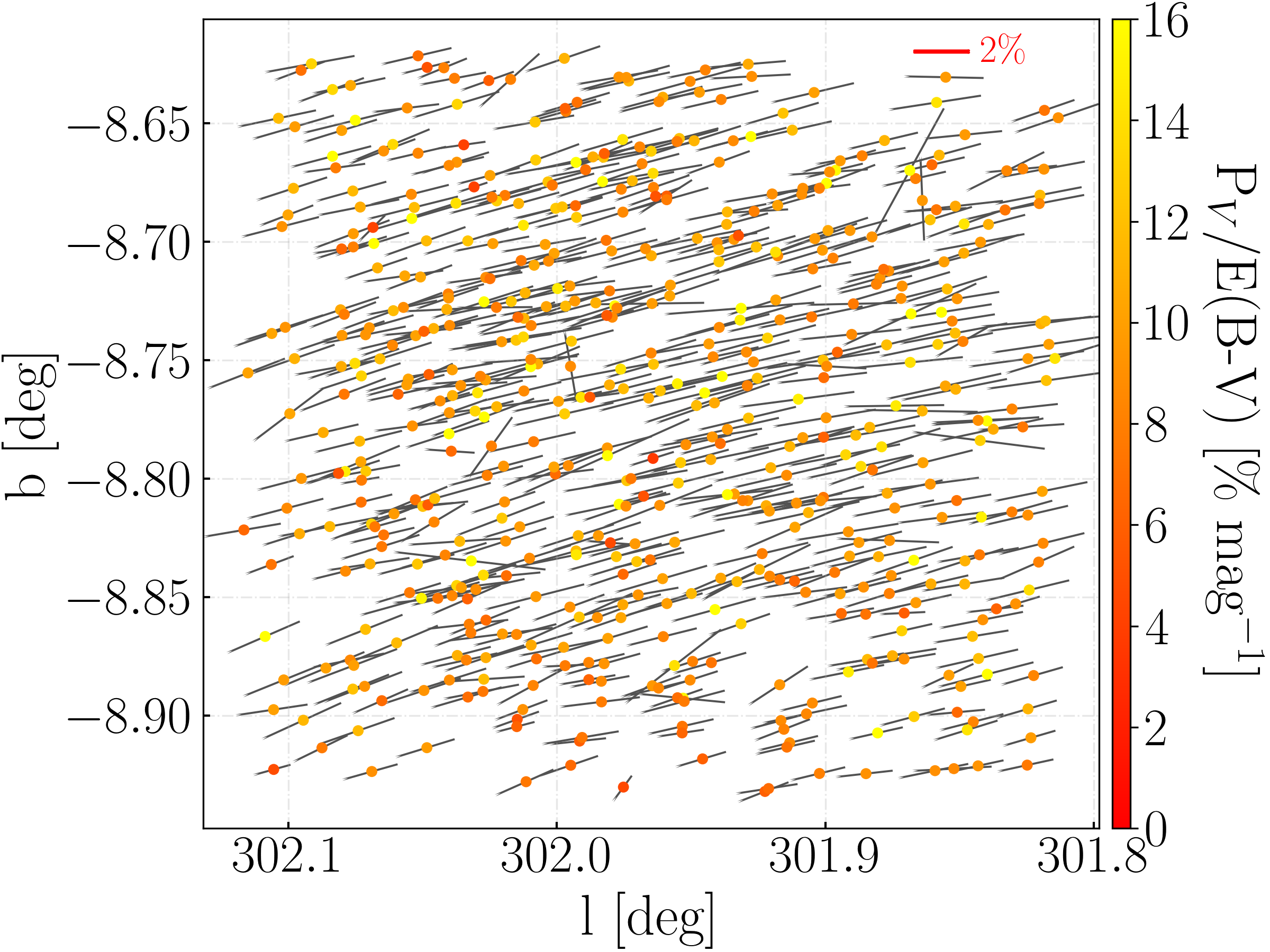}{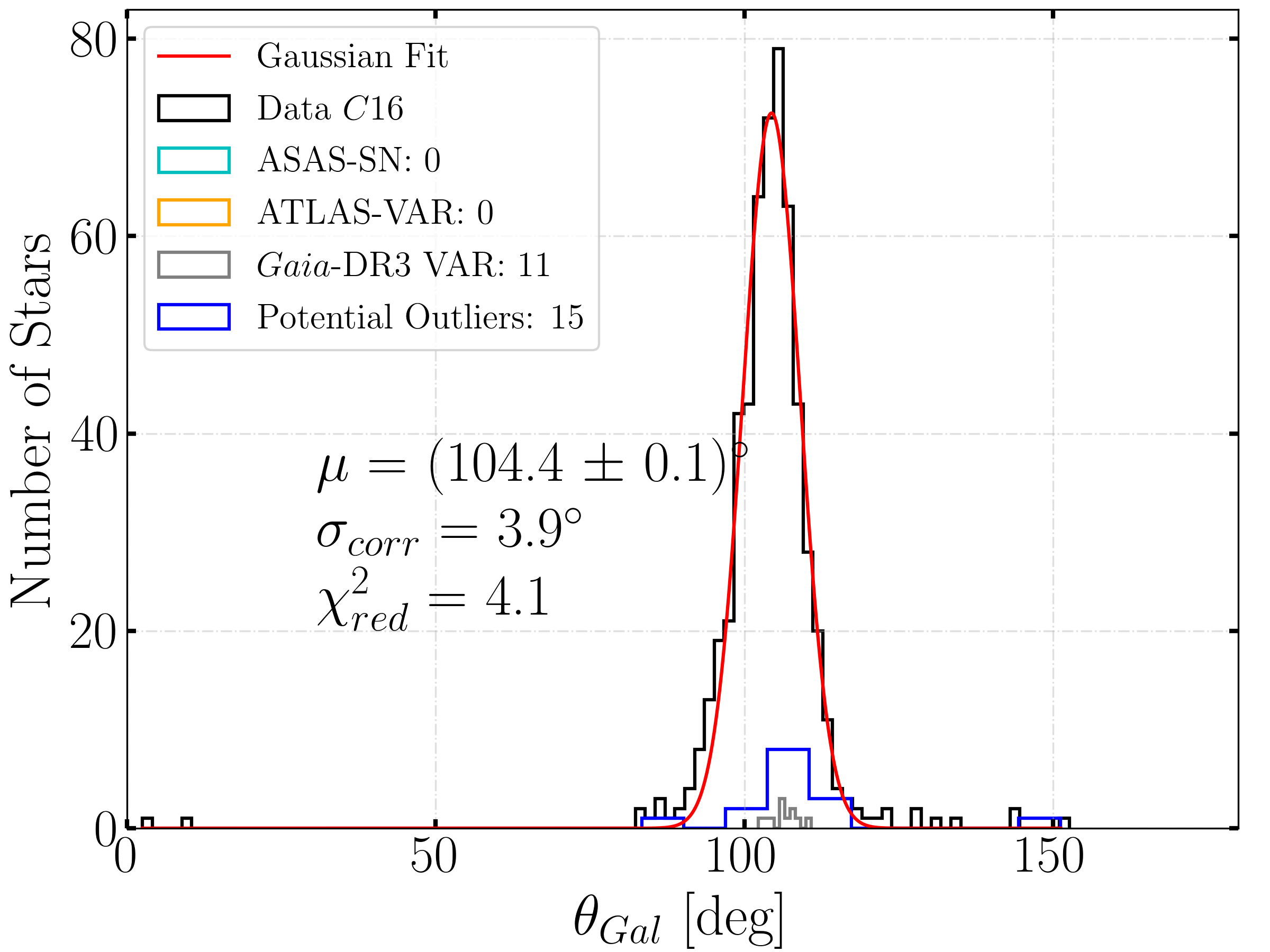}
            \caption{\textit{Top}: Polarization vectors in field \textit{C16} colored by the polarization efficiency. \textit{Bottom}: Polarization angle (in the Galactic frame) of the same field. The dispersion is corrected in quadrature by the measured error average as in \cite{Pereyra_Magalhaes_2007}.
            \label{fig:field_Pvectors}}
        \end{figure}
        %

\section{Discussion} \label{sec:Discu}
        
        \subsection{Variations of P\texorpdfstring{$_V$}{V}/E(\texorpdfstring{\bv}{B-V}) across the Galaxy}\label{subsec:Disc_variations_PE}
        
        \textit{C2}, \textit{C16}, and \textit{C37} are the closest IPS-GI fields to the ideal scenario outlined in Figure~\ref{fig:PvsE_cartoon} (left). The LOS pointing towards $|b|>7.5\degr$ largely avoid the most extinct regions of the Galactic thin disk, as the extinction map of \cite{Vergely_2022} shows (Figure~\ref{fig:int_b_LOS}, top). In a face-on view of the same map (Figure~\ref{fig:int_b_LOS}, bottom), the LOS -- especially field \textit{C16} -- seem to be almost perpendicular to the large Galactic structures. Assuming that the magnetic field lies along those structures, it should be largely perpendicular to these LOS as well. However, the constant reddening and polarization intensity observed from a certain distance (e.g.~see scatter plots in Figure~\ref{fig:field_dist_plot}), as well as the lack of significant structures in the \textit{V}-band extinction density profiles within the same range of distances (e.g.~top panel of the same figure), prove that the LOS leave the Galactic thin disk at some point (see vertical orange dot-dashed lines), and the polarizing dust structure is expected to be within the first \mbox{$\sim$1~kpc}. Hence, the presence of a nearby regular magnetic field in the plane-of-sky and the lack of -- or very little -- additional dust after 1~kpc might explain the high alignment of the polarization vectors within the field-of-view and the remarkably small dispersion of the polarization angle (Figure~\ref{fig:field_Pvectors}). Furthermore, the highly polarized stars observed above 1~kpc in \textit{C2} may be explained by small variations in density within the nearby dust screen (see e.g.~Figures \ref{fig:Fields_PvsE} and \ref{fig:field_dist_plot}). However, we have no evidence of such small structures so far. The properties described above and the high polarization efficiency (Figure~\ref{fig:PE_slopes} and Table~\ref{tab:PE_uplim}) prove that fields \textit{C2},  \textit{C16}, and \textit{C37} are ideal to define the maximum \mbox{P$_V$/E(\bv)}. 

        In the Galactic neighborhood ($d<1$~kpc), intermediate latitude stars have on average a higher degree of polarization (P$_V\sim2.7$\%) than the stars in low-latitude fields (P$_V\sim1.8$\%). Nevertheless, the reddening is almost the same in all directions for all nearby stars due to the low column density built up in short LOS, e.g.~$99\%$ of IPS-GI stars within $d < 1$~kpc have E(\bv) $\leq 0.8$~mag. In consequence, the polarization efficiency is higher in intermediate latitude fields (e.g.~\textit{C2}, \textit{C16}, and \textit{C37}) than low-latitude fields (see the top and bottom-right panels of Figure~\ref{fig:PE_sky}). The above result could indicate either the presence of an increasingly regular magnetic field on the plane-of-sky, less depolarization, a change in polarizing dust properties, or any combination of these conditions towards higher latitudes. However, none of these possible explanations can be proven yet, and the number of intermediate latitude fields in our sample is too low to reach a conclusion. More optical polarization observations at intermediate latitude regions are needed.
        
        At $d > 1$~kpc, intermediate latitude LOS leave the Galactic thin disk (see the vertical orange dot-dashed line in Figure~\ref{fig:field_dist_plot}) and encounter little or no polarizing dust at all (see top panels of Figure~\ref{fig:PE_sky}), then, the starlight polarization of distant stars is produced only in the nearby (foreground) dust structures. Contrary, low-latitude LOS run into more interstellar dust. For instance, $99\%$ of these stars have E(\bv) $\leq 1.6$~mag on average. The starlight polarization may increase but so does the reddening (and perhaps the depolarization), therefore, the polarization efficiency is lower than in intermediate latitude fields. 
        
        \subsection{P\texorpdfstring{$_V$}{V}/E(\texorpdfstring{\bv}{B-V}) models with Galactic longitude}\label{subsec:Disc_PE_model_Jones}
        
            To get an idea of the expected variability of \mbox{P$_V$/E(\bv)} with Galactic longitude, we introduce a simple toy model. We assume that the dust density is constant and dust properties are uniform. We model the large-scale magnetic field in the Solar neighborhood as a uniform field of constant orientation, directed along the local pitch angle at the Sun. For nearby stars, the modeled magnetic field direction corresponds well to a spiral magnetic field,  whereas, for distant stars, the assumption holds less well.
            
            Then, the only aspect that determines the dependence of polarization efficiency on Galactic longitude is the variation of polarizability as a function of the inclination $i$ of the magnetic field with the line of sight. We use two different dust models to determine the dependence on $i$: \cite{Jones_1992} and \cite{Rodrigues_1997}. Both models incorporate dust partially aligned with the magnetic field. 
            
            Firstly, \cite{Jones_1992} define the effective polarization power as $\eta = \kappa_{\perp}/\kappa_{\parallel}$, where $\kappa_{\perp}$ ($\kappa_{\parallel}$) is the extinction coefficient perpendicular (parallel) to the long axis of the grain. The polarization efficiency is proportional to the differential polarization per unit pathlength $dx$ as
            \begin{equation}
                \label{eq:dP_dtau}
                    d\rm P = \frac{1-\eta}{1+\eta}\kappa dx= \frac{1-\eta}{1+\eta} d\tau  ,
            \end{equation}
            and $\eta$ depends on the inclination angle $i$ as
            \begin{equation}
                \label{eq:eta}
                    \eta(i) = \frac{\eta_p+(1-\eta_p)\cos(i) + R}{1+R}  ,
                \end{equation}
            where $\eta_p$ is the polarizing power of aligned grains and $R$ is the ratio of the extinction of the unaligned component to that of the aligned component, $\kappa_u / \kappa_{a\|}$. The polarization efficiency can then be defined as 
            \begin{equation}
                \frac{\rm P}{\rm E(\bv)} = \frac{3.1}{1.086}\frac{\rm P}{\tau} \approx \frac{3.1}{1.086}\frac{1-\eta(i)}{1+\eta(i)}  ,
                \label{eq:pe_jones}
            \end{equation}
            assuming a total to selective extinction ratio R$_V = 3.1$. \citet{Jones_1992} constrain the parameters $\eta_p$ and $R$ by assuming maximum polarization for an inclination angle $i = 90^{\circ}$. They conclude that although $\eta_p$ and $R$ are degenerate, $\eta$ depends very little on their exact values for the allowed parameter ranges. They use $(\eta_p,R) = (0.5,3)$, which we will adopt here as well.
            %
            \begin{figure}[t!]
                \epsscale{1.19}
                \plotone{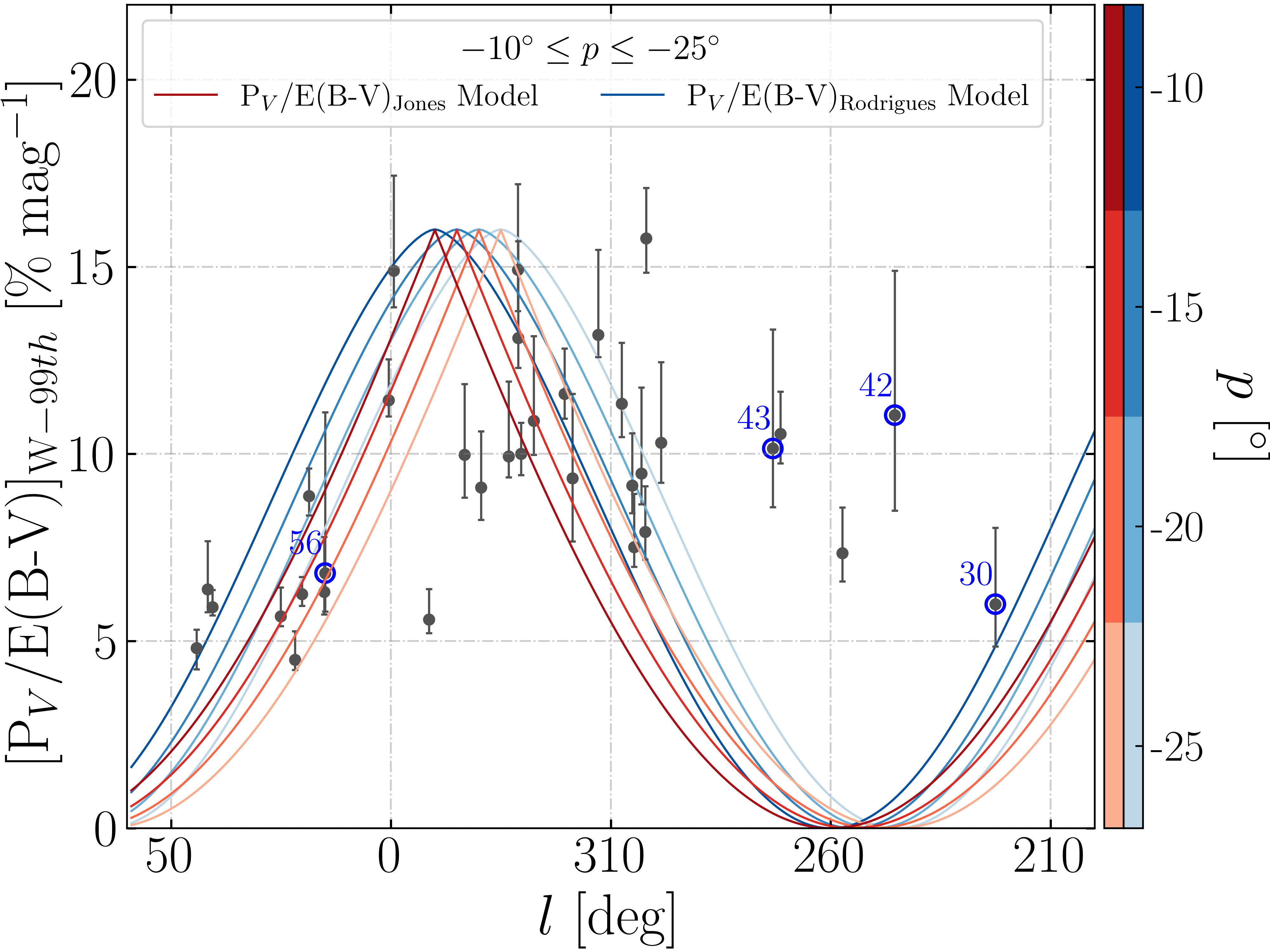}
                \caption{Polarization efficiency upper limits with Galactic longitude. The red and blue curves are the toy models for different pitch angles, $p$.
                \label{fig:PEmodels}}
            \end{figure}
            %
            
            Secondly, \citet{Rodrigues_1997} developed a dust model of spherical and cylindrical grains for the Milky Way and Small Magellanic Cloud data: infrared to ultraviolet extinction and optical polarization in broad bands. The dichroic absorption of the Galactic model was derived for different inclination angles by Santos-Lima  et al (2023, in preparation), who found a best-fit relation of
               \begin{equation}
                \frac{\rm P}{\rm E(\bv)} \propto 1 - \sin^{1.57}(i).
            \label{eq:pe_rodrigues}
            \end{equation}
            This model has an arbitrary scaling, which we took to match the scaling in the previous model. 
            %
            \begin{figure*}[t!]
            \gridline{
                      \leftfig{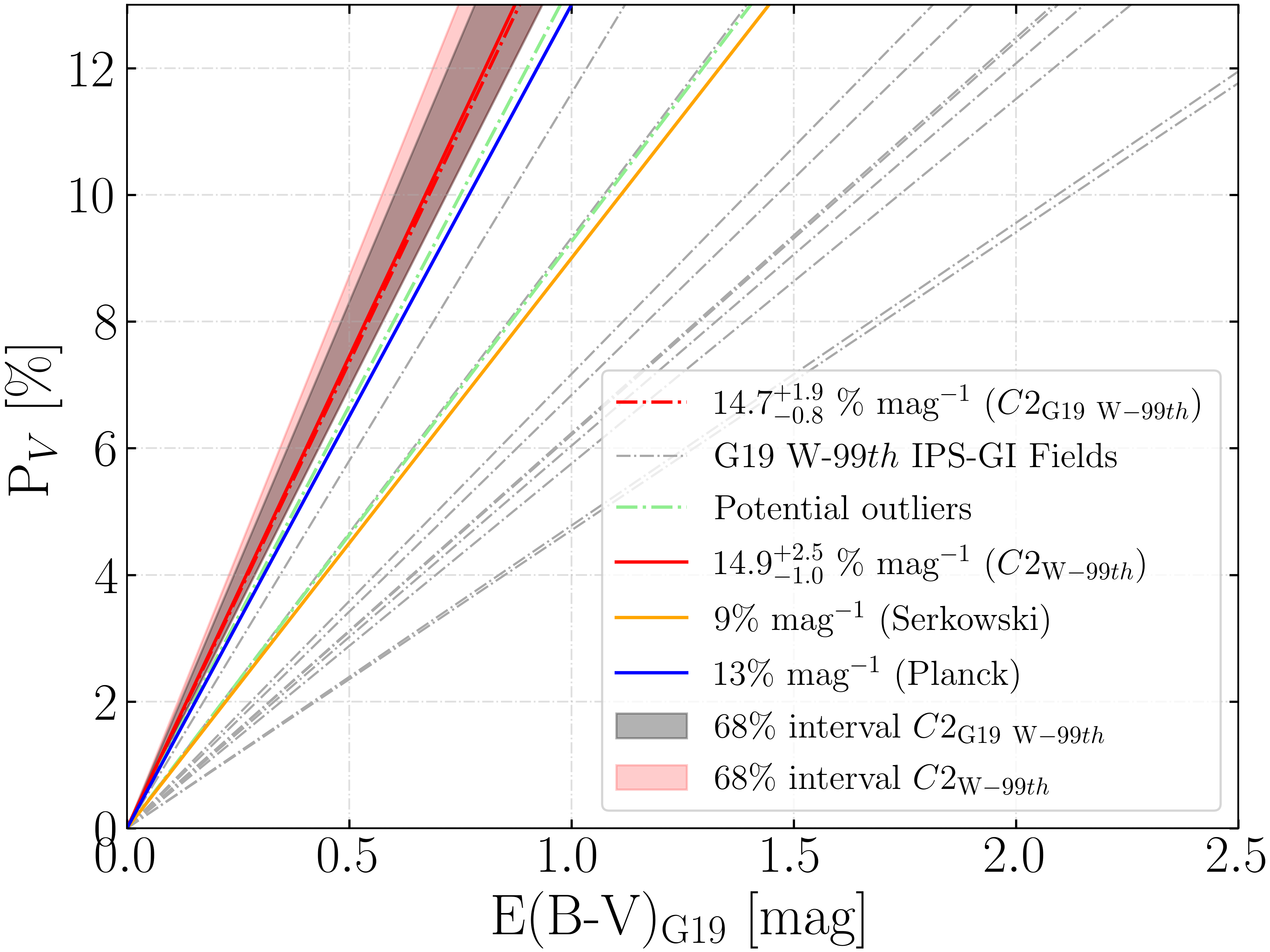}{0.33\linewidth}{}
                      \fig{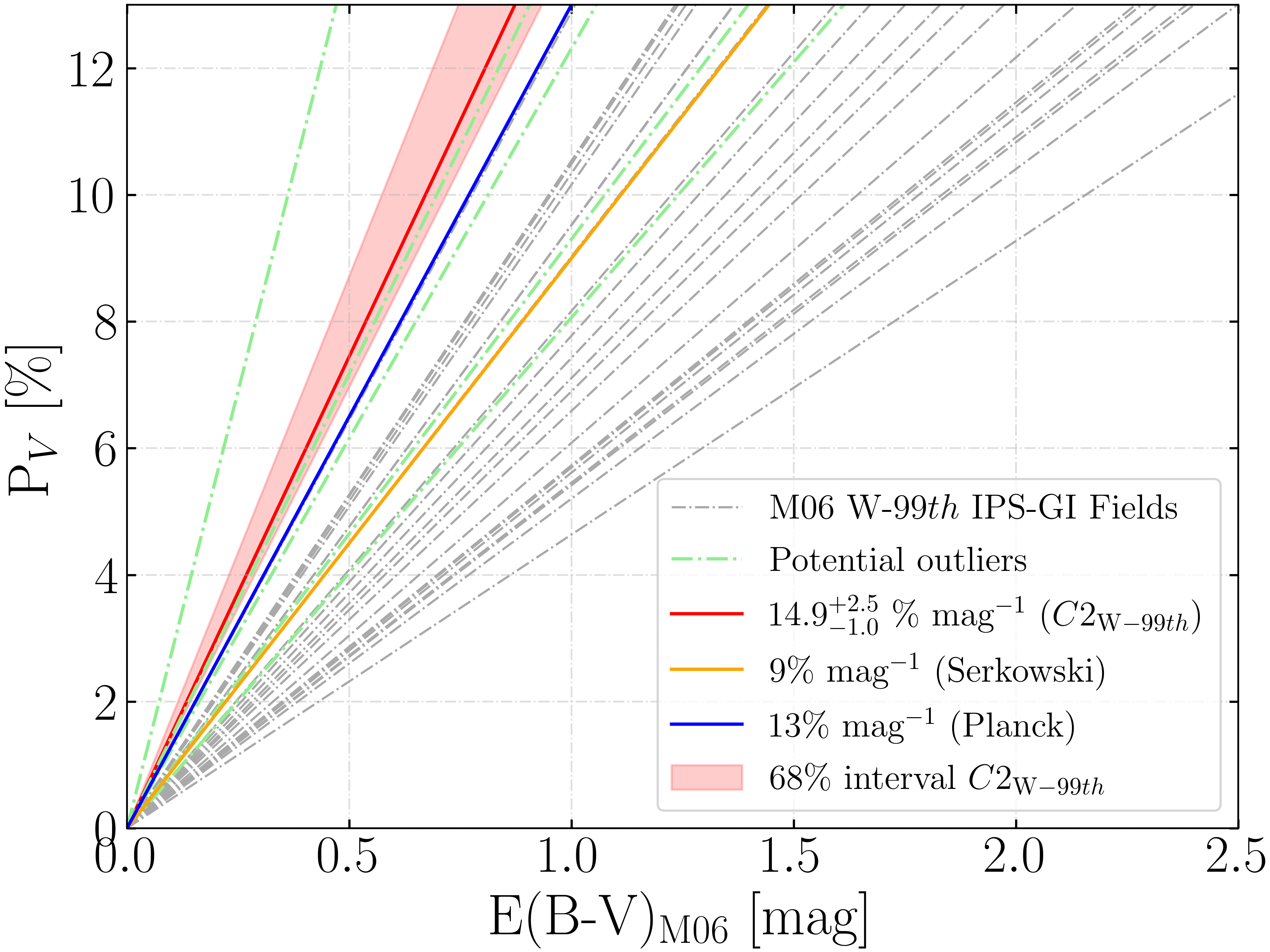}{0.33\linewidth}{}
                      \rightfig{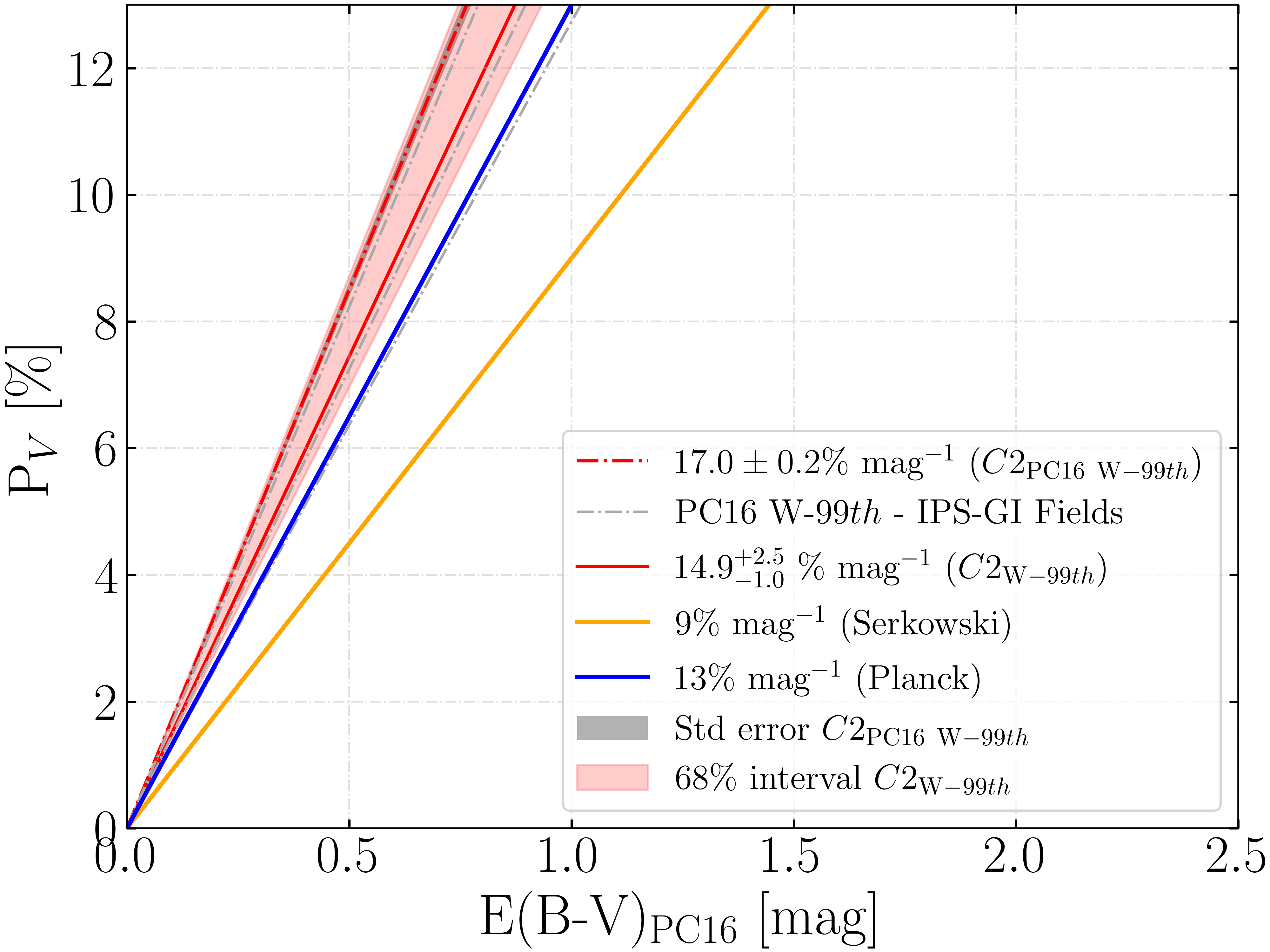}{0.33\linewidth}{}
                      }
            \caption{Similar to Figure~\ref{fig:PE_slopes} but using \citetalias{Green_2019} (left), \citetalias{Marshall_2006} (middle), and \citetalias{Planck-Collaboration_2016} (right) reddening to calculate the upper limits on each IPS-GI field (gray dashed lines). The red dashed lines show the upper limit of field \textit{C2}. Light-green dashed lines are unreliable upper envelopes due to potential outliers or unreliable reddening measurements. For \citetalias{Planck-Collaboration_2016}, only the intermediate-latitude fields are plotted, see Section~\ref{subsec:Ebv_comp}.
            \label{fig:PE_slopes_G19_M06_P16}}
            \end{figure*}
            
            In our simple picture, the inclination angle is related to Galactic longitude as $i = \pi/2 - l + p$, where $p$ is the pitch angle, which we vary between -10$\degr$ and -25$\degr$ \citep{vallee_2015}. Figure~\ref{fig:PEmodels} presents the final models of polarization efficiency with Galactic longitude. The toy models indeed reproduce the high values at $330\degr \leq l \leq 360\degr$, and low \mbox{P$_V$/E(\bv)} at $l\sim 30\degr-80\degr$ and possibly $l\sim 235\degr-285\degr$, where the LOS are expected to be parallel to the magnetic field. High-pitch angles appear to best reproduce our observations of high polarization efficiency, also agreeing with the local pitch angle of the large-scale dust structures observed in \cite{Vergely_2022} maps (Figure~\ref{fig:int_b_LOS}). Nevertheless, the models fail to describe our observations around $l \sim 285\degr$. These are the cases of fields  \textit{C35}, \textit{C41}, \textit{C42}, and \textit{C43}, in which the polarization efficiencies are not the lowest. Besides potential outliers, it is clear proof that the magnetic field is not parallel to the LOS in these particular fields. Toward these Galactic longitudes, we find the Gum nebula, which radio polarimetry properties were studied in detail by \citet{Purcell_2015}, who showed that the entire structure has a polarization angle between 43$\degr$ and 55$\degr$ at an adopted distance of $\sim450$~pc. Although the magnetic field orientation in field \textit{C41} agrees with the values found by \citet{Purcell_2015}, there is no evidence of dust inside the nebula. In fact, the polarizing dust structures are likely behind it (see e.g. the lower panel of Figure~\ref{fig:int_b_LOS} in the South-South-West direction).

            Despite all simplifications and assumptions made, the toy models are close to IPS-GI observations. This remains true even when considering other widely used dust models \citep[see e.g.~][and references therein]{Andersson_2015}, such as the approximation of the observed polarization to $\propto \mathrm{cos}^2(\gamma)$ from \cite{Lee&Draine_1985} dust grain models, which gives a slightly broader curve. Nevertheless, our data show that \mbox{P$_V$/E(\bv)} is also a function of the latitude as well as of the longitude. Additionally, nearby ($d<1$~kpc) small-scale and meso-scale structures are critical in GMF modeling. A future version of these models must consider the different inclinations of the magnetic field along the sight-line and variations with Galactic latitude.
            
    \subsection{Implications for dust models} 
    \label{subsec:Discu_implic_dust_models}

        A comprehensive interstellar dust model should be able to reproduce the observed extinction, emission, and polarization in the entire spectral domain. It should therefore account for dust grains composition, size distributions, and alignment properties. (\citealt{Rodrigues_1997, Jones_2015, Hensley_2021}, and references therein). For instance, \cite{Hensley_2021} synthesized the most important observational constraints for multi-wavelength dust models on the diffuse ISM, showing the importance of polarization efficiency due to the origin of optical polarization (short axis of elongated interstellar dust grains aligned preferably parallel to the local magnetic field). \cite{Rodrigues_1997} studied the interstellar ultraviolet extinction and optical polarization of the Small Magellanic Cloud using dust models based on the \cite{Mathis_1977} prescription, with different grain size distributions and composition. They also fitted the ``standard'' Galactic extinction and polarization curves. The latter was built assuming a maximum polarization efficiency of 9\%~mag$^{-1}$. The best fit for the wavelength dependencies of polarization and extinction determines the grain size ranges, alignment properties, composition distribution, and the depletion of \ch{Si} and \ch{C} in the interstellar medium. Interestingly, the resulting maximum polarization efficiency of these models is greater than 9\%~mag$^{-1}$. \cite{Rodrigues_1997} presented models for three specific angles (10$\degr$, 30$\degr$, and 60$\degr$) between the magnetic field and the plane of the sky. Extrapolating their results for an angle of 0$\degr$, using Equation~\ref{eq:pe_rodrigues}, we obtained polarizing efficiencies between 14.4\%~mag$^{-1}$ and 16.5\%~mag$^{-1}$ (depending on the model), in close agreement with our results for the maximum polarization efficiency of $15.8$\%~mag$^{-1}$. However, a more careful prediction for the grain properties that could produce such high polarizing efficiency requires new modeling.

    \subsection{Dependency of P\texorpdfstring{$_V$}{V}/E(\texorpdfstring{\bv}{B-V}) on dust maps} \label{subsec:Discu_PE_SH22_G19_M06}
     
        Previous studies have shown that the polarization efficiency upper limit may be highly dependent on the dust map used \citep[see e.g.][]{Panopoulou_2019}. It is noteworthy that the weighted polarization efficiency upper envelopes calculated (for the sake of comparison) with the reddening of \citetalias{Green_2019}, [P$_V$/E(\bv)]$_{\mathrm{G19_{W-99th}}}$, and \citetalias{Planck-Collaboration_2016}, [P$_V$/E(\bv)]$_{\mathrm{PC16}_{W-99th}}$  (Figure~\ref{fig:PE_slopes_G19_M06_P16}), also show values higher than 13$\%$~mag$^{-1}$, and very consistent with the upper limits observed in fields \textit{C2}, \textit{C16}, and \textit{C37} using \citetalias{Anders_2022} reddening. Nonetheless, one should be aware that low-resolution dust maps may lead to polarization efficiency overestimation since multiple polarization measurements can have the same pixel value, and the information on the small-dense structures is lost. 
        
        On the other hand, the weighted upper limits calculated with \citetalias{Marshall_2006} reddening, [P$_V$/E(\bv)]$_{\mathrm{M16_{W-99th}}}$ (Figure~\ref{fig:PE_slopes_G19_M06_P16}, middle panel), are lower in average. Few exceptions must be carefully compared (light-green slopes in Figure~\ref{fig:PE_slopes_G19_M06_P16}: fields \textit{C42} and \textit{C56} in the left, and \textit{C14}, \textit{C43}, \textit{C55}, \textit{C56} and \textit{C61} in the middle) since poor statistics, and sometimes poor measurements (see e.g.~the discussion about the dust maps quality in Section~\ref{subsec:Ebv_comp}), still affect the weighted \mbox{P$_V$/E(\bv)} upper limit calculation leading to large uncertainties (Section~\ref{subsec:P_vs_E_Cal_PE_uplim}). Unfortunately, we are missing information on fields that are not covered by the aforementioned databases; not to mention the limitations of dust maps explained in Section~\ref{subsec:Ebv_comp}.
        %
        \begin{figure}[t!]
            \epsscale{1.15}
            \plottwo{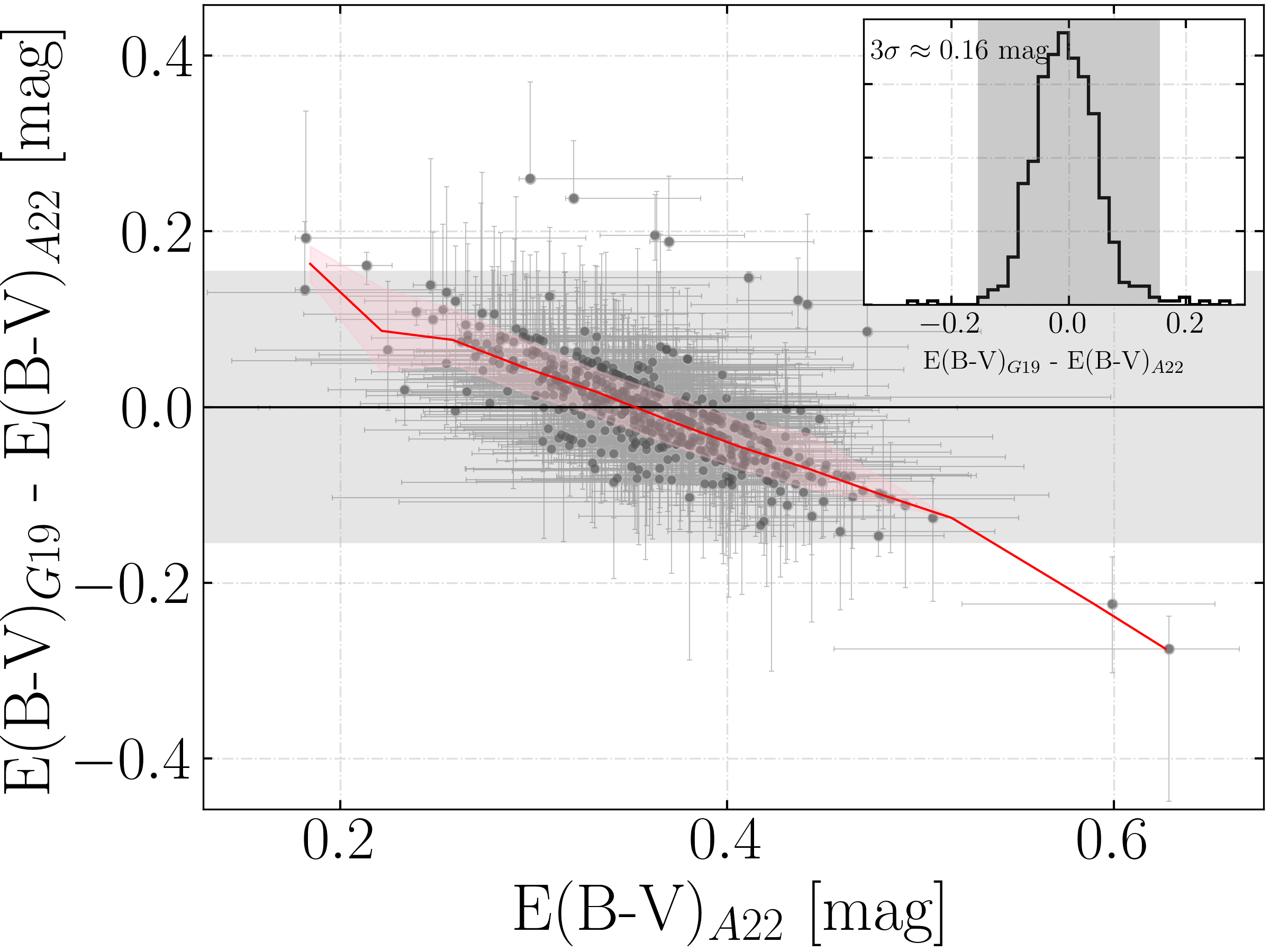}{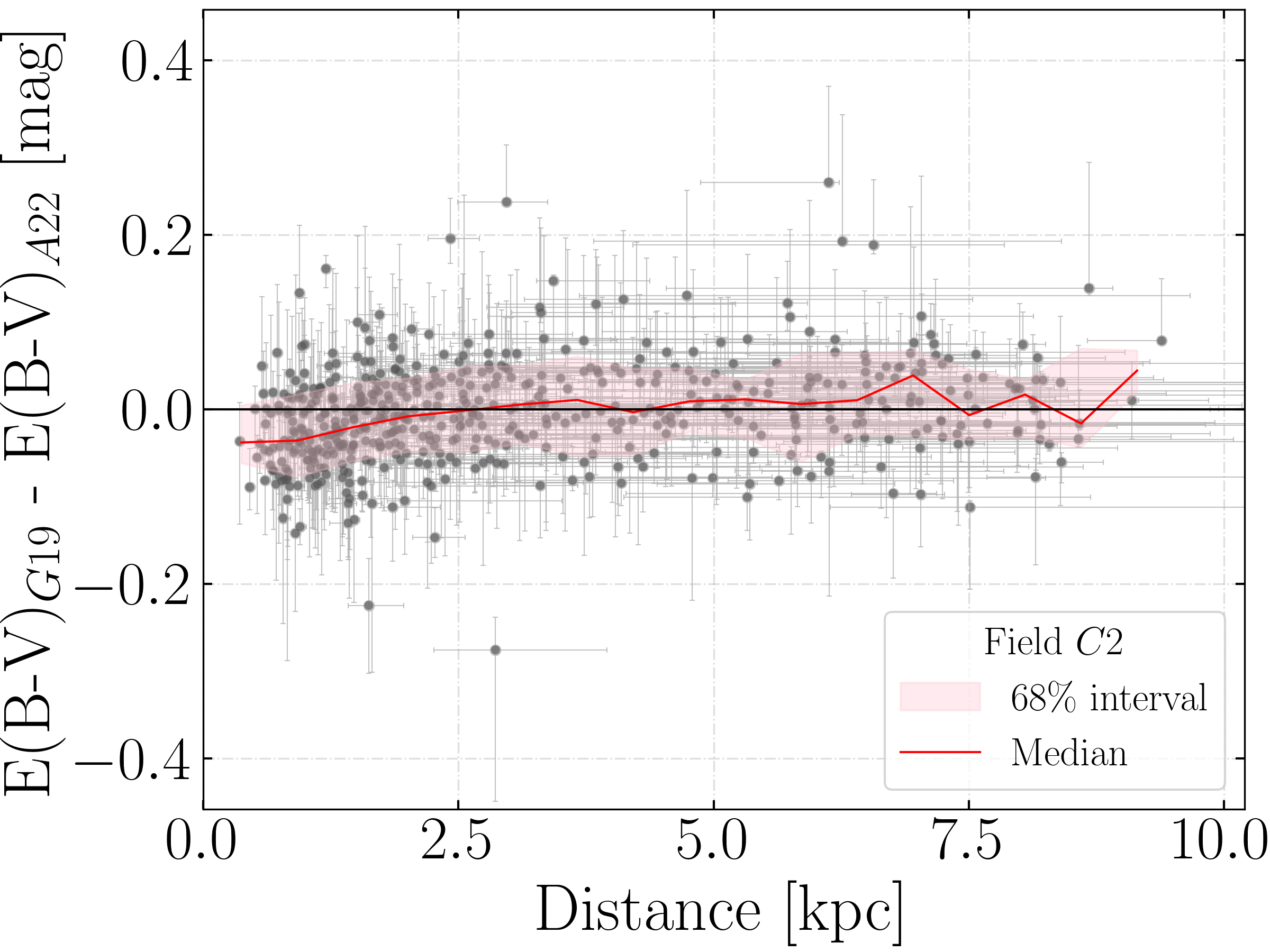}
            \caption{Similar to the systematic reddening differences of Figure~\ref{fig:Green-Stil_vs_SH} but only for field \textit{C2}. The histogram shows the distribution of the difference and the values under the $3\sigma$ limit (the gray-shaded area).
            \label{fig:Green_vs_SH_field2}}
        \end{figure}
        
        In particular, \citetalias{Green_2019} and \citetalias{Planck-Collaboration_2016} have maximum weighted upper limits in field \textit{C2} ($14.7^{+1.9}_{-0.8}\%$~mag$^{-1}$ and $17.0\pm0.2\%$~mag$^{-1}$, respectively) remarkably consistent (within the confidence interval) with the maximum value \mbox{$15.8^{+1.3}_{-0.9}\%$~mag$^{-1}$} (Section~\ref{subsec:Resul_PEff_max}). In the range of \mbox{0.2~mag $\leq$ E(\bv) $\leq$ 0.5~mag} of field \textit{C2}, a systematic errors of $\sim$0.16~mag and $\sim$0.14~mag are expected with respect to \citetalias{Green_2019} and \citetalias{Planck-Collaboration_2016} respectively (see e.g.~Figure~\ref{fig:Green_vs_SH_field2}). Meanwhile, at near distances ($d < 2$~kpc) \citetalias{Anders_2022} reddening are $\sim$0.05~mag higher than \citetalias{Green_2019}'s. Given that the highest polarization efficiency is measured within the first kiloparsec, the systematic error in reddening between \citetalias{Green_2019} and \citetalias{Anders_2022} would contribute towards lower values of \citetalias{Anders_2022} reddening and, therefore, towards higher polarization efficiency.
        
\begin{figure}[t!]
            \epsscale{1.15}
            \plotone{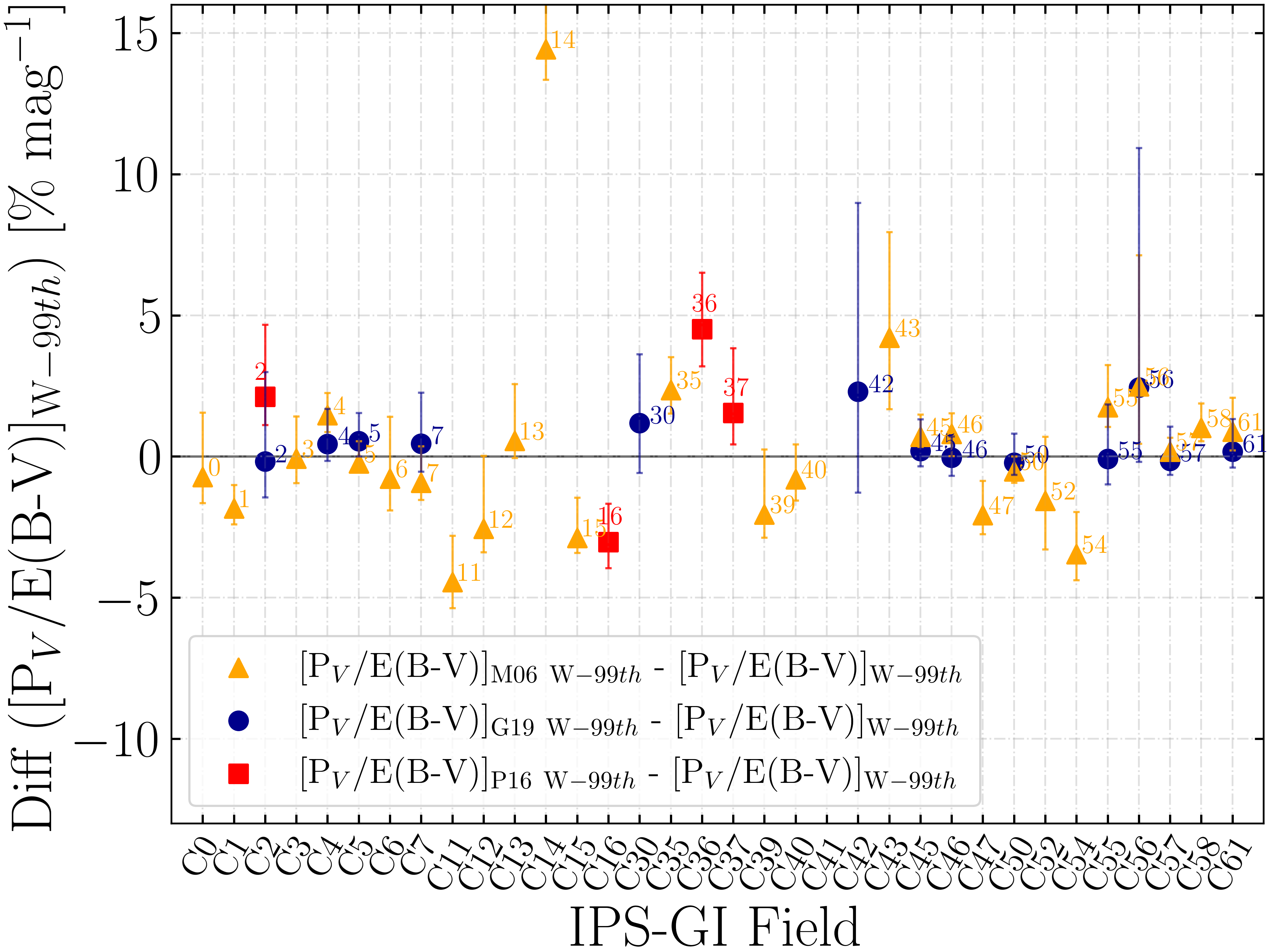}
            \caption{Systematic difference between the weighted polarization efficiency upper envelopes calculated with \citetalias{Green_2019} (Dark blue circles), \citetalias{Marshall_2006} (orange triangles), and \citetalias{Planck-Collaboration_2016} (red squares) reddening, and those calculated with \citetalias{Anders_2022} reddening.
            \label{fig:PE_SH-G19-M06_diff-comp}}
        \end{figure}

        Figure~\ref{fig:PE_SH-G19-M06_diff-comp} shows the difference between the weighted polarization efficiency upper limits calculated with the different dust maps considered in this work (Section \ref{subsec:dustmaps}). The difference between the \citetalias{Marshall_2006} and \citetalias{Anders_2022} weighted upper limits have a large scatter, often below $\sim$5\%~mag$^{-1}$, which may reflect the limitations of \citetalias{Marshall_2006} dust map. The difference between \mbox{[P$_V$/E(\bv)]$_{\mathrm{G19_{W-99th}}}$} and \mbox{[P$_V$/E(\bv)]$_{\mathrm{W-99th}}$}, on the other hand, is lower and very close to zero with very small scatter. In Sections~\ref{subsec:Ebv_comp} and~\ref{subsec:Resul_PE_up_fields}, and Figure~\ref{fig:PE_SH-G19-M06_diff-comp}, we showed that \citetalias{Anders_2022} and \citetalias{Green_2019} are the most consistent, which is somewhat expected since \citetalias{Green_2019} dust map was used in the construction of \citetalias{Anders_2022}'s extinction prior. Even in fields with large uncertainties, both calculations are equally affected by the potential outliers, i.e.~\textit{C30}, \textit{C42}, and \textit{C56}, except \textit{C43} where \citetalias{Green_2019} does not have measurements. Due to the small difference, it is very likely that the trend would be the same in fields where there is no data from \citetalias{Green_2019} yet. In fact, high polarization efficiency calculated with an updated version of \citetalias{Green_2019} dust map would be expected in the intermediate Galactic latitude regions, where \mbox{[P$_V$/E(\bv)]$_{\mathrm{W-99th}}$} is high as well.
        %
        %

        The intermediate latitude fields, where multiple dust maps are available, show high polarization efficiency regardless of the dust map used, i.e.~\citetalias{Green_2019} and \citetalias{Planck-Collaboration_2016}. Recall that \citetalias{Planck-Collaboration_2016} is only comparable with \citetalias{Anders_2022} at intermediate and high latitudes, i.e.~in diffuse LOS, as explained in Section~\ref{subsec:Ebv_comp}. The values found in this work for the \mbox{[P$_V$/E(\bv)]} upper limits, which are higher than past estimations, are unlikely to be atypical measurements and are very consistent with the factors compared here. 
        
        %
        \begin{figure}[t!]
            \epsscale{2.35}
            \plottwo{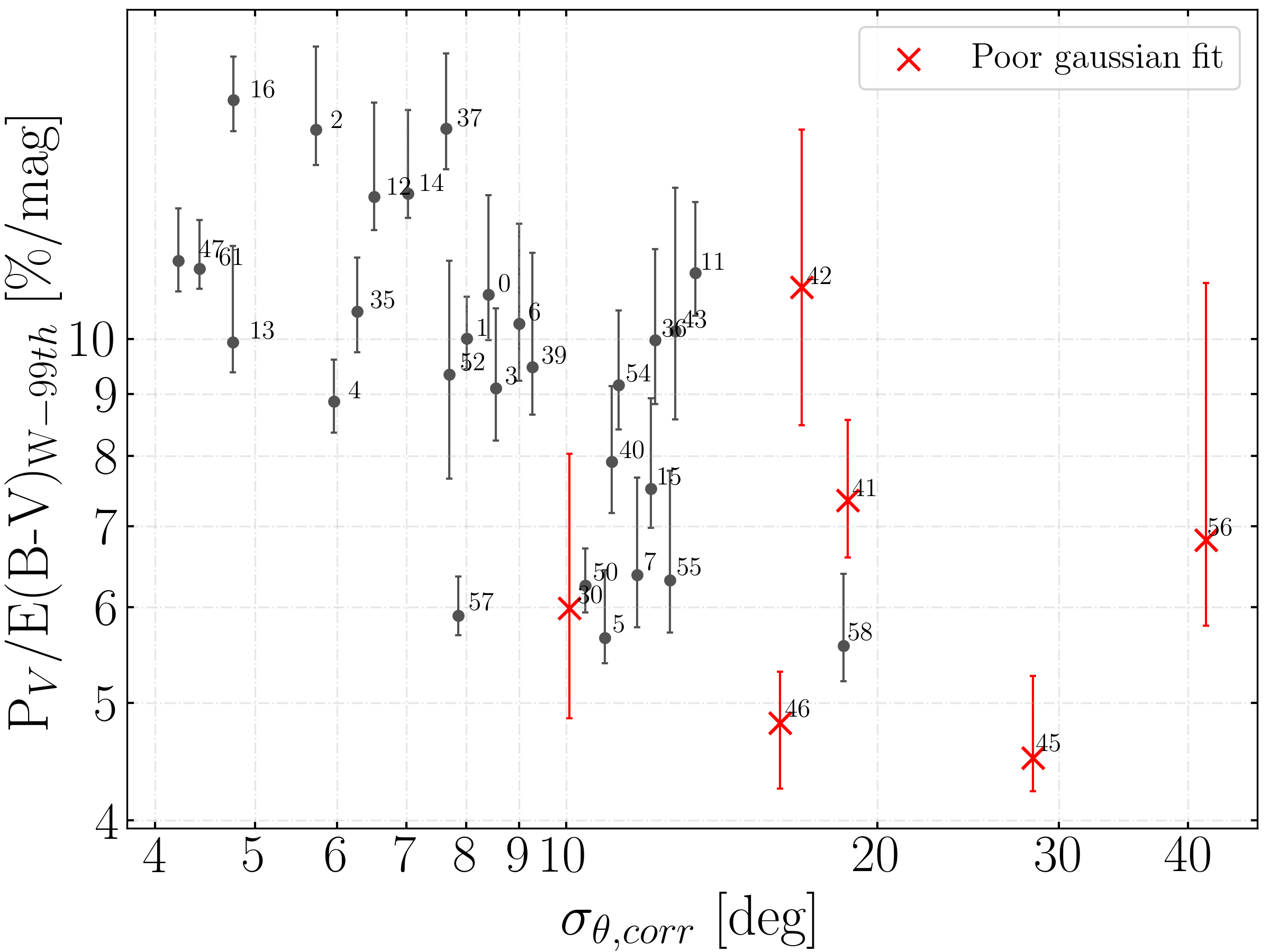}{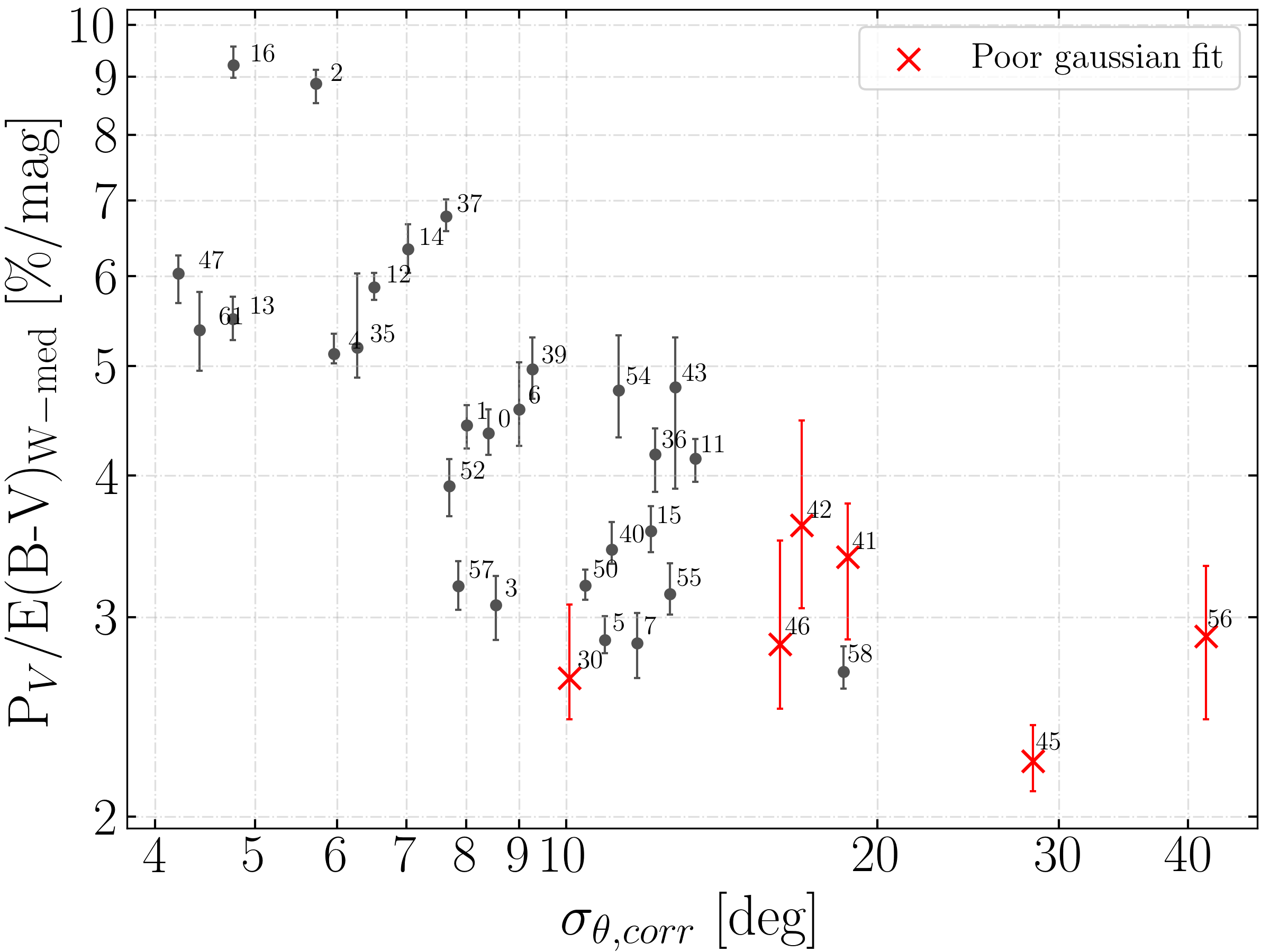}
            \caption{Weighted polarization efficiency upper envelope (top) and weighted median polarization efficiency (bottom) as a function of the dispersion of the polarization angle corrected by the average of the measured errors as in \cite{Pereyra_Magalhaes_2007}. Bad Gaussian fits of the polarization angle distribution are indicated with red ``X'' markers.
            \label{fig:PE_vs_dispPA}}
        \end{figure}

    \subsection{Correlation with the polarization angle dispersion} \label{subsec:PEff_vs_dispPA}

        The polarization angle in most of the IPS-GI fields is remarkably coherent (see e.g.~Figure~\ref{fig:field_Pvectors}); i.e.~approximately 84\% of the fields showed a dispersion of the polarization angle below $15\degr$ and 50\% is below  $10\degr$ \citep{Versteeg_2023}. The dispersion in polarization angle indicates the fluctuations in the orientation of the plane-of-sky component of the magnetic field. For nearby stars (\mbox{$d < 1$~kpc}), the projected size of the IPS-GI fields is smaller than 5 pc. Both scales are consistent with interstellar turbulence and fluctuations due to small-scale structures (e.g.~supernova remnants, \ion{H}{2} regions, cold dark clouds, among others). However, for distant objects (\mbox{$d > 1$~kpc}),
        and considering the small size of the IPS-GI fields in relation to the long pathlengths, these variations in magnetic field orientation are expected to be dominated by meso-scale fluctuations along the LOS.

        Figure~\ref{fig:PE_vs_dispPA} shows that fields with high polarization efficiency often have a lower dispersion in polarization angle than low polarization efficiency fields. The anti-correlation between the polarization angle dispersion and the weighted polarization efficiency upper envelope is marginal (the Spearman correlation coefficient is $-0.60$, above $3\sigma$, and the two-tailed \textit{p}-value is \mbox{$2.0\times 10^{-4}$}). However, the anti-correlation with the weighted median of \mbox{P$_V$/E(\bv)} is more significant (the correlation coefficient is $-0.74$ and the two-tailed \textit{p}-value is \mbox{$5.2\times 10^{-7}$}). 
        
        This result strengthens the idea of a highly regular magnetic field in the plane-of-sky, low depolarization, and a small random magnetic field component in IPS-GI fields with high polarization efficiencies. It is also consistent with the anti-correlation found by \cite{Planck-Collaboration_2018_20} between the polarization angle dispersion and the polarization fraction, which they attribute to depolarization by the turbulent magnetic fields of different structures along the sight-line. Nevertheless, Doi et al. (2023, in preparation) demonstrated that such depolarization may be also due to multiple, local regular magnetic field structures with very different orientations ($90\degr$ offset).

\section{Summary} \label{sec:Conclu}

    We used optical polarization maps of 34 sky fields sampling the general interstellar medium. Each field covers an area of $0.3\degr \times 0.3\degr$ square and has polarization measurements of typically hundreds of stars. These measurements were combined with the optical extinction and \textit{Gaia}-EDR3 distances from \citetalias{Anders_2022} to study the polarization efficiency in the interstellar medium. 

    We propose a new method to estimate the polarization efficiency upper limit considering the asymmetric uncertainties of the measurements. This method is robust against outliers and avoids misleading estimations of \mbox{[P$_V$/E(\bv)]$_{\mathrm{max}}$}. The upper limit carries information on dust and interstellar magnetic field properties. It is a powerful diagnostic when comparing regions in the sky and an important parameter for interpreting starlight polarization measurements to numerical simulations of the magnetized ISM (e.g., see Santos-Lima et al. 2023, in preparation). Moreover, \mbox{[P$_V$/E(\bv)]$_{\mathrm{max}}$} represents an important observational constraint for interstellar dust models (Section~\ref{subsec:Discu_implic_dust_models}). 

    The maximum polarization efficiency in our sample is \mbox{[P$_V$/E(\bv)]$_{C16} = 15.8^{+1.3}_{-0.9}\%$~mag$^{-1}$}. This high polarization efficiency, observed in a number of  intermediate latitude fields ($b>7.5\degr$), is higher than \textit{Planck}'s upper limit (13$\%$~mag$^{-1}$), which is averaged over the whole high-latitude sky, independently of the dust map used (\citetalias{Anders_2022, Green_2019, Marshall_2006, Planck-Collaboration_2016}). The high \mbox{P$_V$/E(\bv)} upper envelope is consistent with a nearby (\mbox{$d<1$~kpc}) highly regular GMF on the plane-of-sky and the lack of depolarization due to additional dust structures beyond \mbox{1~kpc}, as it was demonstrated by the observed degree of polarization and reddening with distance, and the anti-correlation with the polarization angle dispersion. 
    
    The IPS-GI data show variable polarization efficiencies in the general ISM with Galactic coordinates broadly consistent with toy models, in Galactic longitude, that assume a uniform local magnetic field with a reasonable pitch angle consistent with the literature. However, our observations demonstrate that interstellar optical polarization is highly dependent on nearby small-scale and meso-scale structures. Moreover, the toy models do not account for \mbox{P$_V$/E(\bv)} variations with Galactic latitude. Therefore, comprehensive modeling is needed to accurately explain variations of the polarization efficiency across the Galaxy.

\begin{acknowledgments}
    The authors thank the careful reading of the anonymous referee whose review helped improve this paper.

    YAA acknowledges Dr. M. Riello for his contributions in discussions about the photometric validation of \citetalias{Gaia_Collaboration_2021b} and the determination of the \textit{variability proxy}, a key \textit{Gaia} metric to find potential variable sources.
    
    YAA is also grateful to Dr. Andrey Akinshin for his insights and the mathematical development of a weighted quantile estimator.
    
    The authors acknowledge the Interstellar Institute's program ``With Two Eyes'' and the Paris-Saclay University's Institut Pascal for hosting discussions that nourished the development of the ideas behind this work.
    
    Over the years, IPS data have been gathered by a number of dedicated observers, to whom the authors are very grateful: Flaviane C. F. Benedito, Alex Carciofi, Cassia Fernandez, Tib\'erio Ferrari, Livia S. C. A. Ferreira, Viviana S. Gabriel, Aiara Lobo-Gomes, Luciana de Matos, Rocio Melgarejo, Antonio Pereyra, Nadili Ribeiro, Marcelo Rubinho, Daiane B. Seriacopi, Fernando Silva, Rodolfo Valentim, and Aline Vidotto.
    
    YAA, MH, and MJFV acknowledge funding from the European Research Council (ERC) under the European Union’s Horizon 2020 research and innovation programme (grant agreement No 772663).
    
    CVR acknowledges support from {\it Conselho Nacional de Desenvolvimento Científico e Tecnológico} - CNPq (Brazil) (Proc.~310930/2021-9).
    
    AMM's work and optical/NIR polarimetry at IAG have been supported over the years by several grants from S\~ao Paulo state funding agency FAPESP, especially 01/12589-1 and 10/19694-4. AMM has also been partially supported by the Brazilian agency CNPq (grant 310506/2015-8). AMM graduate students have been provided with grants over the years from the Brazilian agency CAPES.
    
    Finally, this work has made use of data from the European Space Agency (ESA) mission \textit{Gaia} (\url{https://www.cosmos.esa.int/gaia}), processed by the \textit{Gaia} Data Processing and Analysis Consortium (DPAC, \url{https://www.cosmos.esa.int/web/gaia/dpac/consortium}). Funding for the DPAC has been provided by national institutions, in particular, the institutions participating in the \textit{Gaia} Multilateral Agreement.
\end{acknowledgments}

\facility{LNA:BC0.6m}

\software{Astropy \citep{Astropy_Collaboration_2013,Astropy_Collaboration_2018}, Matplotlib \citep{Hunter_Matplotlib_2007}, NumPy \citep{Harris_numpy_2020}, SciPy \citep{Virtanen_SciPy_2020}}.

\newpage

\appendix
\restartappendixnumbering
\section{Variable Sources} \label{appex:variables}

    \subsection{Known Variable Sources in IPS-GI catalog} \label{appex:known_varia}
    
\begin{figure}[ht!]
            \epsscale{1.15}
            \plottwo{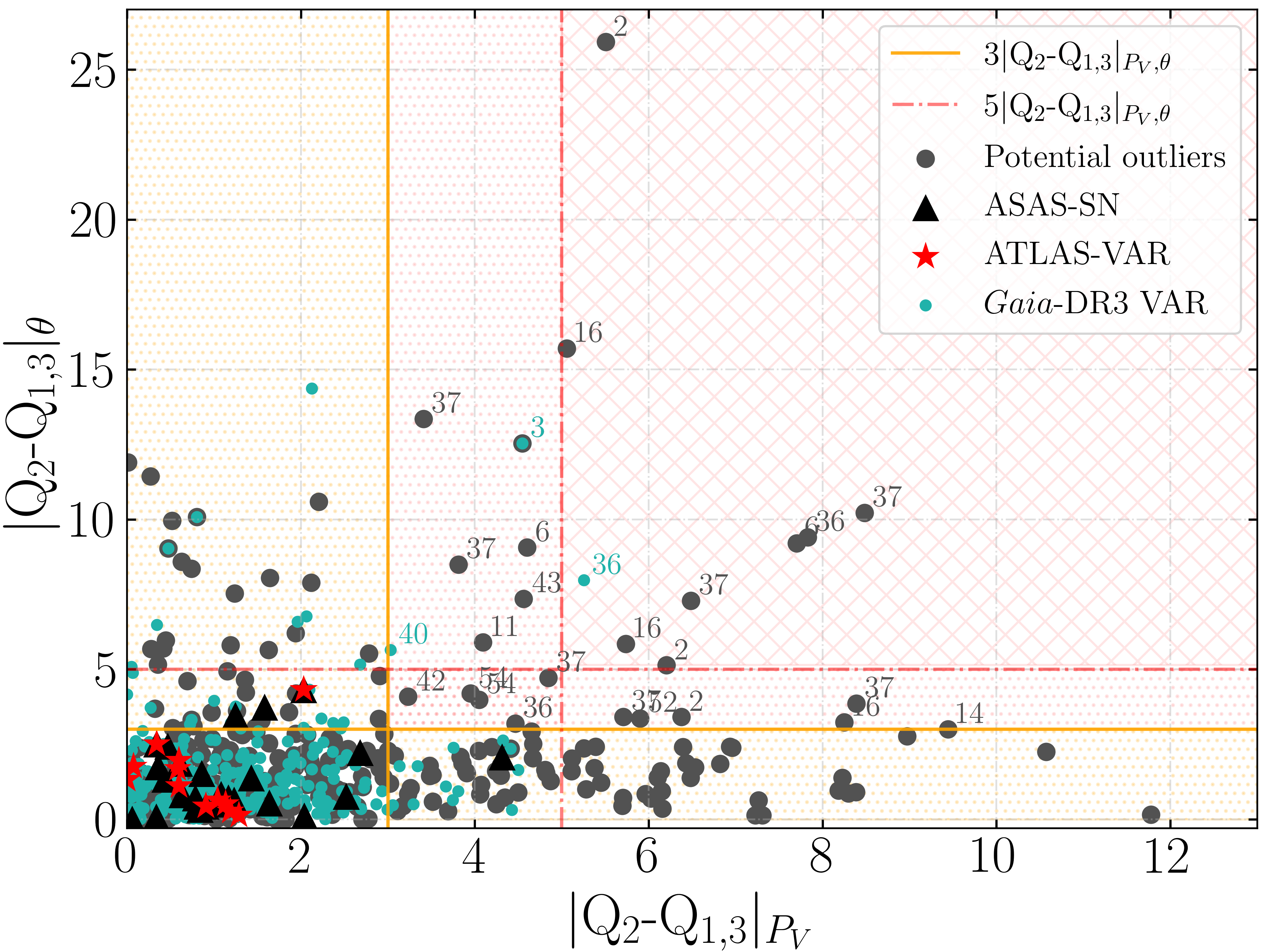}{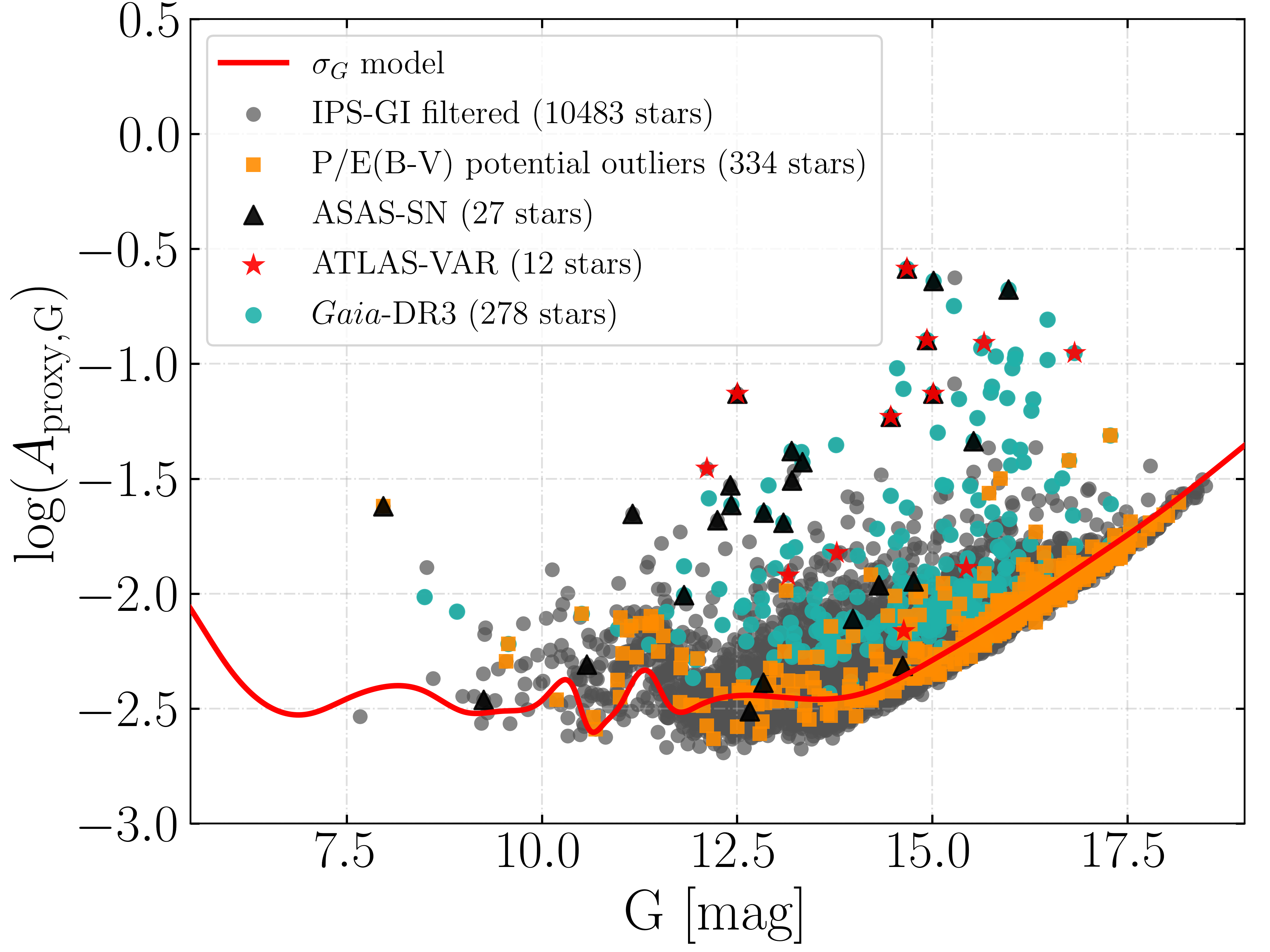}
            \caption{\textit{Left}: Deviations from the median value in polarization fraction and polarization angle of potential outliers in polarization efficiency (Section~\ref{subsec:P_vs_E_Cal_PE_uplim}) and variable stars known in the IPS-GI database (Appendix~\ref{appex:known_varia}). The deviations from $P$ and $\theta$ are the modulus of the difference between the median and the first or the third quantile in the distributions. \textit{Right:} The logarithm of the \textit{variability proxy} as a function of the $G$-band magnitude of IPS-GI data (blue dots). The red solid line is the logarithm of the expected mean $G$ magnitude error. The orange squares are potential outliers in polarization efficiency (Section~\ref{subsec:wQ_asym_err}). The black triangles and red stars are variable sources identified in the filtered cross-matched lists with ASAS-SN and ATLAS-VAR databases (Section~\ref{appex:known_varia}).
            \label{fig:dev_P_theta_all_IPS}}
        \end{figure}

        Variable stars may have a polarization component that is intrinsic or produced by circumstellar phenomena other than the dichroic extinction of interstellar dust. In essence, variable stars may be an issue for our polarimetry analysis in the general ISM since they can add unreliable measurements (see Section~\ref{subsec:Outliers}). Hence, we cross-matched the IPS-GI catalog with 3 different variable stars catalogs to identify these sources in our data set. The catalogs considered are: the ATLAS-VAR \citep{Heinze_2018} using a $2"$ margin, the ASAS-SN \citep{Jayasinghe_2018, Jayasinghe_2019a, Jayasinghe_2019b, Jayasinghe_2020, Christy_2022}  with $2"$ and $G=2$~mag margins, and the \textit{Gaia}-DR3 \citep[][]{GaiaDR3_Collaboration_2022} variable stars using the \textit{Gaia}-DR3 \texttt{source\_id} parameter. We found 79 matches with ASAS-SN, 37 with ATLAS-VAR (only dubious probability below 0.6, see section 4.2 in \citealt{Heinze_2018}), and 807 with \textit{Gaia}-DR3 variable stars catalog (only ``real'' variables, see \texttt{vari\_classifier\_result} parameter explanation in \citealt{GaiaDR3_Collaboration_2022}). A total of 294 variable stars passed all quality filters described in Section~\ref{subsec:Qflags}, 19 of which are in more than one of the above databases (see Table~\ref{tab:variable_stars}). The variables reported by \cite{Versteeg_2023} within the IPS-GI catalog did not pass our quality filters (Section~\ref{subsec:Qflags}).
        The final 294 variable sources do not seem to affect our results. Only 21 of them coincide with some potential outliers in polarization efficiency (see Section~\ref{subsec:wQ_asym_err}), and few (3) have deviations in polarization angle and polarization fraction above $3 \sigma$ from the median distribution on each IPS-GI field (Figure~\ref{fig:dev_P_theta_all_IPS}, left).
        
        \startlongtable
        \begin{deluxetable}{cccccc}
            \tablecaption{Variable stars in IPS-GI catalog. \label{tab:variable_stars}}
            \tablehead{
            \colhead{\textit{Gaia}-EDR3 ID} & \colhead{RA J2000} & \colhead{DEC J2000} & \colhead{IPS-GI Field} & \colhead{Class} & \colhead{Ref.}\\
            \colhead{} & \colhead{[deg]} & \colhead{[deg]} & \colhead{} & \colhead{}  & \colhead{} 
            }
            \colnumbers
            \startdata
                5884688765646295808 & 238.93509 & -54.55096  & \textit{0}  & (EA) Detached Algol-type binaries                  & 1, 3 \\
                5986010102181385472 & 236.85535 & -48.541686 & \textit{1}  & Eclipsing binary    & 3 \\
                4126224902887846912 & 253.74885 & -21.981876 & \textit{2}  & (ROT) Rotational modulation                 & 1, 3 \\
                4126267100963020928 & 253.72496 & -21.763218 & \textit{2}  & (RRAB) Fundamental Mode RR Lyrae            & 1, 2, 3 \\
                4126253799432188544 & 253.74466 & -21.792309 & \textit{2}  & Compact companion   & 3 \\
                4103505354883822464 & 280.27874 & -14.817128 & \textit{4}  & (EW) W Ursae Majoris type binaries                  & 1, 2, 3 \\
                4103502258157032960 & 280.36007 & -14.84001  & \textit{4}  & (DBH) Distant binary, half period                 & 2, 3 \\
                4252660532440869248 & 279.78211 & -7.357264  & \textit{5}  & (CBH) Close binary, half period                 & 2, 3 \\
                5868675615759940224 & 202.50417 & -61.089469 & \textit{11} & (YSO) Young Stellar Objects                 & 1 \\
                5868656649182615040 & 202.7103  & -61.337477 & \textit{11} & (GCAS) $\gamma$ Cassiopeiae variables                & 1 \\
                5868651662671280768 & 202.51935 & -61.348392 & \textit{11} & (EB) $\beta$ Lyrae-type binaries                  & 1, 3 \\
                5928955550488284416 & 247.94232 & -54.966693 & \textit{12} & Eclipsing binary    & 3 \\
                5872309948348868480 & 210.90681 & -56.057494 & \textit{14} & Rotation Modulation & 3 \\
                6022987502845965568 & 243.16885 & -34.587574 & \textit{36} & (RRAB) Fundamental Mode RR Lyrae                & 1, 3 \\
                6055566650847818368 & 192.89953 & -61.246276 & \textit{39} & (ROT) Rotational modulation                 & 1, 3 \\
                5313359983854184448 & 140.63481 & -51.170444 & \textit{43} & (L) Irregular Variables                   & 1 \\
                5876965589814204160 & 228.79787 & -58.944918 & \textit{47} & Main Sequence (MS) oscillators     & 3 \\
                5893111467061840640 & 222.32234 & -56.301996 & \textit{52} & (EB) $\beta$ Lyrae-type binaries                  & 1, 3 \\
                4097791983182736128 & 275.45374 & -16.457443 & \textit{56} & (EW) W Ursae Majoris type binaries                 & 1, 2, 3 \\
                4311099919049980416 & 282.79372 & 8.816396   & \textit{57} & (EW) W Ursae Majoris type binaries                & 1, 2, 3 \\
            \enddata
            \tablecomments{Example list of cross-matched variable sources that passed our quality filters (Section~\ref{subsec:Qflags}). The complete table is available upon request to the author. Columns: \textit{1)} \textit{Gaia}-EDR3 source id, \textit{2)} RA J2000 coordinate, \textit{3)} DEC J2000 coordinate, \textit{4)} IPS-GI field id, \textit{5)} Database classification (of the first reference in column 6), and \textit{6)} Source database reference. The full Table~\ref{tab:variable_stars} is available in the machine-readable format. A portion is shown here for guidance regarding its form and content.}
            \tablerefs{(1) ASAS-SN, \citealt{Jayasinghe_2018, Jayasinghe_2019a, Jayasinghe_2019b, Jayasinghe_2020}, (2) ATLAS-VAR, \citealt{Heinze_2018}, (3) \textit{Gaia}-DR3 variable stars, \citealt[][and references therein]{Eyer_2022}.}
        \end{deluxetable}
        
        %

    \subsection{The \textit{Gaia}-EDR3 variability proxy} \label{appex:varia_proxy}
        
        \citet{Riello_2021} introduced a new \textit{Gaia}-EDR3 metric called the \textit{variability proxy} which essentially corresponds to the estimated fractional error of a specific band on a single astrometric field (see their definition in appendix F, page 32). This metric performs very well to identify, for instance, extended sources such as galaxies (see e.g.~Figure~F.2. of \citealt{Riello_2021}) and it could be useful to identify potential variable sources as well. We calculated the \textit{variability proxy} in the $G$-band using \citetalias{Gaia_Collaboration_2021b} parameters as follows: 
        
        \begin{equation}
            A_{proxy\ G} = \sqrt{\tt\string phot\_g\_n\_obs}\ \ \frac{\tt\string phot\_g\_mean\_flux\_error}{\tt\string phot\_g\_mean\_flux} ,
        \end{equation}
        where $\tt\string phot\_g\_n\_obs$ is the number of \textit{Gaia} observations in the $G$-band, $\tt\string phot\_g\_mean\_flux$ is the mean $G$-band flux and $\tt\string phot\_g\_mean\_flux\_error$ is the corresponding mean flux error. We compare in Figure~\ref{fig:dev_P_theta_all_IPS} (right) the \textit{variability proxy} as a function of the G magnitude with the expected mean magnitude error in the $G$-band (see e.g.~top panel of Figure~14, \citealt{Riello_2021}), for a single \textit{Gaia} observation. Additionally, we included the variable stars identified in the cross-matched lists with ASAS-SN, ATLAS, and \textit{Gaia}-DR3 databases (see Section~\ref{appex:known_varia}). 
        
        The fractional errors in magnitude of the IPS-GI stars regularly follow the expected mean error for the $G$-band, except for some deviations that scale from 0.5~mag for the brightest sources to 1.5~mag for the faintest sources. High fractional errors due to low flux measurements and noisy-crowded backgrounds are expected in faint stars. The variable stars have high fractional errors and are far from the mean magnitude error model, proving that the \textit{variability proxy} is good to confirm variable sources or identify candidates. Furthermore, the potential outliers in polarization efficiency (see Section~\ref{subsec:wQ_asym_err}) are mostly faint stars consistent with the expected mean \textit{Gaia} magnitude error. These sources do not have fractional errors deviating above 0.5~mag except for one particular source (black triangle at $G \approx 8$), which turns out to be one of the few potential outliers confirmed to be a variable star (Figure~\ref{fig:dev_P_theta_all_IPS}, right).  In conclusion, we cannot clearly identify variable stars as IPS-GI outliers in the degree of polarization per reddening unit, nor say what exactly causes these outliers.

\clearpage
%
%
\bibliography{IPSIII_PE_gnal_ISM}{}

\newcommand{\noop}[1]{}
\begin{thebibliography}{}
\expandafter\ifx\csname natexlab\endcsname\relax\def\natexlab#1{#1}\fi
\providecommand{\url}[1]{\href{#1}{#1}}
\providecommand{\dodoi}[1]{doi:~\href{http://doi.org/#1}{\nolinkurl{#1}}}
\providecommand{\doeprint}[1]{\href{http://ascl.net/#1}{\nolinkurl{http://ascl.net/#1}}}
\providecommand{\doarXiv}[1]{\href{https://arxiv.org/abs/#1}{\nolinkurl{https://arxiv.org/abs/#1}}}

\bibitem[{{Akinshin}(2022)}]{Akinshin_2022}
{Akinshin}, A. 2022, arXiv e-prints, arXiv:2208.13459.
\newblock \doarXiv{2208.13459}

\bibitem[{{Akinshin}(2023)}]{Akinshin_2023}
---. 2023, arXiv e-prints, arXiv:2304.07265, \dodoi{10.48550/arXiv.2304.07265}

\bibitem[{{Anders} {et~al.}(2022){Anders}, {Khalatyan}, {Queiroz}, {Chiappini},
  {Ard{\`e}vol}, {Casamiquela}, {Figueras}, {Jim{\'e}nez-Arranz}, {Jordi},
  {Mongui{\'o}}, {Romero-G{\'o}mez}, {Altamirano}, {Antoja}, {Assaad},
  {Cantat-Gaudin}, {Castro-Ginard}, {Enke}, {Girardi}, {Guiglion}, {Khan},
  {Luri}, {Miglio}, {Minchev}, {Ramos}, {Santiago}, \&
  {Steinmetz}}]{Anders_2022}
{Anders}, F., {Khalatyan}, A., {Queiroz}, A.~B.~A., {et~al.} 2022, \aap, 658,
  A91, \dodoi{10.1051/0004-6361/202142369}

\bibitem[{{Andersson} {et~al.}(2015){Andersson}, {Lazarian}, \&
  {Vaillancourt}}]{Andersson_2015}
{Andersson}, B.~G., {Lazarian}, A., \& {Vaillancourt}, J.~E. 2015, \araa, 53,
  501, \dodoi{10.1146/annurev-astro-082214-122414}

\bibitem[{{Andersson} \& {Potter}(2007)}]{Andersson_Potter_2007}
{Andersson}, B.~G., \& {Potter}, S.~B. 2007, \apj, 665, 369,
  \dodoi{10.1086/519755}

\bibitem[{{Astropy Collaboration} {et~al.}(2013){Astropy Collaboration},
  {Robitaille}, {Tollerud}, {Greenfield}, {Droettboom}, {Bray}, {Aldcroft},
  {Davis}, {Ginsburg}, {Price-Whelan}, {Kerzendorf}, {Conley}, {Crighton},
  {Barbary}, {Muna}, {Ferguson}, {Grollier}, {Parikh}, {Nair}, {Unther},
  {Deil}, {Woillez}, {Conseil}, {Kramer}, {Turner}, {Singer}, {Fox}, {Weaver},
  {Zabalza}, {Edwards}, {Azalee Bostroem}, {Burke}, {Casey}, {Crawford},
  {Dencheva}, {Ely}, {Jenness}, {Labrie}, {Lim}, {Pierfederici}, {Pontzen},
  {Ptak}, {Refsdal}, {Servillat}, \& {Streicher}}]{Astropy_Collaboration_2013}
{Astropy Collaboration}, {Robitaille}, T.~P., {Tollerud}, E.~J., {et~al.} 2013,
  \aap, 558, A33, \dodoi{10.1051/0004-6361/201322068}

\bibitem[{{Astropy Collaboration} {et~al.}(2018){Astropy Collaboration},
  {Price-Whelan}, {Sip{\H{o}}cz}, {G{\"u}nther}, {Lim}, {Crawford}, {Conseil},
  {Shupe}, {Craig}, {Dencheva}, {Ginsburg}, {VanderPlas}, {Bradley},
  {P{\'e}rez-Su{\'a}rez}, {de Val-Borro}, {Aldcroft}, {Cruz}, {Robitaille},
  {Tollerud}, {Ardelean}, {Babej}, {Bach}, {Bachetti}, {Bakanov}, {Bamford},
  {Barentsen}, {Barmby}, {Baumbach}, {Berry}, {Biscani}, {Boquien}, {Bostroem},
  {Bouma}, {Brammer}, {Bray}, {Breytenbach}, {Buddelmeijer}, {Burke},
  {Calderone}, {Cano Rodr{\'\i}guez}, {Cara}, {Cardoso}, {Cheedella}, {Copin},
  {Corrales}, {Crichton}, {D'Avella}, {Deil}, {Depagne}, {Dietrich}, {Donath},
  {Droettboom}, {Earl}, {Erben}, {Fabbro}, {Ferreira}, {Finethy}, {Fox},
  {Garrison}, {Gibbons}, {Goldstein}, {Gommers}, {Greco}, {Greenfield},
  {Groener}, {Grollier}, {Hagen}, {Hirst}, {Homeier}, {Horton}, {Hosseinzadeh},
  {Hu}, {Hunkeler}, {Ivezi{\'c}}, {Jain}, {Jenness}, {Kanarek}, {Kendrew},
  {Kern}, {Kerzendorf}, {Khvalko}, {King}, {Kirkby}, {Kulkarni}, {Kumar},
  {Lee}, {Lenz}, {Littlefair}, {Ma}, {Macleod}, {Mastropietro}, {McCully},
  {Montagnac}, {Morris}, {Mueller}, {Mumford}, {Muna}, {Murphy}, {Nelson},
  {Nguyen}, {Ninan}, {N{\"o}the}, {Ogaz}, {Oh}, {Parejko}, {Parley}, {Pascual},
  {Patil}, {Patil}, {Plunkett}, {Prochaska}, {Rastogi}, {Reddy Janga},
  {Sabater}, {Sakurikar}, {Seifert}, {Sherbert}, {Sherwood-Taylor}, {Shih},
  {Sick}, {Silbiger}, {Singanamalla}, {Singer}, {Sladen}, {Sooley},
  {Sornarajah}, {Streicher}, {Teuben}, {Thomas}, {Tremblay}, {Turner},
  {Terr{\'o}n}, {van Kerkwijk}, {de la Vega}, {Watkins}, {Weaver}, {Whitmore},
  {Woillez}, {Zabalza}, \& {Astropy Contributors}}]{Astropy_Collaboration_2018}
{Astropy Collaboration}, {Price-Whelan}, A.~M., {Sip{\H{o}}cz}, B.~M., {et~al.}
  2018, \aj, 156, 123, \dodoi{10.3847/1538-3881/aabc4f}

\bibitem[{{Barlow}(2003)}]{Barlow_2003}
{Barlow}, R. 2003, arXiv e-prints, physics/0306138.
\newblock \doarXiv{physics/0306138}

\bibitem[{{Beck} \& {Wielebinski}(2013)}]{Beck_2013}
{Beck}, R., \& {Wielebinski}, R. 2013, in Planets, Stars and Stellar Systems.
  Volume 5: Galactic Structure and Stellar Populations, ed. T.~D. {Oswalt} \&
  G.~{Gilmore}, Vol.~5 (Springer Netherlands), 641,
  \dodoi{10.1007/978-94-007-5612-0_13}

\bibitem[{{Beresnyak} \& {Lazarian}(2019)}]{Beresnyak_2019}
{Beresnyak}, A., \& {Lazarian}, A. 2019, {Turbulence in Magnetohydrodynamics}

\bibitem[{{Capitanio} {et~al.}(2017){Capitanio}, {Lallement}, {Vergely},
  {Elyajouri}, \& {Monreal-Ibero}}]{Capitanio_2017}
{Capitanio}, L., {Lallement}, R., {Vergely}, J.~L., {Elyajouri}, M., \&
  {Monreal-Ibero}, A. 2017, \aap, 606, A65, \dodoi{10.1051/0004-6361/201730831}

\bibitem[{{Christy} {et~al.}(2022){Christy}, {Jayasinghe}, {Stanek},
  {Kochanek}, {Thompson}, {Shappee}, {Holoien}, {Prieto}, {Dong}, \&
  {Giles}}]{Christy_2022}
{Christy}, C.~T., {Jayasinghe}, T., {Stanek}, K.~Z., {et~al.} 2022, arXiv
  e-prints, arXiv:2205.02239.
\newblock \doarXiv{2205.02239}

\bibitem[{{Clarke} \& {Stewart}(1986)}]{Clarke_Stewart_1986}
{Clarke}, D., \& {Stewart}, B.~G. 1986, Vistas in Astronomy, 29, 27,
  \dodoi{10.1016/0083-6656(86)90013-9}

\bibitem[{{Draine} \& {Fraisse}(2009)}]{Draine&Fraisse_2009}
{Draine}, B.~T., \& {Fraisse}, A.~A. 2009, \apj, 696, 1,
  \dodoi{10.1088/0004-637X/696/1/1}

\bibitem[{{Draine} \& {Hensley}(2021)}]{Draine_2021}
{Draine}, B.~T., \& {Hensley}, B.~S. 2021, arXiv e-prints, arXiv:2101.07277.
\newblock \doarXiv{2101.07277}

\bibitem[{{Erdim} \& {Hudaverdi}(2019)}]{Erdim_SOAD_2019}
{Erdim}, M.~K., \& {Hudaverdi}, M. 2019, in American Institute of Physics
  Conference Series, Vol. 2178, Turkish Physical Society 35th International
  Physics Congress (TPS35), 030023, \dodoi{10.1063/1.5135421}

\bibitem[{{Eyer} {et~al.}(2022{\natexlab{a}}){Eyer}, {Audard}, {Holl},
  {Rimoldini}, {Carnerero}, {Clementini}, {De Ridder}, {Distefano}, {Evans},
  {Gavras}, {Gomel}, {Lebzelter}, {Marton}, {Mowlavi}, {Panahi}, {Ripepi},
  {Wyrzykowski}, {Nienartowicz}, {Jevardat de Fombelle}, {Lecoeur-Taibi},
  {Rohrbasser}, {Riello}, {Garcia-Lario}, {Lanzafame}, {Mazeh}, {Raiteri},
  {Zucker}, {Abraham}, {Aerts}, {Aguado}, {Anderson}, {Bashi}, {Binnenfeld},
  {Faigler}, {Garofalo}, {Karbevska}, {Kospal}, {Kruszynska}, {Kun}, {Lanza},
  {Leccia}, {Marconi}, {Messina}, {Molinaro}, {Molnar}, {Muraveva}, {Musella},
  {Nagy}, {Pagano}, {Palaversa}, {Plachy}, {Rybicki}, {Shahaf}, {Szabados},
  {Szegedi-Elek}, {Trabucchi}, {Barblan}, \&
  {Roelens}}]{GaiaDR3_Collaboration_2022}
{Eyer}, L., {Audard}, M., {Holl}, B., {et~al.} 2022{\natexlab{a}}, arXiv
  e-prints, arXiv:2206.06416.
\newblock \doarXiv{2206.06416}

\bibitem[{{Eyer} {et~al.}(2022{\natexlab{b}}){Eyer}, {Audard}, {Holl},
  {Rimoldini}, {Carnerero}, {Clementini}, {De Ridder}, {Distefano}, {Evans},
  {Gavras}, {Gomel}, {Lebzelter}, {Marton}, {Mowlavi}, {Panahi}, {Ripepi},
  {Wyrzykowski}, {Nienartowicz}, {Jevardat de Fombelle}, {Lecoeur-Taibi},
  {Rohrbasser}, {Riello}, {Garcia-Lario}, {Lanzafame}, {Mazeh}, {Raiteri},
  {Zucker}, {Abraham}, {Aerts}, {Aguado}, {Anderson}, {Bashi}, {Binnenfeld},
  {Faigler}, {Garofalo}, {Karbevska}, {Kospal}, {Kruszynska}, {Kun}, {Lanza},
  {Leccia}, {Marconi}, {Messina}, {Molinaro}, {Molnar}, {Muraveva}, {Musella},
  {Nagy}, {Pagano}, {Palaversa}, {Plachy}, {Rybicki}, {Shahaf}, {Szabados},
  {Szegedi-Elek}, {Trabucchi}, {Barblan}, \& {Roelens}}]{Eyer_2022}
---. 2022{\natexlab{b}}, arXiv e-prints, arXiv:2206.06416.
\newblock \doarXiv{2206.06416}

\bibitem[{{Fitzpatrick}(2004)}]{Fitzpatrick_2004}
{Fitzpatrick}, E.~L. 2004, in Astronomical Society of the Pacific Conference
  Series, Vol. 309, Astrophysics of Dust, ed. A.~N. {Witt}, G.~C. {Clayton}, \&
  B.~T. {Draine}, 33.
\newblock \doarXiv{astro-ph/0401344}

\bibitem[{{Fosalba} {et~al.}(2002){Fosalba}, {Lazarian}, {Prunet}, \&
  {Tauber}}]{Fosalba_2002}
{Fosalba}, P., {Lazarian}, A., {Prunet}, S., \& {Tauber}, J.~A. 2002, \apj,
  564, 762, \dodoi{10.1086/324297}

\bibitem[{{Gaia Collaboration} {et~al.}(2018){Gaia Collaboration}, {Brown},
  {Vallenari}, {Prusti}, {de Bruijne}, {Babusiaux}, {Bailer-Jones}, {Biermann},
  {Evans}, {Eyer}, {Jansen}, {Jordi}, {Klioner}, {Lammers}, {Lindegren},
  {Luri}, {Mignard}, {Panem}, {Pourbaix}, {Randich}, {Sartoretti}, {Siddiqui},
  {Soubiran}, {van Leeuwen}, {Walton}, {Arenou}, {Bastian}, {Cropper},
  {Drimmel}, {Katz}, {Lattanzi}, {Bakker}, {Cacciari}, {Casta{\~n}eda},
  {Chaoul}, {Cheek}, {De Angeli}, {Fabricius}, {Guerra}, {Holl}, {Masana},
  {Messineo}, {Mowlavi}, {Nienartowicz}, {Panuzzo}, {Portell}, {Riello},
  {Seabroke}, {Tanga}, {Th{\'e}venin}, {Gracia-Abril}, {Comoretto},
  {Garcia-Reinaldos}, {Teyssier}, {Altmann}, {Andrae}, {Audard},
  {Bellas-Velidis}, {Benson}, {Berthier}, {Blomme}, {Burgess}, {Busso},
  {Carry}, {Cellino}, {Clementini}, {Clotet}, {Creevey}, {Davidson}, {De
  Ridder}, {Delchambre}, {Dell'Oro}, {Ducourant},
  {Fern{\'a}ndez-Hern{\'a}ndez}, {Fouesneau}, {Fr{\'e}mat}, {Galluccio},
  {Garc{\'\i}a-Torres}, {Gonz{\'a}lez-N{\'u}{\~n}ez}, {Gonz{\'a}lez-Vidal},
  {Gosset}, {Guy}, {Halbwachs}, {Hambly}, {Harrison}, {Hern{\'a}ndez},
  {Hestroffer}, {Hodgkin}, {Hutton}, {Jasniewicz}, {Jean-Antoine-Piccolo},
  {Jordan}, {Korn}, {Krone-Martins}, {Lanzafame}, {Lebzelter}, {L{\"o}ffler},
  {Manteiga}, {Marrese}, {Mart{\'\i}n-Fleitas}, {Moitinho}, {Mora}, {Muinonen},
  {Osinde}, {Pancino}, {Pauwels}, {Petit}, {Recio-Blanco}, {Richards},
  {Rimoldini}, {Robin}, {Sarro}, {Siopis}, {Smith}, {Sozzetti}, {S{\"u}veges},
  {Torra}, {van Reeven}, {Abbas}, {Abreu Aramburu}, {Accart}, {Aerts},
  {Altavilla}, {{\'A}lvarez}, {Alvarez}, {Alves}, {Anderson}, {Andrei},
  {Anglada Varela}, {Antiche}, {Antoja}, {Arcay}, {Astraatmadja}, {Bach},
  {Baker}, {Balaguer-N{\'u}{\~n}ez}, {Balm}, {Barache}, {Barata}, {Barbato},
  {Barblan}, {Barklem}, {Barrado}, {Barros}, {Barstow}, {Bartholom{\'e}
  Mu{\~n}oz}, {Bassilana}, {Becciani}, {Bellazzini}, {Berihuete}, {Bertone},
  {Bianchi}, {Bienaym{\'e}}, {Blanco-Cuaresma}, {Boch}, {Boeche}, {Bombrun},
  {Borrachero}, {Bossini}, {Bouquillon}, {Bourda}, {Bragaglia}, {Bramante},
  {Breddels}, {Bressan}, {Brouillet}, {Br{\"u}semeister}, {Brugaletta},
  {Bucciarelli}, {Burlacu}, {Busonero}, {Butkevich}, {Buzzi}, {Caffau},
  {Cancelliere}, {Cannizzaro}, {Cantat-Gaudin}, {Carballo}, {Carlucci},
  {Carrasco}, {Casamiquela}, {Castellani}, {Castro-Ginard}, {Charlot},
  {Chemin}, {Chiavassa}, {Cocozza}, {Costigan}, {Cowell}, {Crifo}, {Crosta},
  {Crowley}, {Cuypers}, {Dafonte}, {Damerdji}, {Dapergolas}, {David}, {David},
  {de Laverny}, {De Luise}, {De March}, {de Martino}, {de Souza}, {de Torres},
  {Debosscher}, {del Pozo}, {Delbo}, {Delgado}, {Delgado}, {Di Matteo},
  {Diakite}, {Diener}, {Distefano}, {Dolding}, {Drazinos}, {Dur{\'a}n},
  {Edvardsson}, {Enke}, {Eriksson}, {Esquej}, {Eynard Bontemps}, {Fabre},
  {Fabrizio}, {Faigler}, {Falc{\~a}o}, {Farr{\`a}s Casas}, {Federici},
  {Fedorets}, {Fernique}, {Figueras}, {Filippi}, {Findeisen}, {Fonti},
  {Fraile}, {Fraser}, {Fr{\'e}zouls}, {Gai}, {Galleti}, {Garabato},
  {Garc{\'\i}a-Sedano}, {Garofalo}, {Garralda}, {Gavel}, {Gavras}, {Gerssen},
  {Geyer}, {Giacobbe}, {Gilmore}, {Girona}, {Giuffrida}, {Glass}, {Gomes},
  {Granvik}, {Gueguen}, {Guerrier}, {Guiraud}, {Guti{\'e}rrez-S{\'a}nchez},
  {Haigron}, {Hatzidimitriou}, {Hauser}, {Haywood}, {Heiter}, {Helmi}, {Heu},
  {Hilger}, {Hobbs}, {Hofmann}, {Holland}, {Huckle}, {Hypki}, {Icardi},
  {Jan{\ss}en}, {Jevardat de Fombelle}, {Jonker}, {Juh{\'a}sz}, {Julbe},
  {Karampelas}, {Kewley}, {Klar}, {Kochoska}, {Kohley}, {Kolenberg},
  {Kontizas}, {Kontizas}, {Koposov}, {Kordopatis}, {Kostrzewa-Rutkowska},
  {Koubsky}, {Lambert}, {Lanza}, {Lasne}, {Lavigne}, {Le Fustec}, {Le
  Poncin-Lafitte}, {Lebreton}, {Leccia}, {Leclerc}, {Lecoeur-Taibi},
  {Lenhardt}, {Leroux}, {Liao}, {Licata}, {Lindstr{\o}m}, {Lister}, {Livanou},
  {Lobel}, {L{\'o}pez}, {Managau}, {Mann}, {Mantelet}, {Marchal}, {Marchant},
  {Marconi}, {Marinoni}, {Marschalk{\'o}}, {Marshall}, {Martino}, {Marton},
  {Mary}, {Massari}, {Matijevi{\v{c}}}, {Mazeh}, {McMillan}, {Messina},
  {Michalik}, {Millar}, {Molina}, {Molinaro}, {Moln{\'a}r}, {Montegriffo},
  {Mor}, {Morbidelli}, {Morel}, {Morris}, {Mulone}, {Muraveva}, {Musella},
  {Nelemans}, {Nicastro}, {Noval}, {O'Mullane}, {Ord{\'e}novic},
  {Ord{\'o}{\~n}ez-Blanco}, {Osborne}, {Pagani}, {Pagano}, {Pailler},
  {Palacin}, {Palaversa}, {Panahi}, {Pawlak}, {Piersimoni}, {Pineau}, {Plachy},
  {Plum}, {Poggio}, {Poujoulet}, {Pr{\v{s}}a}, {Pulone}, {Racero}, {Ragaini},
  {Rambaux}, {Ramos-Lerate}, {Regibo}, {Reyl{\'e}}, {Riclet}, {Ripepi}, {Riva},
  {Rivard}, {Rixon}, {Roegiers}, {Roelens}, {Romero-G{\'o}mez}, {Rowell},
  {Royer}, {Ruiz-Dern}, {Sadowski}, {Sagrist{\`a} Sell{\'e}s}, {Sahlmann},
  {Salgado}, {Salguero}, {Sanna}, {Santana-Ros}, {Sarasso}, {Savietto},
  {Schultheis}, {Sciacca}, {Segol}, {Segovia}, {S{\'e}gransan}, {Shih},
  {Siltala}, {Silva}, {Smart}, {Smith}, {Solano}, {Solitro}, {Sordo}, {Soria
  Nieto}, {Souchay}, {Spagna}, {Spoto}, {Stampa}, {Steele},
  {Steidelm{\"u}ller}, {Stephenson}, {Stoev}, {Suess}, {Surdej}, {Szabados},
  {Szegedi-Elek}, {Tapiador}, {Taris}, {Tauran}, {Taylor}, {Teixeira},
  {Terrett}, {Teyssandier}, {Thuillot}, {Titarenko}, {Torra Clotet}, {Turon},
  {Ulla}, {Utrilla}, {Uzzi}, {Vaillant}, {Valentini}, {Valette}, {van Elteren},
  {Van Hemelryck}, {van Leeuwen}, {Vaschetto}, {Vecchiato}, {Veljanoski},
  {Viala}, {Vicente}, {Vogt}, {von Essen}, {Voss}, {Votruba}, {Voutsinas},
  {Walmsley}, {Weiler}, {Wertz}, {Wevers}, {Wyrzykowski}, {Yoldas},
  {{\v{Z}}erjal}, {Ziaeepour}, {Zorec}, {Zschocke}, {Zucker}, {Zurbach}, \&
  {Zwitter}}]{Gaia_Collaboration_2018}
{Gaia Collaboration}, {Brown}, A.~G.~A., {Vallenari}, A., {et~al.} 2018, \aap,
  616, A1, \dodoi{10.1051/0004-6361/201833051}

\bibitem[{{Gaia Collaboration} {et~al.}(2021){Gaia Collaboration}, {Brown},
  {Vallenari}, {Prusti}, {de Bruijne}, {Babusiaux}, {Biermann}, {Creevey},
  {Evans}, {Eyer}, {Hutton}, {Jansen}, {Jordi}, {Klioner}, {Lammers},
  {Lindegren}, {Luri}, {Mignard}, {Panem}, {Pourbaix}, {Randich}, {Sartoretti},
  {Soubiran}, {Walton}, {Arenou}, {Bailer-Jones}, {Bastian}, {Cropper},
  {Drimmel}, {Katz}, {Lattanzi}, {van Leeuwen}, {Bakker}, {Cacciari},
  {Casta{\~n}eda}, {De Angeli}, {Ducourant}, {Fabricius}, {Fouesneau},
  {Fr{\'e}mat}, {Guerra}, {Guerrier}, {Guiraud}, {Jean-Antoine Piccolo},
  {Masana}, {Messineo}, {Mowlavi}, {Nicolas}, {Nienartowicz}, {Pailler},
  {Panuzzo}, {Riclet}, {Roux}, {Seabroke}, {Sordo}, {Tanga}, {Th{\'e}venin},
  {Gracia-Abril}, {Portell}, {Teyssier}, {Altmann}, {Andrae}, {Bellas-Velidis},
  {Benson}, {Berthier}, {Blomme}, {Brugaletta}, {Burgess}, {Busso}, {Carry},
  {Cellino}, {Cheek}, {Clementini}, {Damerdji}, {Davidson}, {Delchambre},
  {Dell'Oro}, {Fern{\'a}ndez-Hern{\'a}ndez}, {Galluccio}, {Garc{\'\i}a-Lario},
  {Garcia-Reinaldos}, {Gonz{\'a}lez-N{\'u}{\~n}ez}, {Gosset}, {Haigron},
  {Halbwachs}, {Hambly}, {Harrison}, {Hatzidimitriou}, {Heiter},
  {Hern{\'a}ndez}, {Hestroffer}, {Hodgkin}, {Holl}, {Jan{\ss}en}, {Jevardat de
  Fombelle}, {Jordan}, {Krone-Martins}, {Lanzafame}, {L{\"o}ffler}, {Lorca},
  {Manteiga}, {Marchal}, {Marrese}, {Moitinho}, {Mora}, {Muinonen}, {Osborne},
  {Pancino}, {Pauwels}, {Petit}, {Recio-Blanco}, {Richards}, {Riello},
  {Rimoldini}, {Robin}, {Roegiers}, {Rybizki}, {Sarro}, {Siopis}, {Smith},
  {Sozzetti}, {Ulla}, {Utrilla}, {van Leeuwen}, {van Reeven}, {Abbas}, {Abreu
  Aramburu}, {Accart}, {Aerts}, {Aguado}, {Ajaj}, {Altavilla}, {{\'A}lvarez},
  {{\'A}lvarez Cid-Fuentes}, {Alves}, {Anderson}, {Anglada Varela}, {Antoja},
  {Audard}, {Baines}, {Baker}, {Balaguer-N{\'u}{\~n}ez}, {Balbinot}, {Balog},
  {Barache}, {Barbato}, {Barros}, {Barstow}, {Bartolom{\'e}}, {Bassilana},
  {Bauchet}, {Baudesson-Stella}, {Becciani}, {Bellazzini}, {Bernet}, {Bertone},
  {Bianchi}, {Blanco-Cuaresma}, {Boch}, {Bombrun}, {Bossini}, {Bouquillon},
  {Bragaglia}, {Bramante}, {Breedt}, {Bressan}, {Brouillet}, {Bucciarelli},
  {Burlacu}, {Busonero}, {Butkevich}, {Buzzi}, {Caffau}, {Cancelliere},
  {C{\'a}novas}, {Cantat-Gaudin}, {Carballo}, {Carlucci}, {Carnerero},
  {Carrasco}, {Casamiquela}, {Castellani}, {Castro-Ginard}, {Castro Sampol},
  {Chaoul}, {Charlot}, {Chemin}, {Chiavassa}, {Cioni}, {Comoretto}, {Cooper},
  {Cornez}, {Cowell}, {Crifo}, {Crosta}, {Crowley}, {Dafonte}, {Dapergolas},
  {David}, {David}, {de Laverny}, {De Luise}, {De March}, {De Ridder}, {de
  Souza}, {de Teodoro}, {de Torres}, {del Peloso}, {del Pozo}, {Delbo},
  {Delgado}, {Delgado}, {Delisle}, {Di Matteo}, {Diakite}, {Diener},
  {Distefano}, {Dolding}, {Eappachen}, {Edvardsson}, {Enke}, {Esquej}, {Fabre},
  {Fabrizio}, {Faigler}, {Fedorets}, {Fernique}, {Fienga}, {Figueras},
  {Fouron}, {Fragkoudi}, {Fraile}, {Franke}, {Gai}, {Garabato},
  {Garcia-Gutierrez}, {Garc{\'\i}a-Torres}, {Garofalo}, {Gavras}, {Gerlach},
  {Geyer}, {Giacobbe}, {Gilmore}, {Girona}, {Giuffrida}, {Gomel}, {Gomez},
  {Gonzalez-Santamaria}, {Gonz{\'a}lez-Vidal}, {Granvik},
  {Guti{\'e}rrez-S{\'a}nchez}, {Guy}, {Hauser}, {Haywood}, {Helmi}, {Hidalgo},
  {Hilger}, {H{\l}adczuk}, {Hobbs}, {Holland}, {Huckle}, {Jasniewicz},
  {Jonker}, {Juaristi Campillo}, {Julbe}, {Karbevska}, {Kervella}, {Khanna},
  {Kochoska}, {Kontizas}, {Kordopatis}, {Korn}, {Kostrzewa-Rutkowska},
  {Kruszy{\'n}ska}, {Lambert}, {Lanza}, {Lasne}, {Le Campion}, {Le Fustec},
  {Lebreton}, {Lebzelter}, {Leccia}, {Leclerc}, {Lecoeur-Taibi}, {Liao},
  {Licata}, {Lindstr{\o}m}, {Lister}, {Livanou}, {Lobel}, {Madrero Pardo},
  {Managau}, {Mann}, {Marchant}, {Marconi}, {Marcos Santos}, {Marinoni},
  {Marocco}, {Marshall}, {Martin Polo}, {Mart{\'\i}n-Fleitas}, {Masip},
  {Massari}, {Mastrobuono-Battisti}, {Mazeh}, {McMillan}, {Messina},
  {Michalik}, {Millar}, {Mints}, {Molina}, {Molinaro}, {Moln{\'a}r},
  {Montegriffo}, {Mor}, {Morbidelli}, {Morel}, {Morris}, {Mulone}, {Munoz},
  {Muraveva}, {Murphy}, {Musella}, {Noval}, {Ord{\'e}novic}, {Orr{\`u}},
  {Osinde}, {Pagani}, {Pagano}, {Palaversa}, {Palicio}, {Panahi}, {Pawlak},
  {Pe{\~n}alosa Esteller}, {Penttil{\"a}}, {Piersimoni}, {Pineau}, {Plachy},
  {Plum}, {Poggio}, {Poretti}, {Poujoulet}, {Pr{\v{s}}a}, {Pulone}, {Racero},
  {Ragaini}, {Rainer}, {Raiteri}, {Rambaux}, {Ramos}, {Ramos-Lerate}, {Re
  Fiorentin}, {Regibo}, {Reyl{\'e}}, {Ripepi}, {Riva}, {Rixon}, {Robichon},
  {Robin}, {Roelens}, {Rohrbasser}, {Romero-G{\'o}mez}, {Rowell}, {Royer},
  {Rybicki}, {Sadowski}, {Sagrist{\`a} Sell{\'e}s}, {Sahlmann}, {Salgado},
  {Salguero}, {Samaras}, {Sanchez Gimenez}, {Sanna}, {Santove{\~n}a},
  {Sarasso}, {Schultheis}, {Sciacca}, {Segol}, {Segovia}, {S{\'e}gransan},
  {Semeux}, {Shahaf}, {Siddiqui}, {Siebert}, {Siltala}, {Slezak}, {Smart},
  {Solano}, {Solitro}, {Souami}, {Souchay}, {Spagna}, {Spoto}, {Steele},
  {Steidelm{\"u}ller}, {Stephenson}, {S{\"u}veges}, {Szabados}, {Szegedi-Elek},
  {Taris}, {Tauran}, {Taylor}, {Teixeira}, {Thuillot}, {Tonello}, {Torra},
  {Torra}, {Turon}, {Unger}, {Vaillant}, {van Dillen}, {Vanel}, {Vecchiato},
  {Viala}, {Vicente}, {Voutsinas}, {Weiler}, {Wevers}, {Wyrzykowski}, {Yoldas},
  {Yvard}, {Zhao}, {Zorec}, {Zucker}, {Zurbach}, \&
  {Zwitter}}]{Gaia_Collaboration_2021b}
---. 2021, \aap, 649, A1, \dodoi{10.1051/0004-6361/202039657}

\bibitem[{{Green}(2018)}]{Green_dustmaps_2018}
{Green}, G. 2018, The Journal of Open Source Software, 3, 695,
  \dodoi{10.21105/joss.00695}

\bibitem[{{Green} {et~al.}(2019){Green}, {Schlafly}, {Zucker}, {Speagle}, \&
  {Finkbeiner}}]{Green_2019}
{Green}, G.~M., {Schlafly}, E., {Zucker}, C., {Speagle}, J.~S., \&
  {Finkbeiner}, D. 2019, \apj, 887, 93, \dodoi{10.3847/1538-4357/ab5362}

\bibitem[{{Harris} {et~al.}(2020){Harris}, Millman, van~der Walt, Gommers,
  Virtanen, Cournapeau, Wieser, Taylor, Berg, Smith, Kern, Picus, Hoyer, van
  Kerkwijk, Brett, Haldane, del R{\'{i}}o, Wiebe, Peterson,
  G{\'{e}}rard-Marchant, Sheppard, Reddy, Weckesser, Abbasi, Gohlke, \&
  Oliphant}]{Harris_numpy_2020}
{Harris}, C.~R., Millman, K.~J., van~der Walt, S.~J., {et~al.} 2020, Nature,
  585, 357, \dodoi{10.1038/s41586-020-2649-2}

\bibitem[{{Heiles}(2000)}]{Heiles_2000}
{Heiles}, C. 2000, \aj, 119, 923, \dodoi{10.1086/301236}

\bibitem[{{Heinze} {et~al.}(2018){Heinze}, {Tonry}, {Denneau}, {Flewelling},
  {Stalder}, {Rest}, {Smith}, {Smartt}, \& {Weiland}}]{Heinze_2018}
{Heinze}, A.~N., {Tonry}, J.~L., {Denneau}, L., {et~al.} 2018, \aj, 156, 241,
  \dodoi{10.3847/1538-3881/aae47f}

\bibitem[{{Hensley} \& {Draine}(2021)}]{Hensley_2021}
{Hensley}, B.~S., \& {Draine}, B.~T. 2021, \apj, 906, 73,
  \dodoi{10.3847/1538-4357/abc8f1}

\bibitem[{{Hunter}(2007)}]{Hunter_Matplotlib_2007}
{Hunter}, J.~D. 2007, Computing in Science \& Engineering, 9, 90,
  \dodoi{10.1109/MCSE.2007.55}

\bibitem[{Hyndman \& Fan(1996)}]{Hyndman_1996}
Hyndman, R.~J., \& Fan, Y. 1996, The American Statistician, 50, 361.
\newblock \url{http://www.jstor.org/stable/2684934}

\bibitem[{{Jayasinghe} {et~al.}(2018){Jayasinghe}, {Kochanek}, {Stanek},
  {Shappee}, {Holoien}, {Thompson}, {Prieto}, {Dong}, {Pawlak}, {Shields},
  {Pojmanski}, {Otero}, {Britt}, \& {Will}}]{Jayasinghe_2018}
{Jayasinghe}, T., {Kochanek}, C.~S., {Stanek}, K.~Z., {et~al.} 2018, \mnras,
  477, 3145, \dodoi{10.1093/mnras/sty838}

\bibitem[{{Jayasinghe} {et~al.}(2019{\natexlab{a}}){Jayasinghe}, {Stanek},
  {Kochanek}, {Shappee}, {Holoien}, {Thompson}, {Prieto}, {Dong}, {Pawlak},
  {Pejcha}, {Shields}, {Pojmanski}, {Otero}, {Britt}, \&
  {Will}}]{Jayasinghe_2019a}
{Jayasinghe}, T., {Stanek}, K.~Z., {Kochanek}, C.~S., {et~al.}
  2019{\natexlab{a}}, \mnras, 486, 1907, \dodoi{10.1093/mnras/stz844}

\bibitem[{{Jayasinghe} {et~al.}(2019{\natexlab{b}}){Jayasinghe}, {Stanek},
  {Kochanek}, {Shappee}, {Holoien}, {Thompson}, {Prieto}, {Dong}, {Pawlak},
  {Pejcha}, {Shields}, {Pojmanski}, {Otero}, {Hurst}, {Britt}, \&
  {Will}}]{Jayasinghe_2019b}
---. 2019{\natexlab{b}}, \mnras, 485, 961, \dodoi{10.1093/mnras/stz444}

\bibitem[{{Jayasinghe} {et~al.}(2020){Jayasinghe}, {Stanek}, {Kochanek},
  {Shappee}, {Holoien}, {Thompson}, {Prieto}, {Dong}, {Pawlak}, {Pejcha},
  {Shields}, {Pojmanski}, {Otero}, {Hurst}, {Britt}, \&
  {Will}}]{Jayasinghe_2020}
---. 2020, \mnras, 491, 13, \dodoi{10.1093/mnras/stz2711}

\bibitem[{{Jones}(1989)}]{Jones_1989}
{Jones}, T.~J. 1989, \apj, 346, 728, \dodoi{10.1086/168054}

\bibitem[{{Jones} {et~al.}(2015){Jones}, {Bagley}, {Krejny}, {Andersson}, \&
  {Bastien}}]{Jones_2015}
{Jones}, T.~J., {Bagley}, M., {Krejny}, M., {Andersson}, B.~G., \& {Bastien},
  P. 2015, \aj, 149, 31, \dodoi{10.1088/0004-6256/149/1/31}

\bibitem[{{Jones} {et~al.}(1992){Jones}, {Klebe}, \& {Dickey}}]{Jones_1992}
{Jones}, T.~J., {Klebe}, D., \& {Dickey}, J.~M. 1992, \apj, 389, 602,
  \dodoi{10.1086/171233}

\bibitem[{{Jones} \& {Whittet}(2015)}]{jones_whittet_2015}
{Jones}, T.~J., \& {Whittet}, D. C.~B. 2015, Interstellar polarization, ed.
  L.~Kolokolova, J.~Hough, \& A.-C. Levasseur-Regourd (Cambridge University
  Press), 147–161, \dodoi{10.1017/CBO9781107358249.009}

\bibitem[{{Kim} \& {Martin}(1995)}]{Kim_Martin_1995}
{Kim}, S.-H., \& {Martin}, P.~G. 1995, \apj, 444, 293, \dodoi{10.1086/175604}

\bibitem[{{Lallement} {et~al.}(2019){Lallement}, {Babusiaux}, {Vergely},
  {Katz}, {Arenou}, {Valette}, {Hottier}, \& {Capitanio}}]{Lallement_2019}
{Lallement}, R., {Babusiaux}, C., {Vergely}, J.~L., {et~al.} 2019, \aap, 625,
  A135, \dodoi{10.1051/0004-6361/201834695}

\bibitem[{{Lallement} {et~al.}(2022){Lallement}, {Vergely}, {Babusiaux}, \&
  {Cox}}]{Lallement_2022}
{Lallement}, R., {Vergely}, J.~L., {Babusiaux}, C., \& {Cox}, N.~L.~J. 2022,
  \aap, 661, A147, \dodoi{10.1051/0004-6361/202142846}

\bibitem[{{Lallement} {et~al.}(2018){Lallement}, {Capitanio}, {Ruiz-Dern},
  {Danielski}, {Babusiaux}, {Vergely}, {Elyajouri}, {Arenou}, \&
  {Leclerc}}]{Lallement_2018}
{Lallement}, R., {Capitanio}, L., {Ruiz-Dern}, L., {et~al.} 2018, \aap, 616,
  A132, \dodoi{10.1051/0004-6361/201832832}

\bibitem[{{Lee} \& {Draine}(1985)}]{Lee&Draine_1985}
{Lee}, H.~M., \& {Draine}, B.~T. 1985, \apj, 290, 211, \dodoi{10.1086/162974}

\bibitem[{{Magalh{\~a}es} {et~al.}(1984){Magalh{\~a}es}, {Benedetti}, \&
  {Roland}}]{Magalhaes_1984}
{Magalh{\~a}es}, A.~M., {Benedetti}, E., \& {Roland}, E.~H. 1984, \pasp, 96,
  383, \dodoi{10.1086/131351}

\bibitem[{{Magalh{\~a}es} {et~al.}(1996){Magalh{\~a}es}, {Rodrigues},
  {Margoniner}, {Pereyra}, \& {Heathcote}}]{Magalhaes_1996}
{Magalh{\~a}es}, A.~M., {Rodrigues}, C.~V., {Margoniner}, V.~E., {Pereyra}, A.,
  \& {Heathcote}, S. 1996, in Astronomical Society of the Pacific Conference
  Series, Vol.~97, Polarimetry of the Interstellar Medium, ed. W.~G. {Roberge}
  \& D.~C.~B. {Whittet}, 118

\bibitem[{{Magalh{\~a}es} {et~al.}(2005){Magalh{\~a}es}, {Pereyra},
  {Melgarejo}, {de Matos}, {Carciofi}, {Benedito}, {Valentim}, {Vidotto}, {da
  Silva}, {de Souza}, {Faria}, \& {Gabriel}}]{Magalhaes_2005}
{Magalh{\~a}es}, A.~M., {Pereyra}, A., {Melgarejo}, R., {et~al.} 2005, in
  Astronomical Society of the Pacific Conference Series, Vol. 343, Astronomical
  Polarimetry: Current Status and Future Directions, ed. A.~{Adamson},
  C.~{Aspin}, C.~{Davis}, \& T.~{Fujiyoshi}, 305

\bibitem[{{Magalh{\~a}es} {et~al.}(2012){Magalh{\~a}es}, {de Oliveira},
  {Carciofi}, {Costa}, {Dal Pino}, {Diaz}, {Ferrari}, {Fernandez}, {Gomes},
  {Marrara}, {Pereyrac}, {Ribeiro}, {Rodrigues}, {Rubinho}, {Seriacopi}, \&
  {Taylor}}]{Magalhaes_2012}
{Magalh{\~a}es}, A.~M., {de Oliveira}, C.~M., {Carciofi}, A., {et~al.} 2012, in
  American Institute of Physics Conference Series, Vol. 1429, Stellar
  Polarimetry: from Birth to Death, ed. J.~L. {Hoffman}, J.~{Bjorkman}, \&
  B.~{Whitney}, 244--247, \dodoi{10.1063/1.3701933}

\bibitem[{{Majewski} {et~al.}(2017){Majewski}, {Schiavon}, {Frinchaboy},
  {Allende Prieto}, {Barkhouser}, {Bizyaev}, {Blank}, {Brunner}, {Burton},
  {Carrera}, {Chojnowski}, {Cunha}, {Epstein}, {Fitzgerald}, {Garc{\'\i}a
  P{\'e}rez}, {Hearty}, {Henderson}, {Holtzman}, {Johnson}, {Lam}, {Lawler},
  {Maseman}, {M{\'e}sz{\'a}ros}, {Nelson}, {Nguyen}, {Nidever}, {Pinsonneault},
  {Shetrone}, {Smee}, {Smith}, {Stolberg}, {Skrutskie}, {Walker}, {Wilson},
  {Zasowski}, {Anders}, {Basu}, {Beland}, {Blanton}, {Bovy}, {Brownstein},
  {Carlberg}, {Chaplin}, {Chiappini}, {Eisenstein}, {Elsworth}, {Feuillet},
  {Fleming}, {Galbraith-Frew}, {Garc{\'\i}a}, {Garc{\'\i}a-Hern{\'a}ndez},
  {Gillespie}, {Girardi}, {Gunn}, {Hasselquist}, {Hayden}, {Hekker}, {Ivans},
  {Kinemuchi}, {Klaene}, {Mahadevan}, {Mathur}, {Mosser}, {Muna}, {Munn},
  {Nichol}, {O'Connell}, {Parejko}, {Robin}, {Rocha-Pinto}, {Schultheis},
  {Serenelli}, {Shane}, {Silva Aguirre}, {Sobeck}, {Thompson}, {Troup},
  {Weinberg}, \& {Zamora}}]{Majewski_2017}
{Majewski}, S.~R., {Schiavon}, R.~P., {Frinchaboy}, P.~M., {et~al.} 2017, \aj,
  154, 94, \dodoi{10.3847/1538-3881/aa784d}

\bibitem[{{Marshall} {et~al.}(2006){Marshall}, {Robin}, {Reyl{\'e}},
  {Schultheis}, \& {Picaud}}]{Marshall_2006}
{Marshall}, D.~J., {Robin}, A.~C., {Reyl{\'e}}, C., {Schultheis}, M., \&
  {Picaud}, S. 2006, \aap, 453, 635, \dodoi{10.1051/0004-6361:20053842}

\bibitem[{{Mathis} {et~al.}(1977){Mathis}, {Rumpl}, \&
  {Nordsieck}}]{Mathis_1977}
{Mathis}, J.~S., {Rumpl}, W., \& {Nordsieck}, K.~H. 1977, \apj, 217, 425,
  \dodoi{10.1086/155591}

\bibitem[{{Naghizadeh-Khouei} \& {Clarke}(1993)}]{Naghizadeh_Khouei_1993}
{Naghizadeh-Khouei}, J., \& {Clarke}, D. 1993, \aap, 274, 968

\bibitem[{{Panopoulou} {et~al.}(2019){Panopoulou}, {Hensley}, {Skalidis},
  {Blinov}, \& {Tassis}}]{Panopoulou_2019}
{Panopoulou}, G.~V., {Hensley}, B.~S., {Skalidis}, R., {Blinov}, D., \&
  {Tassis}, K. 2019, \aap, 624, L8, \dodoi{10.1051/0004-6361/201935266}

\bibitem[{{Pereyra} \& {Magalh{\~a}es}(2007)}]{Pereyra_Magalhaes_2007}
{Pereyra}, A., \& {Magalh{\~a}es}, A.~M. 2007, \apj, 662, 1014,
  \dodoi{10.1086/517906}

\bibitem[{{Planck Collaboration Int. XLVIII} {et~al.}(2016){Planck
  Collaboration Int. XLVIII}, {Aghanim}, {Ashdown}, {Aumont}, {Baccigalupi},
  {Ballardini}, {Banday}, {Barreiro}, {Bartolo}, {Basak}, {Benabed}, {Bernard},
  {Bersanelli}, {Bielewicz}, {Bonavera}, {Bond}, {Borrill}, {Bouchet},
  {Boulanger}, {Burigana}, {Calabrese}, {Cardoso}, {Carron}, {Chiang},
  {Colombo}, {Comis}, {Couchot}, {Coulais}, {Crill}, {Curto}, {Cuttaia}, {de
  Bernardis}, {de Zotti}, {Delabrouille}, {Di Valentino}, {Dickinson}, {Diego},
  {Dor{\'e}}, {Douspis}, {Ducout}, {Dupac}, {Dusini}, {Elsner}, {En{\ss}lin},
  {Eriksen}, {Falgarone}, {Fantaye}, {Finelli}, {Forastieri}, {Frailis},
  {Fraisse}, {Franceschi}, {Frolov}, {Galeotta}, {Galli}, {Ganga},
  {G{\'e}nova-Santos}, {Gerbino}, {Ghosh}, {Giraud-H{\'e}raud},
  {Gonz{\'a}lez-Nuevo}, {G{\'o}rski}, {Gruppuso}, {Gudmundsson}, {Hansen},
  {Helou}, {Henrot-Versill{\'e}}, {Herranz}, {Hivon}, {Huang}, {Jaffe},
  {Jones}, {Keih{\"a}nen}, {Keskitalo}, {Kiiveri}, {Kisner}, {Krachmalnicoff},
  {Kunz}, {Kurki-Suonio}, {Lamarre}, {Langer}, {Lasenby}, {Lattanzi},
  {Lawrence}, {Le Jeune}, {Levrier}, {Lilje}, {Lilley}, {Lindholm},
  {L{\'o}pez-Caniego}, {Ma}, {Mac{\'\i}as-P{\'e}rez}, {Maggio}, {Maino},
  {Mandolesi}, {Mangilli}, {Maris}, {Martin}, {Mart{\'\i}nez-Gonz{\'a}lez},
  {Matarrese}, {Mauri}, {McEwen}, {Melchiorri}, {Mennella}, {Migliaccio},
  {Miville-Desch{\^e}nes}, {Molinari}, {Moneti}, {Montier}, {Morgante}, {Moss},
  {Natoli}, {Oxborrow}, {Pagano}, {Paoletti}, {Patanchon}, {Perdereau},
  {Perotto}, {Pettorino}, {Piacentini}, {Plaszczynski}, {Polastri}, {Polenta},
  {Puget}, {Rachen}, {Racine}, {Reinecke}, {Remazeilles}, {Renzi}, {Rocha},
  {Rosset}, {Rossetti}, {Roudier}, {Rubi{\~n}o-Mart{\'\i}n}, {Ruiz-Granados},
  {Salvati}, {Sandri}, {Savelainen}, {Scott}, {Sirignano}, {Sirri}, {Soler},
  {Spencer}, {Suur-Uski}, {Tauber}, {Tavagnacco}, {Tenti}, {Toffolatti},
  {Tomasi}, {Tristram}, {Trombetti}, {Valiviita}, {Van Tent}, {Vielva},
  {Villa}, {Vittorio}, {Wandelt}, {Wehus}, {Zacchei}, \&
  {Zonca}}]{Planck-Collaboration_2016}
{Planck Collaboration Int. XLVIII}, {Aghanim}, N., {Ashdown}, M., {et~al.}
  2016, \aap, 596, A109, \dodoi{10.1051/0004-6361/201629022}

\bibitem[{{Planck Collaboration Int. XXI} {et~al.}(2015){Planck Collaboration
  Int. XXI}, {Ade}, {Aghanim}, {Alina}, {Aniano}, {Armitage-Caplan}, {Arnaud},
  {Ashdown}, {Atrio-Barandela}, {Aumont}, {Baccigalupi}, {Banday}, {Barreiro},
  {Battaner}, {Beichman}, {Benabed}, {Benoit-L{\'e}vy}, {Bernard},
  {Bersanelli}, {Bielewicz}, {Bock}, {Bond}, {Borrill}, {Bouchet}, {Boulanger},
  {Burigana}, {Cardoso}, {Catalano}, {Chamballu}, {Chary}, {Chiang},
  {Christensen}, {Colombi}, {Colombo}, {Combet}, {Couchot}, {Coulais}, {Crill},
  {Curto}, {Cuttaia}, {Danese}, {Davies}, {Davis}, {de Bernardis}, {de Rosa},
  {de Zotti}, {Delabrouille}, {D{\'e}sert}, {Dickinson}, {Diego}, {Donzelli},
  {Dor{\'e}}, {Douspis}, {Dunkley}, {Dupac}, {Efstathiou}, {En{\ss}lin},
  {Eriksen}, {Falgarone}, {Fanciullo}, {Finelli}, {Forni}, {Frailis},
  {Fraisse}, {Franceschi}, {Galeotta}, {Ganga}, {Ghosh}, {Giard},
  {Giraud-H{\'e}raud}, {Gonz{\'a}lez-Nuevo}, {G{\'o}rski}, {Gregorio},
  {Gruppuso}, {Guillet}, {Hansen}, {Harrison}, {Helou},
  {Hern{\'a}ndez-Monteagudo}, {Hildebrandt}, {Hivon}, {Hobson}, {Holmes},
  {Hornstrup}, {Huffenberger}, {Jaffe}, {Jaffe}, {Jones}, {Juvela},
  {Keih{\"a}nen}, {Keskitalo}, {Kisner}, {Kneissl}, {Knoche}, {Kunz},
  {Kurki-Suonio}, {Lagache}, {L{\"a}hteenm{\"a}ki}, {Lamarre}, {Lasenby},
  {Lawrence}, {Leonardi}, {Levrier}, {Liguori}, {Lilje}, {Linden-V{\o}rnle},
  {L{\'o}pez-Caniego}, {Lubin}, {Mac{\'\i}as-P{\'e}rez}, {Maffei},
  {Magalh{\~a}es}, {Maino}, {Mandolesi}, {Maris}, {Marshall}, {Martin},
  {Mart{\'\i}nez-Gonz{\'a}lez}, {Masi}, {Matarrese}, {Mazzotta}, {Melchiorri},
  {Mendes}, {Mennella}, {Migliaccio}, {Miville-Desch{\^e}nes}, {Moneti},
  {Montier}, {Morgante}, {Mortlock}, {Munshi}, {Murphy}, {Naselsky}, {Nati},
  {Natoli}, {Netterfield}, {Noviello}, {Novikov}, {Novikov}, {Oxborrow},
  {Pagano}, {Pajot}, {Paladini}, {Paoletti}, {Pasian}, {Perdereau}, {Perotto},
  {Perrotta}, {Piacentini}, {Piat}, {Pietrobon}, {Plaszczynski}, {Poidevin},
  {Pointecouteau}, {Polenta}, {Popa}, {Pratt}, {Prunet}, {Puget}, {Rachen},
  {Reach}, {Rebolo}, {Reinecke}, {Remazeilles}, {Renault}, {Ricciardi},
  {Riller}, {Ristorcelli}, {Rocha}, {Rosset}, {Roudier}, {Rusholme}, {Sandri},
  {Savini}, {Scott}, {Spencer}, {Stolyarov}, {Stompor}, {Sudiwala}, {Sutton},
  {Suur-Uski}, {Sygnet}, {Tauber}, {Terenzi}, {Toffolatti}, {Tomasi},
  {Tristram}, {Tucci}, {Umana}, {Valenziano}, {Valiviita}, {Van Tent},
  {Vielva}, {Villa}, {Wade}, {Wandelt}, \& {Zonca}}]{Planck-Collaboration_2015}
{Planck Collaboration Int. XXI}, {Ade}, P.~A.~R., {Aghanim}, N., {et~al.} 2015,
  \aap, 576, A106, \dodoi{10.1051/0004-6361/201424087}

\bibitem[{{Planck Collaboration XII} {et~al.}(2020){Planck Collaboration XII},
  {Aghanim}, {Akrami}, {Alves}, {Ashdown}, {Aumont}, {Baccigalupi},
  {Ballardini}, {Banday}, {Barreiro}, {Bartolo}, {Basak}, {Benabed}, {Bernard},
  {Bersanelli}, {Bielewicz}, {Bock}, {Bond}, {Borrill}, {Bouchet}, {Boulanger},
  {Bracco}, {Bucher}, {Burigana}, {Calabrese}, {Cardoso}, {Carron}, {Chary},
  {Chiang}, {Colombo}, {Combet}, {Crill}, {Cuttaia}, {de Bernardis}, {de
  Zotti}, {Delabrouille}, {Delouis}, {Di Valentino}, {Dickinson}, {Diego},
  {Dor{\'e}}, {Douspis}, {Ducout}, {Dupac}, {Efstathiou}, {Elsner},
  {En{\ss}lin}, {Eriksen}, {Falgarone}, {Fantaye}, {Fernandez-Cobos},
  {Ferri{\`e}re}, {Finelli}, {Forastieri}, {Frailis}, {Fraisse}, {Franceschi},
  {Frolov}, {Galeotta}, {Galli}, {Ganga}, {G{\'e}nova-Santos}, {Gerbino},
  {Ghosh}, {Gonz{\'a}lez-Nuevo}, {G{\'o}rski}, {Gratton}, {Green}, {Gruppuso},
  {Gudmundsson}, {Guillet}, {Handley}, {Hansen}, {Helou}, {Herranz}, {Hivon},
  {Huang}, {Jaffe}, {Jones}, {Keih{\"a}nen}, {Keskitalo}, {Kiiveri}, {Kim},
  {Krachmalnicoff}, {Kunz}, {Kurki-Suonio}, {Lagache}, {Lamarre}, {Lasenby},
  {Lattanzi}, {Lawrence}, {Le Jeune}, {Levrier}, {Liguori}, {Lilje},
  {Lindholm}, {L{\'o}pez-Caniego}, {Lubin}, {Ma}, {Mac{\'\i}as-P{\'e}rez},
  {Maggio}, {Maino}, {Mandolesi}, {Mangilli}, {Marcos-Caballero}, {Maris},
  {Martin}, {Mart{\'\i}nez-Gonz{\'a}lez}, {Matarrese}, {Mauri}, {McEwen},
  {Melchiorri}, {Mennella}, {Migliaccio}, {Miville-Desch{\^e}nes}, {Molinari},
  {Moneti}, {Montier}, {Morgante}, {Moss}, {Natoli}, {Pagano}, {Paoletti},
  {Patanchon}, {Perrotta}, {Pettorino}, {Piacentini}, {Polastri}, {Polenta},
  {Puget}, {Rachen}, {Reinecke}, {Remazeilles}, {Renzi}, {Ristorcelli},
  {Rocha}, {Rosset}, {Roudier}, {Rubi{\~n}o-Mart{\'\i}n}, {Ruiz-Granados},
  {Salvati}, {Sandri}, {Savelainen}, {Scott}, {Sirignano}, {Sunyaev},
  {Suur-Uski}, {Tauber}, {Tavagnacco}, {Tenti}, {Toffolatti}, {Tomasi},
  {Trombetti}, {Valiviita}, {Vansyngel}, {Van Tent}, {Vielva}, {Villa},
  {Vittorio}, {Wandelt}, {Wehus}, {Zacchei}, \&
  {Zonca}}]{Planck-Collaboration_2018_20}
{Planck Collaboration XII}, {Aghanim}, N., {Akrami}, Y., {et~al.} 2020, \aap,
  641, A12, \dodoi{10.1051/0004-6361/201833885}

\bibitem[{{Press} {et~al.}(1992){Press}, {Teukolsky}, {Vetterling}, \&
  {Flannery}}]{Press_FortranStats_1992}
{Press}, W.~H., {Teukolsky}, S.~A., {Vetterling}, W.~T., \& {Flannery}, B.~P.
  1992, {Numerical recipes in FORTRAN. The art of scientific computing}

\bibitem[{{Purcell} {et~al.}(2015){Purcell}, {Gaensler}, {Sun}, {Carretti},
  {Bernardi}, {Haverkorn}, {Kesteven}, {Poppi}, {Schnitzeler}, \&
  {Staveley-Smith}}]{Purcell_2015}
{Purcell}, C.~R., {Gaensler}, B.~M., {Sun}, X.~H., {et~al.} 2015, \apj, 804,
  22, \dodoi{10.1088/0004-637X/804/1/22}

\bibitem[{{Queiroz} {et~al.}(2018){Queiroz}, {Anders}, {Santiago}, {Chiappini},
  {Steinmetz}, {Dal Ponte}, {Stassun}, {da Costa}, {Maia}, {Crestani}, {Beers},
  {Fern{\'a}ndez-Trincado}, {Garc{\'\i}a-Hern{\'a}ndez}, {Roman-Lopes}, \&
  {Zamora}}]{Queiroz_2018}
{Queiroz}, A.~B.~A., {Anders}, F., {Santiago}, B.~X., {et~al.} 2018, \mnras,
  476, 2556, \dodoi{10.1093/mnras/sty330}

\bibitem[{{Ram{\'\i}rez} {et~al.}(2017){Ram{\'\i}rez}, {Magalh{\~a}es},
  {Davidson}, {Pereyra}, \& {Rubinho}}]{Ramirez_2017}
{Ram{\'\i}rez}, E.~A., {Magalh{\~a}es}, A.~M., {Davidson}, James~W., J.,
  {Pereyra}, A., \& {Rubinho}, M. 2017, \pasp, 129, 055001,
  \dodoi{10.1088/1538-3873/aa54a7}

\bibitem[{{Riello} {et~al.}(2021){Riello}, {De Angeli}, {Evans}, {Montegriffo},
  {Carrasco}, {Busso}, {Palaversa}, {Burgess}, {Diener}, {Davidson}, {Rowell},
  {Fabricius}, {Jordi}, {Bellazzini}, {Pancino}, {Harrison}, {Cacciari}, {van
  Leeuwen}, {Hambly}, {Hodgkin}, {Osborne}, {Altavilla}, {Barstow}, {Brown},
  {Castellani}, {Cowell}, {De Luise}, {Gilmore}, {Giuffrida}, {Hidalgo},
  {Holland}, {Marinoni}, {Pagani}, {Piersimoni}, {Pulone}, {Ragaini}, {Rainer},
  {Richards}, {Sanna}, {Walton}, {Weiler}, \& {Yoldas}}]{Riello_2021}
{Riello}, M., {De Angeli}, F., {Evans}, D.~W., {et~al.} 2021, \aap, 649, A3,
  \dodoi{10.1051/0004-6361/202039587}

\bibitem[{{Robin} {et~al.}(2003){Robin}, {Reyl{\'e}}, {Derri{\`e}re}, \&
  {Picaud}}]{Robin_2003}
{Robin}, A.~C., {Reyl{\'e}}, C., {Derri{\`e}re}, S., \& {Picaud}, S. 2003,
  \aap, 409, 523, \dodoi{10.1051/0004-6361:20031117}

\bibitem[{{Rodrigues} {et~al.}(1997){Rodrigues}, {Magalh{\~a}es}, {Coyne}, \&
  {Piirola}}]{Rodrigues_1997}
{Rodrigues}, C.~V., {Magalh{\~a}es}, A.~M., {Coyne}, G.~V., \& {Piirola},
  S.~J.~V. 1997, \apj, 485, 618, \dodoi{10.1086/304434}

\bibitem[{{Rybizki} {et~al.}(2022){Rybizki}, {Green}, {Rix}, {El-Badry},
  {Demleitner}, {Zari}, {Udalski}, {Smart}, \& {Gould}}]{Rybizki_2022}
{Rybizki}, J., {Green}, G.~M., {Rix}, H.-W., {et~al.} 2022, \mnras, 510, 2597,
  \dodoi{10.1093/mnras/stab3588}

\bibitem[{{Savage} \& {Mathis}(1979)}]{Savage_1979}
{Savage}, B.~D., \& {Mathis}, J.~S. 1979, \araa, 17, 73,
  \dodoi{10.1146/annurev.aa.17.090179.000445}

\bibitem[{{Schlafly} {et~al.}(2016){Schlafly}, {Meisner}, {Stutz},
  {Kainulainen}, {Peek}, {Tchernyshyov}, {Rix}, {Finkbeiner}, {Covey}, {Green},
  {Bell}, {Burgett}, {Chambers}, {Draper}, {Flewelling}, {Hodapp}, {Kaiser},
  {Magnier}, {Martin}, {Metcalfe}, {Wainscoat}, \& {Waters}}]{Schlafly_2016}
{Schlafly}, E.~F., {Meisner}, A.~M., {Stutz}, A.~M., {et~al.} 2016, \apj, 821,
  78, \dodoi{10.3847/0004-637X/821/2/78}

\bibitem[{{Serkowski} {et~al.}(1975){Serkowski}, {Mathewson}, \&
  {Ford}}]{Serkowski_1975}
{Serkowski}, K., {Mathewson}, D.~S., \& {Ford}, V.~L. 1975, \apj, 196, 261,
  \dodoi{10.1086/153410}

\bibitem[{{Simmons} \& {Stewart}(1985)}]{Simmons&Stewart_1985}
{Simmons}, J.~F.~L., \& {Stewart}, B.~G. 1985, \aap, 142, 100

\bibitem[{{Skalidis} {et~al.}(2018){Skalidis}, {Panopoulou}, {Tassis},
  {Pavlidou}, {Blinov}, {Komis}, \& {Liodakis}}]{Skalidis_2018}
{Skalidis}, R., {Panopoulou}, G.~V., {Tassis}, K., {et~al.} 2018, \aap, 616,
  A52, \dodoi{10.1051/0004-6361/201832827}

\bibitem[{{Skrutskie} {et~al.}(2006){Skrutskie}, {Cutri}, {Stiening},
  {Weinberg}, {Schneider}, {Carpenter}, {Beichman}, {Capps}, {Chester},
  {Elias}, {Huchra}, {Liebert}, {Lonsdale}, {Monet}, {Price}, {Seitzer},
  {Jarrett}, {Kirkpatrick}, {Gizis}, {Howard}, {Evans}, {Fowler}, {Fullmer},
  {Hurt}, {Light}, {Kopan}, {Marsh}, {McCallon}, {Tam}, {Van Dyk}, \&
  {Wheelock}}]{Skrutskie_2MASS_2006}
{Skrutskie}, M.~F., {Cutri}, R.~M., {Stiening}, R., {et~al.} 2006, \aj, 131,
  1163, \dodoi{10.1086/498708}

\bibitem[{{Tassis} {et~al.}(2018){Tassis}, {Ramaprakash}, {Readhead}, {Potter},
  {Wehus}, {Panopoulou}, {Blinov}, {Eriksen}, {Hensley}, {Karakci},
  {Kypriotakis}, {Maharana}, {Ntormousi}, {Pavlidou}, {Pearson}, \&
  {Skalidis}}]{Tassis_2018}
{Tassis}, K., {Ramaprakash}, A.~N., {Readhead}, A. C.~S., {et~al.} 2018, arXiv
  e-prints, arXiv:1810.05652.
\newblock \doarXiv{1810.05652}

\bibitem[{{Taylor}(2005)}]{Taylor_2005}
{Taylor}, M.~B. 2005, in Astronomical Society of the Pacific Conference Series,
  Vol. 347, Astronomical Data Analysis Software and Systems XIV, ed.
  P.~{Shopbell}, M.~{Britton}, \& R.~{Ebert}, 29

\bibitem[{{Tody}(1986)}]{Tody_IRAF1_1986}
{Tody}, D. 1986, in Society of Photo-Optical Instrumentation Engineers (SPIE)
  Conference Series, Vol. 627, Instrumentation in astronomy VI, ed. D.~L.
  {Crawford}, 733, \dodoi{10.1117/12.968154}

\bibitem[{{Tody}(1993)}]{Tody_IRAF2_1993}
{Tody}, D. 1993, in Astronomical Society of the Pacific Conference Series,
  Vol.~52, Astronomical Data Analysis Software and Systems II, ed. R.~J.
  {Hanisch}, R.~J.~V. {Brissenden}, \& J.~{Barnes}, 173

\bibitem[{{Vall{\'e}e}(2015)}]{vallee_2015}
{Vall{\'e}e}, J.~P. 2015, \mnras, 450, 4277, \dodoi{10.1093/mnras/stv862}

\bibitem[{{Vergely} {et~al.}(2022){Vergely}, {Lallement}, \&
  {Cox}}]{Vergely_2022}
{Vergely}, J.~L., {Lallement}, R., \& {Cox}, N.~L.~J. 2022, \aap, 664, A174,
  \dodoi{10.1051/0004-6361/202243319}

\bibitem[{{Versteeg} {et~al.}(2023){Versteeg}, {Magalh{\~a}es}, {Haverkorn},
  {Angarita}, {Rodrigues}, {Santos-Lima}, \& {Kawabata}}]{Versteeg_2023}
{Versteeg}, M.~J.~F., {Magalh{\~a}es}, A.~M., {Haverkorn}, M., {et~al.} 2023,
  \aj, 165, 87, \dodoi{10.3847/1538-3881/aca8fd}

\bibitem[{{Virtanen} {et~al.}(2020){Virtanen}, Gommers, Oliphant, Haberland,
  Reddy, Cournapeau, Burovski, Peterson, Weckesser, Bright, {van der Walt},
  Brett, Wilson, Millman, Mayorov, Nelson, Jones, Kern, Larson, Carey, Polat,
  Feng, Moore, {VanderPlas}, Laxalde, Perktold, Cimrman, Henriksen, Quintero,
  Harris, Archibald, Ribeiro, Pedregosa, {van Mulbregt}, \& {SciPy 1.0
  Contributors}}]{Virtanen_SciPy_2020}
{Virtanen}, P., Gommers, R., Oliphant, T.~E., {et~al.} 2020, Nature Methods,
  17, 261, \dodoi{10.1038/s41592-019-0686-2}

\bibitem[{{Wang} \& {Chen}(2019)}]{Wang_2019}
{Wang}, S., \& {Chen}, X. 2019, \apj, 877, 116,
  \dodoi{10.3847/1538-4357/ab1c61}

\bibitem[{{Whittet} {et~al.}(2008){Whittet}, {Hough}, {Lazarian}, \&
  {Hoang}}]{Whittet_2008}
{Whittet}, D.~C.~B., {Hough}, J.~H., {Lazarian}, A., \& {Hoang}, T. 2008, \apj,
  674, 304, \dodoi{10.1086/525040}

\bibitem[{Wiegand(1968)}]{Wiegand_1968}
Wiegand, H. 1968, Biometrische Zeitschrift, 10, 88,
  \dodoi{https://doi.org/10.1002/bimj.19680100122}

\end{thebibliography}
\bibliographystyle{aasjournal}

\end{document}